\documentclass[usenatbib,usegraphicx]{mn2e}
\usepackage[fleqn]{amsmath}
\usepackage[charter]{mathdesign}

\usepackage{url}
\usepackage[hyperfootnotes, hyperfigures,bookmarks, pdftex]{hyperref}
\hypersetup{bookmarksnumbered, bookmarksopen, colorlinks, citecolor=blue, pdfduplex=DuplexFlipLongEdge}
\hypersetup{pdftitle=Turbulence in Thin Magnetized Disks, pdfauthor=Kris Beckwith, pdfsubject=Astrophysics}
\usepackage{bookmark}
\providecommand{\eprint}[1]{\href{http://arxiv.org/abs/#1}{#1}}
\providecommand{\adsurl}[1]{\href{#1}{ADS}}

\pagerange{\pageref{firstpage}--\pageref{lastpage}} \pubyear{2004}

\def\LaTeX{L\kern-.36em\raise.3ex\hbox{a}\kern-.15em
T\kern-.1667em\lower.7ex\hbox{E}\kern-.125emX}

\newcommand\araa{{ARA\&A}}%
\newcommand\apjs{{ApJS}}%
\newcommand\aapr{{A\&A~Rev.}}%
\newcommand\nar{{New A Rev.}}%
\title[Turbulence in Magnetized Thin Disks]{Turbulence in Global Simulations of  Magnetized Thin Accretion Disks}

\author[Beckwith et al.]
{Kris Beckwith$^{1 \star}$,
Philip J. Armitage$^{1 \dagger}$,
Jacob B. Simon$^{1 \ast}$
\\ $^1$ JILA,
University of Colorado at Boulder,
440 UCB,
Boulder, CO 80309-0440
\\Email: $^\star$ kris.beckwith@jila.colorado.edu,
$^\dagger$ pja@jilau1.colorado.edu,
$^\ast$ jbsimon@jila.colorado.edu}

\begin{document}

\label{firstpage}

\maketitle

%\author{Kris Beckwith, Philip J. Armitage and Jacob B. Simon}
%\affil{JILA\\
%University of Colorado at Boulder\\
%440 UCB\\
%Boulder, CO 80309-0440}
%
%\email{kris.beckwith@jila.colorado.edu; pja@jilau1.colorado.edu; jbsimon@jila.colorado.edu}

\begin{abstract} 
We use a global magnetohydrodynamic simulation of a geometrically thin accretion 
disk to investigate the locality and detailed structure of turbulence driven by the magnetorotational 
instability (MRI). The model disk has an aspect ratio $H / R \simeq 0.07$, and is computed using a higher-order 
Godunov MHD scheme with accurate fluxes. We focus the analysis on late times after the system has lost direct memory 
of its initial magnetic flux state. The disk enters a saturated turbulent state in which the fastest growing modes of the 
MRI are well-resolved, with a relatively high efficiency of angular momentum transport 
$\langle \langle \alpha \rangle \rangle \approx 2.5 \times 10^{-2}$. The accretion stress peaks at the disk 
midplane, above and below which exists a moderately magnetized corona with patches of 
superthermal field. By analyzing the spatial and temporal correlations of the turbulent fields, 
we find that the spatial structure of the magnetic and kinetic energy is moderately well-localized 
(with correlation lengths along the major axis of $2.5H$ and $1.5H$ respectively), and generally consistent with that 
expected from homogenous incompressible turbulence. The density field, conversely, exhibits 
both a longer correlation length and a long correlation time, results which we ascribe to the 
importance of spiral density waves within the flow. Consistent with prior results, we show that the 
mean local stress displays a well-defined correlation with the local vertical flux, and that this relation 
is apparently causal (in the sense of the flux stimulating the stress) during portions of a global 
dynamo cycle. We argue that the observed flux-stress relation supports dynamo models 
in which the structure of coronal magnetic fields plays a central role in determining the 
dynamics of thin-disk accretion.
\end{abstract}

\begin{keywords}
{accretion, accretion discs  - (magnetohydrodynamics) MHD - instabilities}
\end{keywords}

\section{Introduction}\label{intro}
Turbulence associated with the non-linear evolution of the magnetorotational 
instability \citep[MRI:][]{Balbus:1991,Balbus:1998} provides the dominant source of angular 
momentum transport in ionized accretion flows.
This turbulence has been explored using numerical simulations in both the local
approximation, in which a small patch of accretion disk is considered, and the global case, where a large portion of the disk is evolved.   All of these studies have revealed that the MRI
leads to sustained, turbulent angular momentum transport in the outward radial direction.

Despite this progress, basic questions about the physics of the MRI
 and the nature of turbulence within accretion disks remain unanswered. 
The initial numerical work on the MRI focused on local shearing box 
simulations within domains that were a small multiple of the disk 
scale height $H$ \citep{Hawley:1995,Brandenburg:1995,Stone:1996} and
were carried out at moderate resolutions in which the dissipation scale was generally much
larger than in actual disks.  These experiments left unresolved the issue of whether physics on 
the smallest or largest scales is important. On 
small scales, the extent to which the plasma's viscosity and resistivity 
(or their ratio, the magnetic Prandtl number) influence the large-scale 
dynamics of accretion is unclear. Resistivity -- and other non-ideal MHD effects -- 
are indubitably important in weakly ionized protoplanetary disks \citep{Gammie:1996,Simon:2010}, 
and there is some evidence that a dependence on the 
Prandtl number persists even when the dissipative 
scales are much smaller than the disk scale height \citep{Fromang:2007,Lesur:2007,Simon:2009a,Longaretti:2010}.

On larger scales, a critical question is whether global effects qualitatively alter the 
dynamics of accretion mediated by the MRI. It is not obvious that they should. The 
MRI is a local instability, and for geometrically thin accretion disks (with $H / R \ll 1$) 
the fastest growing linear modes are not of global scale. (We exclude here 
radiatively inefficient flows with $H \sim R$, for which the importance of global 
effects is self-evident.) Nonetheless, at least three distinct avenues via which 
global dynamics might affect the evolution of thin magnetized accretion disks have been 
proposed. First, it is possible that long-wavelength fluid or magnetic modes, 
associated with the non-linear state of the MRI itself, could be important in 
determining either the saturated level of the stress, or some other property of 
the accretion flow. Existing numerical investigations into this issue have yielded mixed results. 
\citet{Guan:2009a}, using local simulations, showed that the primary fluid 
variables (${\bf B}$, ${\bf v}$ and $\rho$) de-correlated on a scale 
that was significantly smaller than $H$, supporting a fundamentally 
local picture of MRI-driven turbulence. \citet{Nelson:2010}, on the other hand, 
found that some secondary properties of the turbulence (specifically, the 
amplitude of density fluctuations) converged only for domains whose size -- 
in their simulations $4H \times 16H \times 2H$ -- approached global scales. 
Second, additional instabilities that are not present (or only slowly growing) 
on small scales could become important globally. \citet{Tout:1992}, for example, 
proposed a semi-analytic dynamo model in which the Parker instability played a 
key role. Since the Parker instability grows most rapidly on scales $\lambda \gg H$, 
an implication of their model is that a global simulation of a thin disk ought to 
display quite different dynamics from a local simulation of an otherwise identical 
system. Finally, even if the turbulence originating from the MRI is localized, the 
strength of that turbulence is known to depend upon 
the flux threading the disk \citep{Hawley:1995,Sano:2004}, which is a function of the 
boundary conditions for local simulations. Determining self-consistently 
the distribution of that flux (in the case where the disk as a whole is {\em not} 
threaded by a significant global magnetic field), requires either large local \citep{Guan:2010} 
or global simulations. If the local flux is sufficiently strong, it is possible to envisage 
a limit in which the accretion stress is ultimately determined by the strength 
and connectivity of magnetic fields in the disk corona \citep{Tout:1996,Uzdensky:2008}. 
There is some numerical evidence in favor of such a coronally-anchored dynamo \citep{Sorathia:2010}.

Prior global simulations have attempted to address some of the above questions 
\citep{Armitage:1998,Hawley:2000,Hawley:2001b,Fromang:2006,Flock:2009,Sorathia:2010,ONeill:2010,Flock:2011}, 
but it remains computationally demanding to run global simulations at a resolution high enough 
to allow fair comparison with local calculations. Many uncertainties thus remain. In 
this paper, we present a detailed analysis of the structure of the MHD turbulence 
realized in a global simulation of a geometrically thin disk. Compared to most prior work, the simulation we use has a moderately high resolution on the poloidal plane, but more importantly makes use of a higher-order Godunov scheme that is more accurate on test problems at Þxed spatial resolution. The same numerical scheme is employed in the recent global simulations of \cite{Flock:2011}, but here we use initial conditions designed so that the saturated state is easier to resolve adequately. Our calculations thus represent a step toward the elusive goal of higher accuracy, converged simulations of thin accretion disks \citep{Hawley:2011}.  Furthermore, 
we analyze the turbulence using methods closely related to those employed in recent local simulations, thereby facilitating 
as close a comparison as possible.  Our goal, then, is to elucidate the behavior of MRI-driven turbulence across
a large range in spatial and temporal scales and make a connection between local and global calculations.

The remainder of this work is organized as follows. In \S\ref{numerics} and \S\ref{diagnostics}, we give the details of the method that we use to integrate the equations of ideal Magnetohydrodynamics, along with information about the setup of the simulation presented here and how the simulation data was reduced and analyzed. In \S\ref{evolve} and \S\ref{disk} we describe the evolution of the disk, along with global measures and the spatial structure of the turbulent steady state.  In \S\ref{turbulence} and \S\ref{flux_stress_relation}, we examine the spatial and temporal structure of the turbulence and whether or not a local relationship between vertical flux and the magnetic accretion stress is at work in the simulation. Finally in \S\ref{conclusion}, we summarize our results and point the way to future work.
\section{Numerical Details}\label{numerics}

\subsection{Integration Scheme}

We use the \texttt{Pluto} code for computational astrophysics \citep{Mignone:2007} to simulate the evolution of 
a geometrically thin accretion disk. \texttt{Pluto} implements a higher-order Godunov scheme to integrate general systems of conservation laws of the form:
\begin{equation}
\frac{\partial \mathbf{U}}{\partial t} + \nabla \cdot \mathbf{F}(\mathbf{U})
= \mathbf{S}(\mathbf{U})
\end{equation}
where $\mathbf{U}$ is a vector of conserved variables, $\mathbf{F}(\mathbf{U})$ is a second rank tensor of fluxes derived from the conserved variables and $\mathbf{S}(\mathbf{U})$ are source terms. We utilize \texttt{Pluto} to integrate the equations of ideal MHD written in the Newtonian limit in spherical coordinates $(r,\theta,\phi)$:
\begin{equation}
\begin{split}
\frac{\partial \rho}{\partial t} + \nabla \cdot \left(\rho \mathbf{v} \right) = 0 \;\; ; \;\;
\frac{\partial s}{\partial t} + \nabla \cdot \left(s \mathbf{v} \right) = 0 \\
\frac{\partial \rho \mathbf{v}}{\partial t} +
\nabla \cdot \left( \rho\mathbf{v}\mathbf{v} - \mathbf{B}\mathbf{B} +P^{*}\right)
=-\rho \nabla \Phi + \mathbf{G}\\
\frac{\partial E}{\partial t} +
\nabla \cdot \left[ \left(E + P \right) \mathbf{v} - 
\left( \mathbf{v} \cdot \mathbf{B} \right) \mathbf{B} \right] =
-\rho \mathbf{v} \cdot \nabla \Phi \\
\frac{\partial \mathbf{B}}{\partial t} +
\nabla \cdot \left(\mathbf{v} \mathbf{B} - \mathbf{B} \mathbf{v} \right) = 0
\end{split}
\end{equation}
where $\rho$ is the density, $s = P_g / \rho^\gamma$ is the entropy, $\mathbf{v}$ is the velocity, $\mathbf{B}$ is the magnetic field, $E = P_g / (\gamma - 1) + \rho |\mathbf{v}|^2 + |\mathbf{B}|^2 / 2$ is the total energy, $P^{*}$ is a diagonal tensor, the non-zero components of which are the total pressure, $P = P_g + |\mathbf{B}|^2 / 2$, $P_g$ is the gas pressure, $\gamma = 5/3$ is the ratio of specific heats for an ideal gas equation of state, $\Phi$ is the gravitational potential and $\mathbf{G}$ is a tensor containing geometric source terms appropriate for spherical polar coordinates \cite[the precise form of which can be found in][]{Mignone:2007}.

Most geometrically thin accretion disks are optically thick, and require radiation MHD simulations for a fully consistent numerical treatment \citep[e.g.][]{Hirose:2009}. Since our goal here is to study the structure of disk turbulence itself -- rather than any likely more subtle coupling between the disk's dynamics and thermodynamics -- we adopt an approximate treatment of the energy equation that is designed to maintain a nearly isentropic evolution of the disk. We integrate {\em both} an entropy equation, which we use to determine the pressure in regions of smooth flow, and a total energy equation, which is used at shocks. Physically, this means that we include energy dissipation and entropy generation at shocks (which are however negligible for the run presented here), whereas we discard magnetic or kinetic energy that is dissipated elsewhere. 

The \texttt{Pluto} code has been extensively tested for global simulations of the magnetorotational instability by \cite{Flock:2009}. These authors report that in order to accurately reproduce the linear growth stage of the instability, a particular combination of algorithmic choices is necessary. Specifically, \cite{Flock:2009} recommend the use of second order Runge-Kutta time-integration, second order spatial reconstruction, HLLD or Roe-type Riemann solvers and the upwind-constrained transport `contact' method of \cite{Gardiner:2005} for calculation of the electromotive force's (EMF's) for the induction equation. Failure to follow these recommendations can lead to the presence of instabilities within the evolution.
For these reasons, we follow the algorithmic recommendations of \cite{Flock:2009} in the simulations presented here, choosing to utilize the HLLD Riemann solver for reasons of robustness and computational efficiency.

\subsection{Initial Conditions, Computational Grid \& Boundary Conditions}

The physical inner boundary condition for thin accretion disks can be variously a stellar boundary layer, a stellar 
magnetosphere \citep{Pringle:1972}, or the innermost stable orbit (ISCO) of a relativistic potential. Although in 
principle there are qualitative differences between these systems (for example, the resonance between the 
angular and epicyclic frequencies of a Newtonian potential is not present for a black hole), the basic properties 
of disk turbulence at larger radii are likely to be independent of the inner boundary condition. For simplicity, 
we here use the Pseudo-Newtonian potential due to \cite{Paczynsky:1980}:
\begin{equation}
\Phi = -\frac{1}{r-1}
\end{equation}
This potential is a Newtonian approximation to the gravitational potential of a Schwarzschild black hole, whose
key property is the existence of an ISCO at $r_{ISCO} = 3 r_S$, where $r_S = 2GM/c^2$ is the Schwarzschild radius. Circular orbits within this potential have a specific angular momentum $\ell$ given by,
\begin{equation}
\ell_{kep} = \frac{r \sqrt{r}}{r-1}.
\end{equation}
Outside of $r_{ISCO}$, orbits converge 
toward the standard Keplerian form, $\ell_{kep} = \sqrt{r}$ and so we expect that outside of $r_{ISCO}$, thin accretion disks (i.e. those with $c_s << r \Omega$ where $c_s$ is the gas sound speed) will have fluid elements on Keplerian orbits with $\ell_{kep} = \sqrt{r}$ even for the Pseudo-Newtonian potential given above. Inside of $r_{ISCO}$ no stable circular orbits exist and the fluid plunges into the black hole. Provided that the inner radial boundary is placed deep enough within the potential, the fluid will become supersonic in the radial direction before reaching the inner boundary, and the exterior properties of the flow are independent of the precise details 
of the inner radial boundary \citep{McKinney:2002}. For our simulation, this is accomplished by placing the inner radial boundary at $r=1.5r_S$.

The initial conditions for these simulations utilize a simple configuration, corresponding to a disk with constant $H/R$. The density is initialized to some constant, $\rho_0$ in the midplane between $5r_S \le r \le 15 r_S$ with a standard gaussian distribution in the vertical direction:
\begin{equation}
\rho(Z) = \rho_0 e^{-\frac{Z^2}{2H^2}}
\end{equation}
where $Z = r \cos \theta$ and $H = c_s / \Omega$ is the disk scale height.
The sound speed is specified by the power law:
\begin{equation}
c_s = c_0 \sqrt{\frac{R_0}{R}}
\end{equation}
where $R=r \sin \theta$, $c_0$ is chosen to give the required value of $H/R$ for the disk (here $H/R=0.07$) and $R_0$ is the disk inner edge. The initial pressure distribution is determined by assuming that the disk is isothermal such that $P = c^2_s \rho$. The angular momentum distribution is specified by $\ell_{kep}$. Finally the magnetic field is specified to be toroidal, $B_\phi = cons.$ within the region $7.5r_S \le r \le 12.5 r_S$ and $|Z| \le H$. $B_\phi$ is normalized such that $\left< \beta \right> = 2 \left<P_g\right> / \left<|\mathbf{B}|^2\right> = 10$ (where $\left< Q \right>$ denotes a volume averaged quantity), as was used in \cite{Beckwith:2008a}. Note that the imposition of a moderately strong magnetic field on the hydrodynamic state disrupts any initial hydrostatic equilibrium that we could hope to establish; it is for this reason that we choose such a simple hydrodynamic configuration initially as the disk rapidly relaxes to a new state at the beginning of the evolution. Finally, in order to seed the MRI, random perturbations are applied to the initial pressure distribution at an amplitude of $10\%$.

The simulation presented here utilized $288\times128\times96$ zones in $(r,\theta,\phi)$ covering a domain $1.5r_S \le r \le 20 r_S$, $-5 \le Z/H \le 5$ and $0 \le \phi \le \pi/2$. 
We use a logarithmic grading for the mesh in the radial direction and apply strict outflow boundary conditions at both the inner and outer boundaries (i.e. no fluid or magnetic quantities are allowed to enter the domain). In the vertical direction, we locate the boundaries at $\pm5H$ so that we obtain stresses that are independent of the vertical domain size \citep{Sorathia:2010,Simon:2010}.
The grading in the vertical direction is designed such that we have $32$-zones per $H$ in the region cover $Z\le|H|$ i.e. $64$-zones covering the region $-H\le Z \le H$, with the remaining $64$ zones covering the region $Z> H$. We apply periodic boundary conditions in the vertical direction; whilst these are physically unrealistic for the stratified global simulation that we present here, they have two advantages, namely that they guarantee that magnetic flux cannot enter or leave the domain through the vertical boundaries \citep{Davis:2010} and that they help to eliminate numerical difficulties in low density regions close to the vertical boundary \citep{Reynolds:2008}.
Finally, the use of a quarter-wedge in the $\phi$ domain, significantly reduces the computational cost of the calculation, whilst allowing us to capture the dominant non-axisymmetric modes within the flow \citep{Hawley:2001b}. The grading in this direction is uniform and we apply periodic boundary conditions at $\phi=0,\pi/2$.

%\newpage

\section{Evolution Diagnostics}\label{diagnostics}

\subsection{Reduction and Analysis of Simulation Data}\label{reduce}

Past experience \cite[e.g.][]{Beckwith:2009} shows that it is essential to use high time resolution data in interpreting simulations of magnetized accretion disks. For the simulation presented here, we performed complete three-dimensional data dumps twenty times per orbital period at the ISCO. 
We describe here the procedure for extracting physically relevant quantities from this data. The \texttt{IDL} routines used are available from the authors on request.

The first step of our procedure is to convert the raw data output from the simulation (which is in binary format) into a portable format (specifically, HDF5) that can be queried on any machine \emph{without} prior knowledge of the structure of the data file. We regard such a procedure as essential for archival purposes. 
Quantities that we include in the HDF5 data set are: $\rho, P_g, |\mathbf{B}|^2, \mathbf{V}, \mathbf{B}$. 
HDF5 has the advantage that it can be read by many data analysis and visualization applications, and it is also possible to query individual chunks of data within a given file, rather than being restricted to reading the entire file, as is the case (for example) with raw binary data.

The next step in the data reduction pipeline is to compute diagnostics that provide insight into the state of the magnetized accretion flow described by the simulation. In past works, we have found it useful to consider volume-integrated quantities, $\left<Q(t)\right>$, shell-integrated radial profiles, $\left<Q(r,t)\right>$ and two-dimensional distributions on the $r-\phi$ and $r-\theta$ planes, $\left<Q(r,\phi,t)\right>$ and $\left<Q(r,\theta,t)\right>$, respectively. These are defined through:
\begin{equation}
\label{int_data}
\begin{split}
\left< Q(t) \right> = \int Q(r,\theta,\phi,t) \; r^2 \sin \theta dr d\theta d\phi \\
\left< Q(r,t) \right> = \int Q(r,\theta,\phi,t) \; r^2 \sin \theta d\theta d\phi \\
\left< Q(r,\phi,t) \right> = \int Q(r,\theta,\phi,t) \; r \sin \theta d\theta \\
\left< Q(r,\theta,t) \right> = \int Q(r,\theta,\phi,t) \; r d\phi
\end{split}
\end{equation}
For each quantity $Q$, we compute each of these diagnostics and store them as time-series, again in HDF5 format. In the cases where vertical integrals are performed,  $\left<Q(r,t)\right>$ and $\left<Q(r,\phi,t)\right>$, we compute the vertical integral over three different ranges,
\begin{eqnarray}
 & \vert Z  \vert & <  5 H \nonumber \\
 & \vert Z \vert & < H \\
 H < & \vert Z  \vert & < 5 H \nonumber.
\end{eqnarray} 
We describe these as, respectively, the ``disk+corona", the ``disk body", and the ``corona". Physical motivation for these 
distinctions is given below, although we note that what we call the corona is defined solely in geometric terms, rather 
than via a cut on the plasma $\beta$. 
We proceed similarly for the two-dimensional distributions of the quantity, $Q$ on the $r-\phi$ plane, in this case omitting the integral over $\phi$.  We also consider the perturbations on a quantity, $\delta Q$, which we calculate through:
\begin{equation}
\delta Q(r,\theta,\phi,t) \equiv Q(r,\theta,\phi,t) - \frac{\left< Q(r,\theta,t) \right>}{\int r d\phi}
\end{equation}
with this definition, we have that $\left< \delta Q(r,\theta,t) \right> = 0$. For the purposes of our discussion in this work, we have found it useful to compute these diagnostics for the same quantities as the full three-dimensional data dumps, $\rho, P_g, |\mathbf{B}|^2, \mathbf{|V|}, \mathbf{|B|}$ (with vector quantities stored in component form) along with measures such as the mass-accretion rate $\rho V^r$, the angular momentum $\rho \ell$, the Alfven velocity $v_A$, the sound speed $c_s$, along with the Maxwell and (perturbed) Reynolds stresses.

From these time-histories, it is easy to compute time- and spatially-averaged profiles of interesting quantities. We have found the time-averaged and volume-integral of a quantity, $\left< \left< Q \right> \right>$, the time- and shell-averaged radial profile, $\left< \left< Q(r) \right> \right>$ and the time- and disk surface area-averaged vertical profile, $\left< \left< Q(\theta) \right> \right>$ particularly useful, where:
\begin{equation}
\begin{split}
\left< \left< Q \right> \right> =  \frac{1}{\Delta T}
{\int^{T_2}_{T_1} \left< Q(t) \right> dt} \\
\left< \left< Q(r) \right> \right> = \frac{1}{\Delta T}\frac{\int^{T_2}_{T_1} \left< Q(r,t) \right> dt}
{\int_\phi \int _\theta r^2 \sin \theta d\theta d\phi} \\
\left< \left< Q(\theta) \right> \right> = \frac{1}{\Delta T}
\frac{\int^{T_2}_{T_1} \int^{r_2}_{r_1} \left< Q(r,\theta,t) \right> dr dt}
{\int_{\phi} \int^{r_2}_{r_1} r dr d\phi}
\end{split}
\end{equation}
where $\Delta T = T_2 - T_1$ is the time-interval over which to compute the time-average.

\subsection{Calculation of Power Spectra and Correlation Functions}\label{power_spectra}

We calculate the time-averaged toroidal power spectrum, $\left< \left< |P_{Q}(r,k_\phi)|^2 \right> \right>$:
\begin{equation}
\begin{split}
\left< \left| \left< P_{Q}(r,k_\phi) \right> \right|^2 \right> = \\
\frac{1}{\Delta T} \int^{T_2}_{T_1} \left| \int \left<\delta Q(r,\phi,t)\right> e^{- i k_\phi \phi} d\phi \right|^2 dt
\end{split}
\end{equation}
where $\delta Q(r,\phi,t) =  \left( Q(r,\phi,t) - \left<Q(r,t)\right> \right) / \left<Q(r,t)\right>$. We will often plot this quantity in terms of $k_y = 2 \pi n / \Delta y$ where $\Delta y/H = (H/R)^{-1}$. This power spectrum can be radially-averaged as
\begin{equation}
\left< \left< \left| \left< P_{Q}(k_\phi) \right> \right|^2 \right> \right> =
\frac{1}{\Delta R} \int^{r_2}_{r_1} \left< \left| \left< P_{Q}(r,k_\phi) \right> \right|^2 \right> dr
\end{equation}
where $\Delta R = r_2 - r_1$. Finally, we can use $\left< \left| \left< P_{Q}(r,k_\phi) \right> \right|^2 \right>$ to probe the ratio of power on different spatial scales by computing
\begin{equation}
\label{power_ratio}
\begin{split}
\frac{\left< \left< Q_{k_\phi \ge k_1} (r) \right> \right>}
{\left< \left< Q_{k_\phi < k_1} (r) \right> \right>}
 = \\
 \frac{1}{\Delta T} \int^{T_2}_{T_1} \frac{ \int^{k_{max}}_{k_1}
 \left<\delta Q(r,k_\phi,t)\right> e^{i k_\phi \phi} dk_\phi}
 { \int^{k_{1}}_{k_0}
 \left<\delta Q(r,k_\phi,t)\right> e^{i k_\phi \phi} dk_\phi} dt
 \end{split}
\end{equation}
where $k_{max}$ is the Nyquist critical frequency, $k_{0} = 2/\pi$ is the largest scale in the toroidal domain, $k_1$ is some (specified) break frequency and $\left<\delta Q(r,k_\phi,t)\right> = \int \left<\delta Q(r,\phi,t)\right> e^{- i k_\phi \phi} d\phi$.

The calculation of power spectra and correlation functions from global simulations is well-defined, but their interpretation is complicated by the presence of radial gradients that vary with time. In particular, if we simply took the Fourier Transform of a quantity in the radial direction without accounting for these gradients, then we would essentially be performing a convolution of a flat-window function with a non-periodic function. As a result, the radial power spectrum would contain significant small scale power in order to account for radial gradients. These considerations would then can cause problems when it comes to comparing the results of global simulations against local results. We therefore make use of a well-defined (but non-unique) formalism that allows us to remove (time-varying) gradients on large radial (and in some cases vertical) scales.

In developing this formalism, we have had to make several compromises due to the limited computational resources available for performing this analysis. Ideally, many of these diagnostics should be computed using four-dimensional Fourier Transforms (three spatial dimensions + 1 time dimension) on individual vector components. However, since the typical dimensions of the data set to be Fourier transformed are $100^3 \times 1000$, this is impractical both in terms of the memory required to store the data set and the time required to compute the Fourier transforms. Instead, we have had to utilize spatially-averaged scalar data (e.g. $|\mathbf{B}|^2$) in these calculations and restrict our consideration to correlation functions that either contain information about three-spatial dimensions, or two-spatial dimensions and time.

The first set of power spectra utilized in the discussion of \S\ref{turbulence} are two-dimensional Fourier transforms of the magnetic and kinetic energy densities on the poloidal plane. These are calculated from toroidally-averaged data, $\left< Q(r,\theta,t) \right>$, weighted by the area element $r \sin \theta dr d\theta$. We denote the \emph{volume} weighted quantity resulting from this procedure by $\left< Q_V (r,\theta,t) \right>$. In the discussion of \S\ref{turbulence}, we compute two-dimensional power spectra of this data over some restricted range in radius, $r_1 \le r \le r_2$ and the entire vertical domain, $-5H \le Z \le 5H$. Since the simulation presented here utilizes a graded mesh in $(r,\theta)$, it is necessary to first map data to a uniform mesh, which is defined over the range $r_1 \le r \le r_2$, $-5H \le Z \le 5H$. The coordinates spacings, $(\Delta r)_{reg}$, $(\Delta z)_{reg}$ for the uniform mesh are set so that $(\Delta r)_{reg} \le (\Delta r)_{min}$, $(\Delta z)_{reg} \le (\Delta z)_{min}$ where $(\Delta r)_{min}$, $(\Delta z)_{min}$ are the smallest coordinate spacings in the simulation mesh in the range $r_1 \le r \le r_2$, $-5H \le Z \le 5H$ . We also require that there is an even number of cells in each dimension on the uniform mesh in order to minimize the time taken by the Fourier transform procedure. To map the simulation data onto the uniform mesh, we construct a Delaunay triangulation of the $(r,\theta)$ mesh used in the simulation and then use this triangulation to perform a linear interpolation of simulation data onto the uniform mesh. We denote the data mapped onto the uniform mesh via this procedure $\left< Q_V (r,\theta,t) \right>_{reg}$.

The next step in this procedure is to obtain a normalized measure of the spatial fluctuations in $\left< Q_V (r,\theta,t) \right>_{reg}$:
\begin{equation}
{\cal Q}(r,\theta,t) = \frac{\left< Q_V (r,\theta,t) \right>_{reg}- \left<\left< Q_V (r,\theta,t) \right>_{reg} \right>}{\left<\left< Q_V (r,\theta,t) \right>_{reg} \right>}
\end{equation}
This step is essential as it accounts for both radial and vertical gradients in $\left< Q_V (r,\theta,t) \right>_{reg}$. Without this element of the procedure, application of a Fourier Transform in the radial direction would result in significant enhancements to small scale power in order to account for aperiodicity in these coordinates. \cite{Schnittman:2006} describe a procedure to accomplish this renormalization where the mean at each radius is substracted and then the fluctuations are normalized so that $\int{\cal Q}^{2}(r,\theta,t) d\phi = 1$. This approach to renormalization of the fluctuations has the disadvantage that it removes power from all radial length scales. Our approach is to instead calculate $\left<\left< Q_V (r,\theta,t) \right>_{reg} \right>$ by constructing a two-dimensional polynomial approximation to $\left< Q_V (r,\theta,t) \right>_{reg}$ at each time $t$, ensuring that there are no radial or vertical gradients in ${\cal Q}(r,\theta,t)$ that could influence the shape of the power spectrum in either of these directions.

\newpage

We then take a two-dimensional Fourier transform of ${\cal Q}(r,\theta,t)$:
\begin{equation}
{\cal Q}(k_r,k_\theta,t) = \int \int {\cal Q}(r,\theta,t)
e^{- i \left(k_r r + k_\theta \theta \right)} dr d\theta
\end{equation}
We will often plot this quantity in terms of $k_x = 2\pi l / \Delta x$ and $k_z = 2\pi l / \Delta z$, where $\Delta x/H = 2 (r_2 - r_1) (r_1 + r_2)^{-1} (H/R)^{-1} \cos(\pi/4)$ and $\Delta z/H = 10$. The time- and shell-averaged power spectrum $\left< \left< |P(|k|)|^2 \right> \right>$ is then obtained from
\begin{equation}
\left< \left| \left< P_{\cal Q}(|k|) \right> \right|^2 \right> = \frac{1}{\Delta T} \int^{T_2}_{T_1} \left| 
\left< {\cal Q}(|k|,t) \right> \right|^2 dt
\end{equation}
where $|k| = \sqrt{k^2_x + k^2_z}$ and $\left< {\cal Q}(|k|,t) \right>$ is averaged over shells of constant $|k|$. As emphasized above, this approach to investigating the spatial power spectrum of the turbulence is well-defined, but non-unique. An alternative approach would be to calculate the power-spectrum in the radial direction using (for example) Bessel functions as a basis, avoiding the need to remove radial gradients on large spatial scales. We note however, that procedures equivalent to that described here were used extensively to investigate turbulence in global accretion disk simulations by (for example) \cite{Schnittman:2006,Nelson:2010} and so we use a similar approach here for consistency with prior works in the literature.

A further probe of the structure of the turbulence is the temporal variability of fluctuations on spatial scales $|k|$. This measure is calculated from the normalized measure of the spatial fluctuations, ${\cal Q}(r,\theta,t)$, via a three-dimensional Fourier Transform:
\begin{equation}
{\cal P}(k_r,k_\theta,f) = \int \int \int {\cal Q}(r,\theta,t)
e^{- i \left(k_r r + k_\theta \theta + f t \right)} dr d\theta dt
\end{equation}
As above, we average ${\cal P}(k_r,k_\theta,f)$ over shells of constant $|k| = \sqrt{k^2_x + k^2_z}$ to obtain  ${\cal P}(|k|,f)$. We use this data to obtain estimates of the variability of {\cal Q} on large ($|k| H / 2\pi < 1$) versus small ($|k| H / 2\pi$) spatial scales by averaging $\left| \left< P_{\cal Q}(|k|,f) \right> \right|^2$ over these ranges in $|k|$-space. To improve statistics, we rebin the resulting power spectra onto a grid in frequency space that is a factor of ten coarser than the original data. When doing so, we rebin the power in logarithmic units in order to avoid biasing that can occur when rebinning data that is characterized by (e.g.) red noise.

The discussion of \S\ref{turbulence} and \S\ref{flux_stress_relation} makes use of auto- and cross-correlation functions. These measures are calculated using Fourier transform techniques and so we proceed as above by mapping data on the graded simulation mesh in the region $r_1 \le r \le r_2$ and the vertical domain covering the body of the disk, $-H \le Z \le H$. We denote the volume-weighted data arising from this procedure $Q_V (r,\theta,\phi,t)_{reg}$ and the vertical integral of this data $\left< Q_V (r,\phi,t) \right>_{reg}$. We again consider the normalized fluctuations in these quantities
\begin{equation}
\begin{split}
{\cal Q}(r,\theta,\phi,t) = \frac{ Q_V (r,\theta,\phi,t)_{reg}- \left< Q_V (r,\theta,\phi,t)_{reg} \right>}{\left< Q_V (r,\theta,\phi,t)_{reg} \right>} \\
{\cal Q}(r,\phi,t) = \frac{\left< Q_V (r,\phi,t) \right>_{reg}- \left<\left< Q_V (r,\phi,t) \right>_{reg} \right>}{\left<\left< Q_V (r,\phi,t) \right>_{reg} \right>}
\end{split}
\end{equation}
Here, $\left< Q_V (r,\theta,\phi,t)_{reg} \right>$ and $\left<\left< Q_V (r,\phi,t) \right>_{reg} \right>$ are constructed from one-dimensional polynomial approximations in $r$ to $\left< Q_V (r,\theta,t)_{reg} \right>$ and $\left<\left< Q_V (r,t) \right>_{reg} \right>$ at each $\theta$ and $t$ independently. These mean subtracted and normalized data are then Fourier transformed according to
\begin{equation}
\begin{split}
{\cal Q}(k_r,k_\theta,k_\phi,t) = \\
\int \int \int {\cal Q}(r,\theta,\phi,t) 
e^{- i \left( k_r r + k_\theta \theta + k_\phi \phi \right)}
dr d\theta d\phi \\
{\cal Q}(k_r,k_\phi,f) = \\
\int \int \int {\cal Q}(r,\phi,t)
e^{- i \left( k_r r + k_\phi \phi+ f t \right)}
dr d\phi dt
\end{split}
\end{equation}
The auto-correlation functions, $\left< C_{{\cal Q}}(\Delta r, \Delta \theta, \Delta \phi) \right>$ and $\left< \left< C_{{\cal Q}}(\Delta r, \Delta \phi, \Delta t) \right>\right>$ are then calculated as
\begin{equation}
\begin{split}
\left< C_{{\cal Q}}(\Delta r, \Delta \theta, \Delta \phi) \right> = \\
 \frac{1}{\Delta T} \int^{T_2}_{T_1} \int \int \int 
 \left| {\cal Q} (k_r,k_\theta,k_\phi,t) \right|^{2} 
e^{i \left( k_r r + k_\theta \theta + k_\phi \phi \right)}
 dk_r dk_\theta dk_\phi dt \\
\left< C_{{\cal Q}}(\Delta r, \Delta \phi, \Delta t) \right> = \\
\int \int \left| {\cal Q}(k_r,k_\phi,f) \right|^2
 e^{i \left( k_r r + k_\phi \phi+ f t \right)}
dk_r dk_\phi df
\end{split}
\end{equation}
To construct a time-average of $\left< C_{{\cal Q}}(\Delta r, \Delta \phi, \Delta t) \right>$, we compute the correlation function over some period $\delta t = n P_{orb} (r = 0.5(r_1 + r_2)$ (where $n=2$ typically). Denote this auto-correlation function measured at some time $T$ during the evolution as $\left< C_{{\cal Q}}(\Delta r, \Delta \phi, \Delta t, T) \right>$. The time-average, $\left< \left< C_{{\cal Q}}(\Delta r, \Delta \phi, \Delta t) \right> \right>$ is then computed as
\begin{equation}
\left< \left< C_{{\cal Q}}(\Delta r, \Delta \phi, \Delta t) \right> \right> =
 \frac{1}{\Delta T} \int^{T_2}_{T_1} \left< C_{{\cal Q}}(\Delta r, \Delta \phi, \Delta t, T) \right> dT
 \end{equation}
where $\Delta T = T_2-T_1 = N P_{orb} (r > r_2)$ is the time-interval over which the auto-correlation function is averaged, $N = 2$ typically and the orbital period is typically measured at $r=15r_S$.
 
The cross-correlation function between two quantities, ${\cal Q}_1(r,\phi,t)$ and ${\cal Q}_2(r,\phi,t)$ is calculated as
 \begin{equation}
 \begin{split}
\left< C_{{\cal Q}_1 {\cal Q}_2}(\Delta r, \Delta \phi, \Delta t) \right> = \\
\int \int \int {\cal Q}^{*}_1(k_r,k_\phi,f) {\cal Q}_2(k_r,k_\phi,f)
  e^{i \left( k_r r + k_\phi \phi+ f t \right)}
 dk_r dk_\phi df
 \end{split}
\end{equation}
Here, ${\cal Q}^{*}_1(k_r,k_\theta,k_\phi,t)$ denotes the complex conjugate of ${\cal Q}_1(k_r,k_\theta,k_\phi,t)$. The interpretation of $\left< \left< C_{{\cal Q}_1 {\cal Q}_2}(\Delta r, \Delta \phi, \Delta t) \right> \right>$ calculated in this fashion in that negative (positive) offsets in $\Delta t$ represent fluctuations in ${\cal Q}_1(r,\phi,t)$ leading (trailing) fluctuations in ${\cal Q}_2(r,\phi,t)$. The time-average of $\left< C_{{\cal Q}_1 {\cal Q}_2}(\Delta r, \Delta \phi, \Delta t) \right>$ is calculated in the same way as for $\left< C_{{\cal Q}}(\Delta r, \Delta \phi, \Delta t) \right>$, i.e.
\begin{equation}
\begin{split}
\left< \left< C_{{\cal Q}_1 {\cal Q}_2}(\Delta r, \Delta \phi, \Delta t) \right> \right> = \\
 \frac{1}{\Delta T} \int^{T_2}_{T_1} \left< C_{{\cal Q}_1 {\cal Q}_2}(\Delta r, \Delta \phi, \Delta t, T) \right> dT
 \end{split}
 \end{equation}
where $\Delta T = T_2-T_1 = N P_{orb} (r > r_2)$ is the time-interval over which the cross-correlation function is averaged. We note that we will often plot both the auto- and cross-correlation functions in terms of $\Delta x, \Delta y, \Delta z$ defined as discussed above and $\Delta t$ in units of the orbital period at $r = 0.5 (r_1 + r_2)$.

In order to track the evolution of the auto- and cross-correlation functions in time, we use two approaches. The first, which we have found most useful when the correlation function is centrally concentrated (as is the case for the auto-correlation function) tracks the time-evolution of the maxima in the correlation function. At each $\Delta t$ in the correlation function, we find the maximum amplitude of the correlation function associated with contributions from large ($|k| H / 2\pi < 1$) versus small ($|k| H / 2\pi \ge 1$) where $|k| = \sqrt{k^2_x + k^2_y}$ (where this distinction is made by calculating the correlation functions only including contributions from these scales) and plot the amplitude of this maxima as a function of time. The definition of the auto-correlation function means that this quantity must take its maximum value at $\Delta t = 0$ and so the width of the correlation function in $\Delta t$ provides an estimate of the lifetime of modes on large versus small spatial scales. The second approach, which we have found to be more useful when the correlation function does not have a well-defined shape (as is the case for the cross-correlation function) tracks the total amplitude of the correlation function as a function of time, e.g.
\begin{equation}
\begin{split}
\left< \left< \left< C_{{\cal Q}_1 {\cal Q}_2}(\Delta t) \right> \right> \right> = \\
\int \int \left< \left< C_{{\cal Q}_1 {\cal Q}_2}(\Delta r, \Delta \phi, \Delta t) \right> \right> d \Delta r d \Delta \phi
\end{split}
\end{equation}
Because of the definition of the cross-correlation function, if $\left< \left< \left< C_{{\cal Q}_1 {\cal Q}_2}(\Delta t) \right> \right> \right>$ is maximized at negative (positive) $\Delta t$, then fluctuations in ${\cal Q}_1(r,\phi,t)$ lead (trail) fluctuations in ${\cal Q}_2(r,\phi,t)$.
\section{Global Characteristics of the Disk}\label{evolve}

\subsection{Evolution of the Disk}\label{dynamo}

\begin{figure}
\begin{center}
\leavevmode
$
\begin{array}{cc}
\includegraphics[width=0.48\columnwidth, viewport=30 10 345 325,clip]
{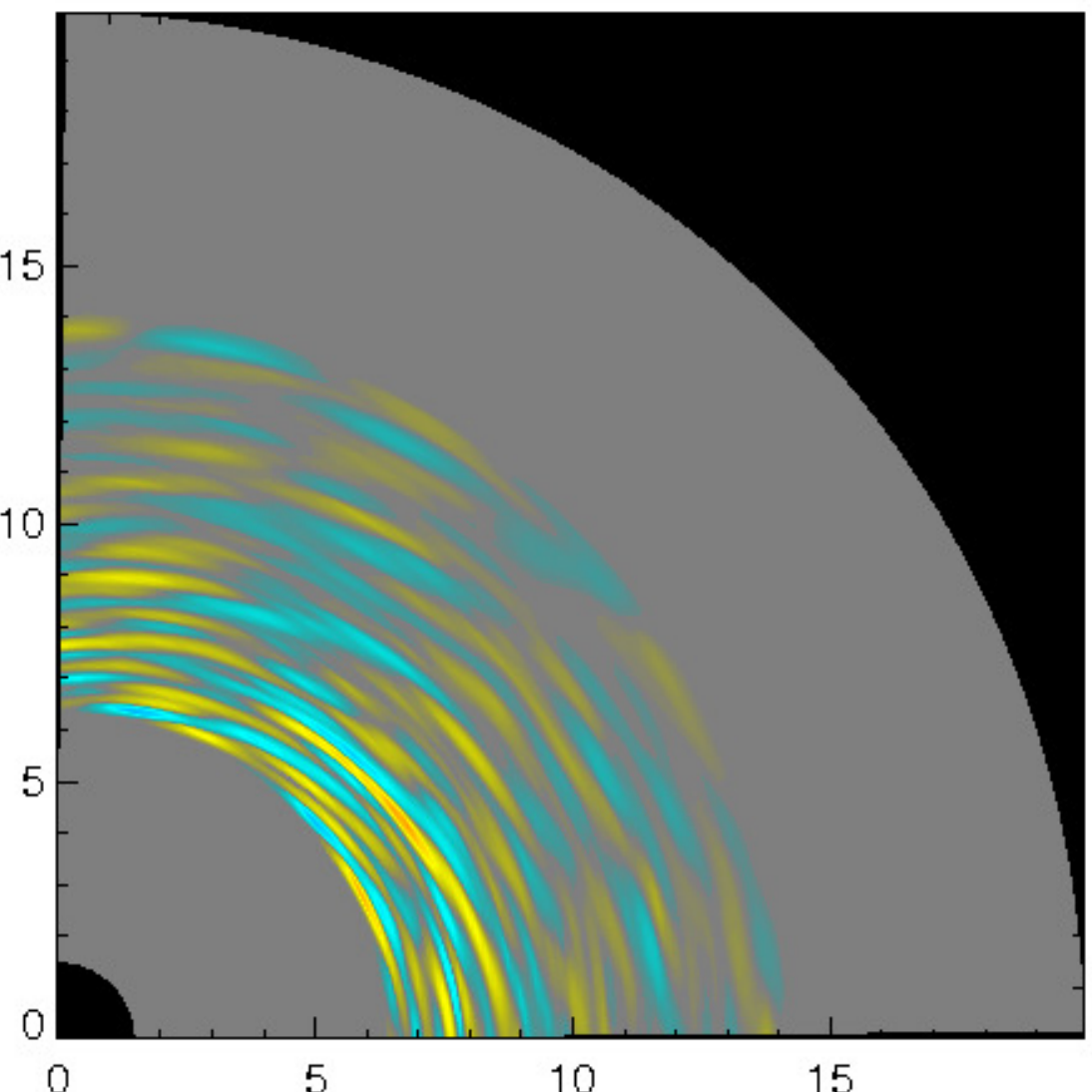} &
\includegraphics[width=0.48\columnwidth, viewport=30 10 345 325,clip]
{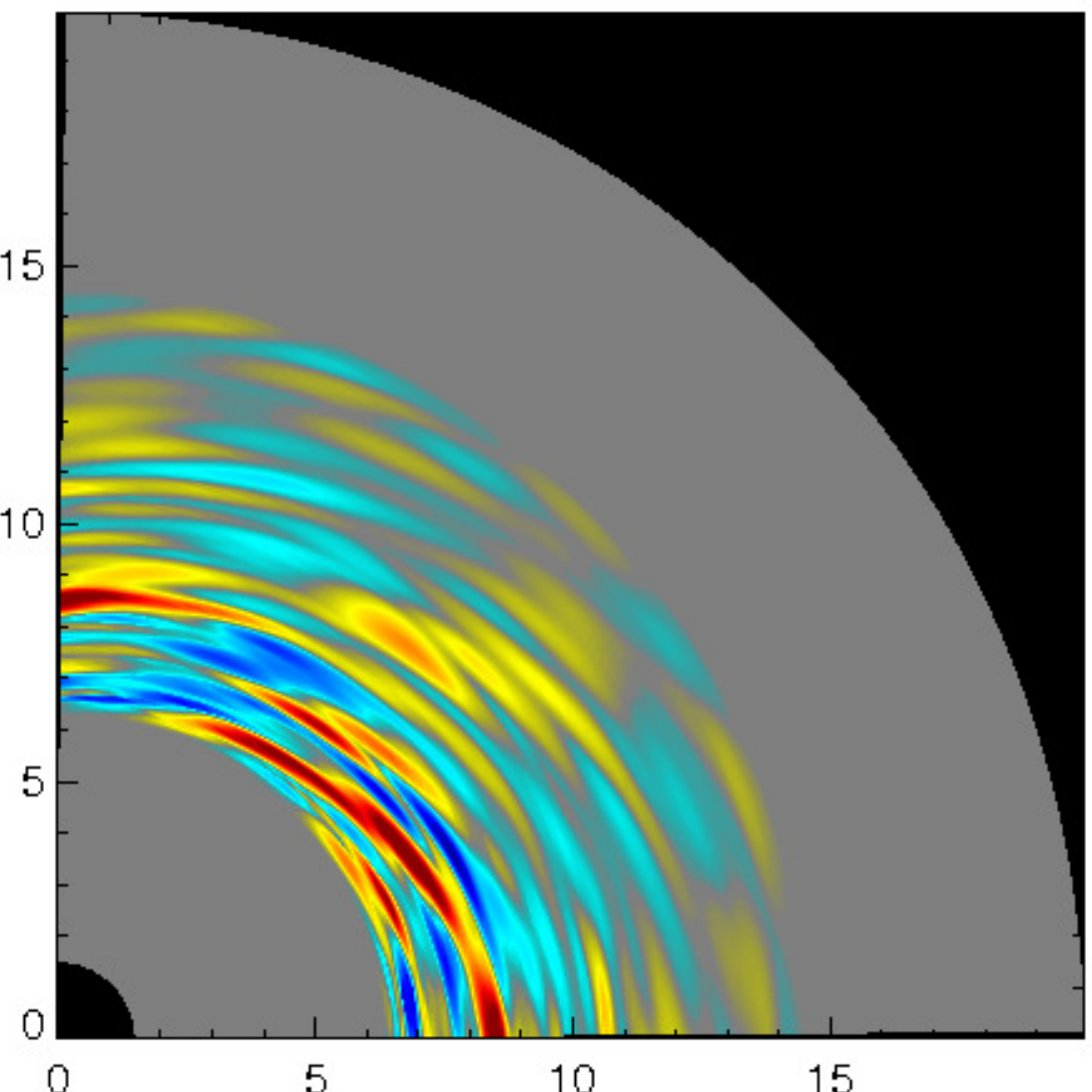} \\
\multicolumn{2}{c}
{\includegraphics[width=0.48\textwidth]{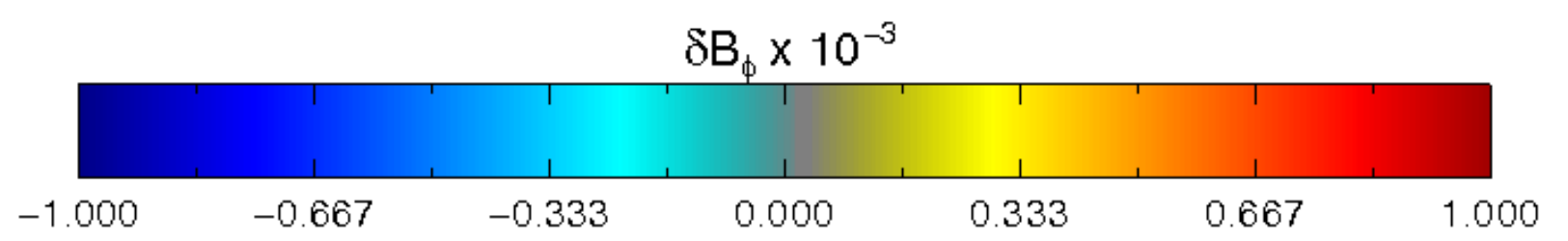}} \\
\includegraphics[width=0.48\columnwidth, viewport=30 10 345 325,clip]
{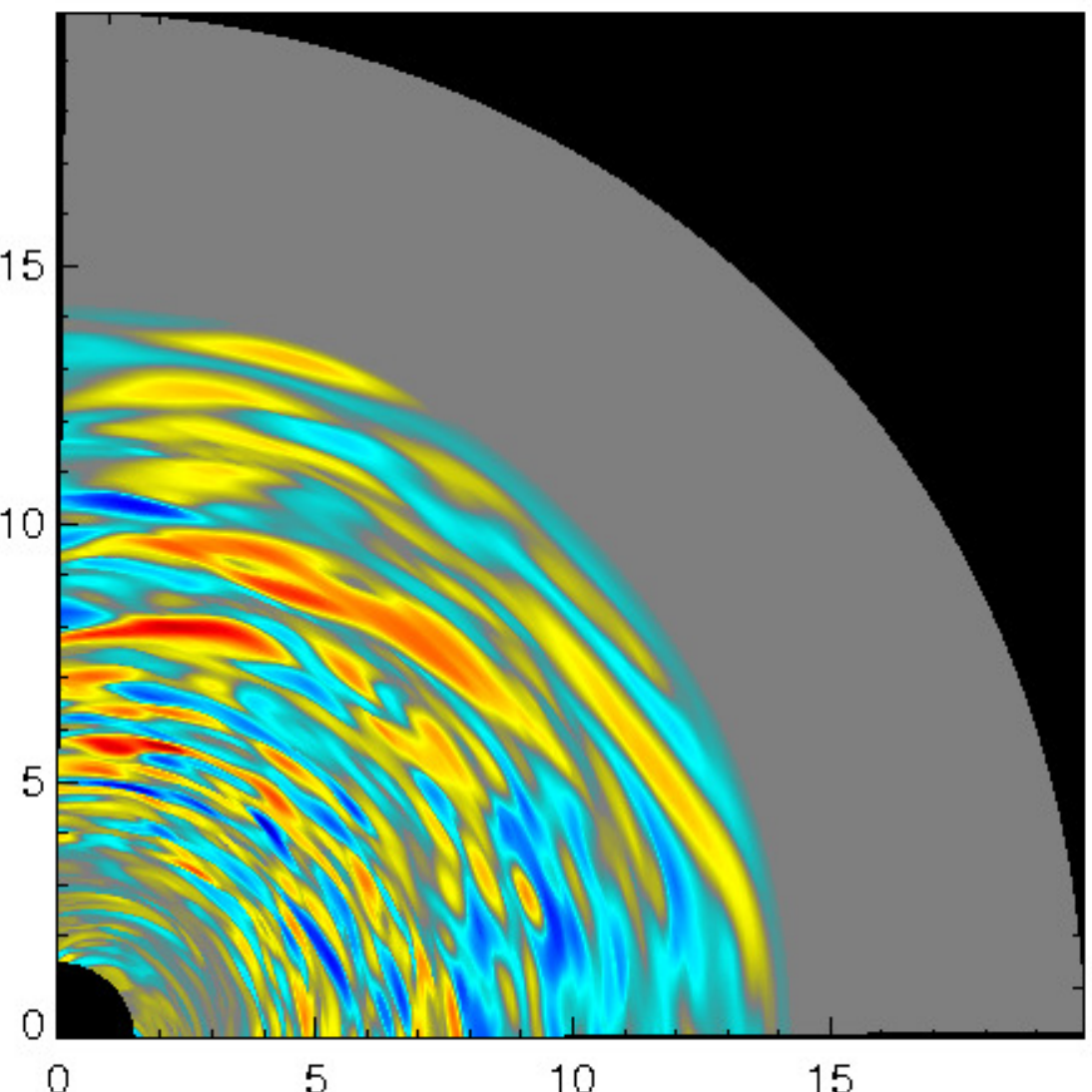} &
\includegraphics[width=0.48\columnwidth, viewport=30 10 345 325,clip]
{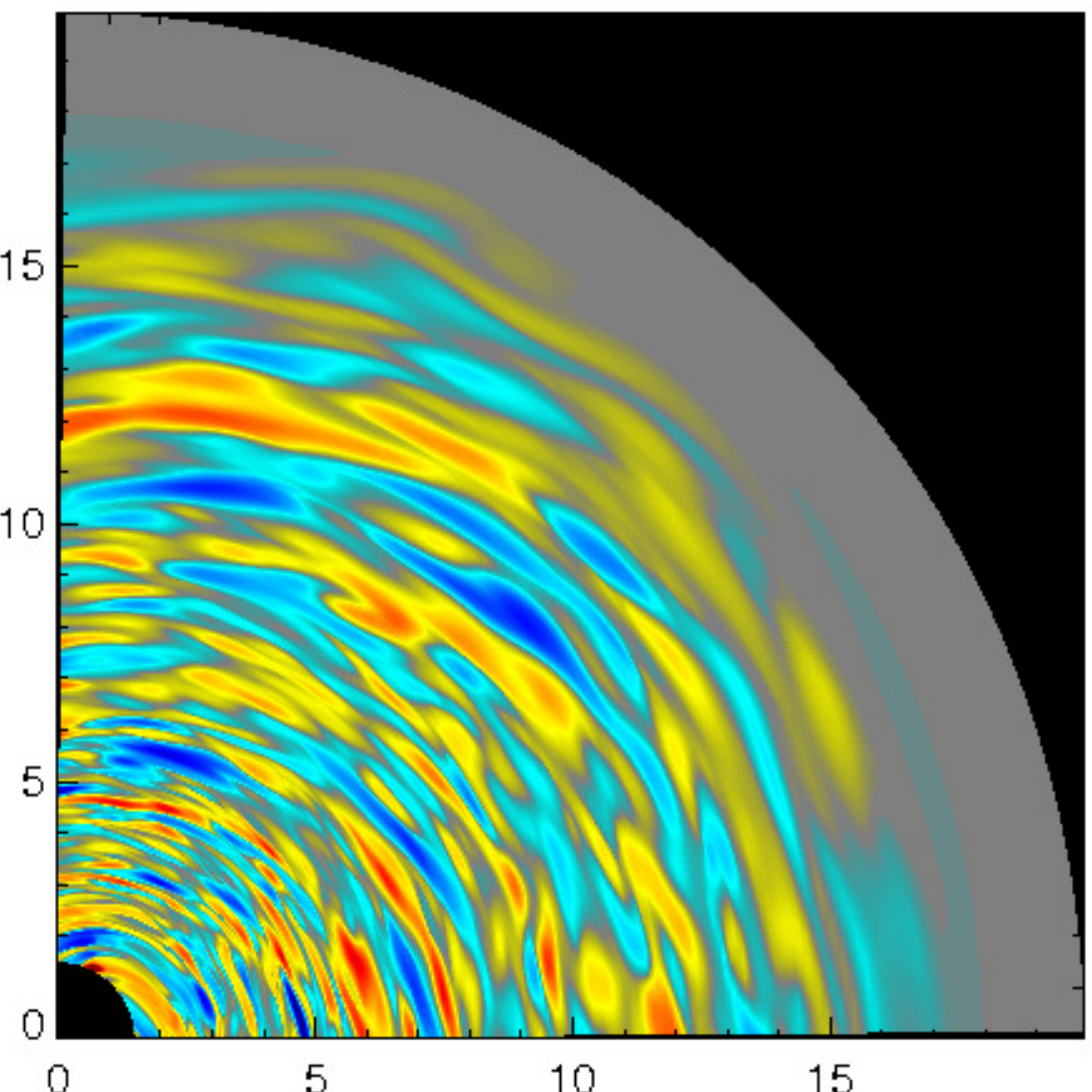} \\
\multicolumn{2}{c}
{\includegraphics[width=0.48\textwidth]{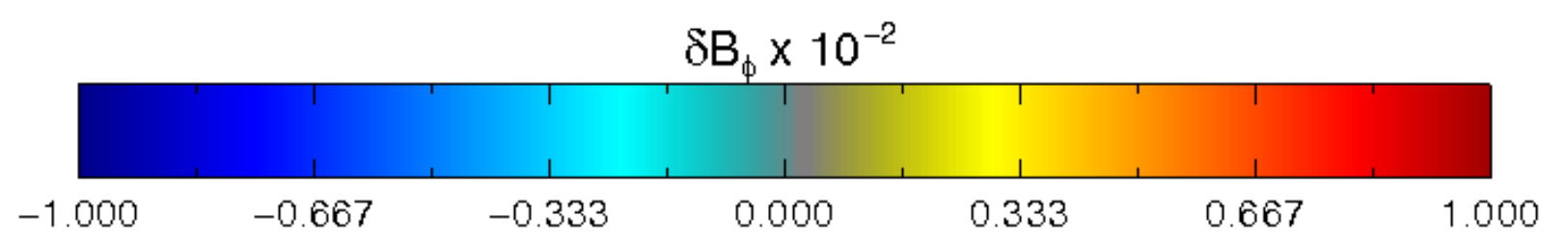}} \\
\end{array}
$
\end{center}
\caption[]{Structure of the toroidal magnetic field on the $r-\phi$ plane, $B_\phi (r,\theta=\pi/2,\phi)$, at (from left to right) $t=0.3,\;0.5$(top row) $t=3.0,\;5.0$ (bottom row) $T_{orb}(r=15r_S)$.}
\label{bphi_evolve} 
\end{figure}

\begin{figure}
\leavevmode
\begin{center}
\includegraphics[width=0.45\textwidth]{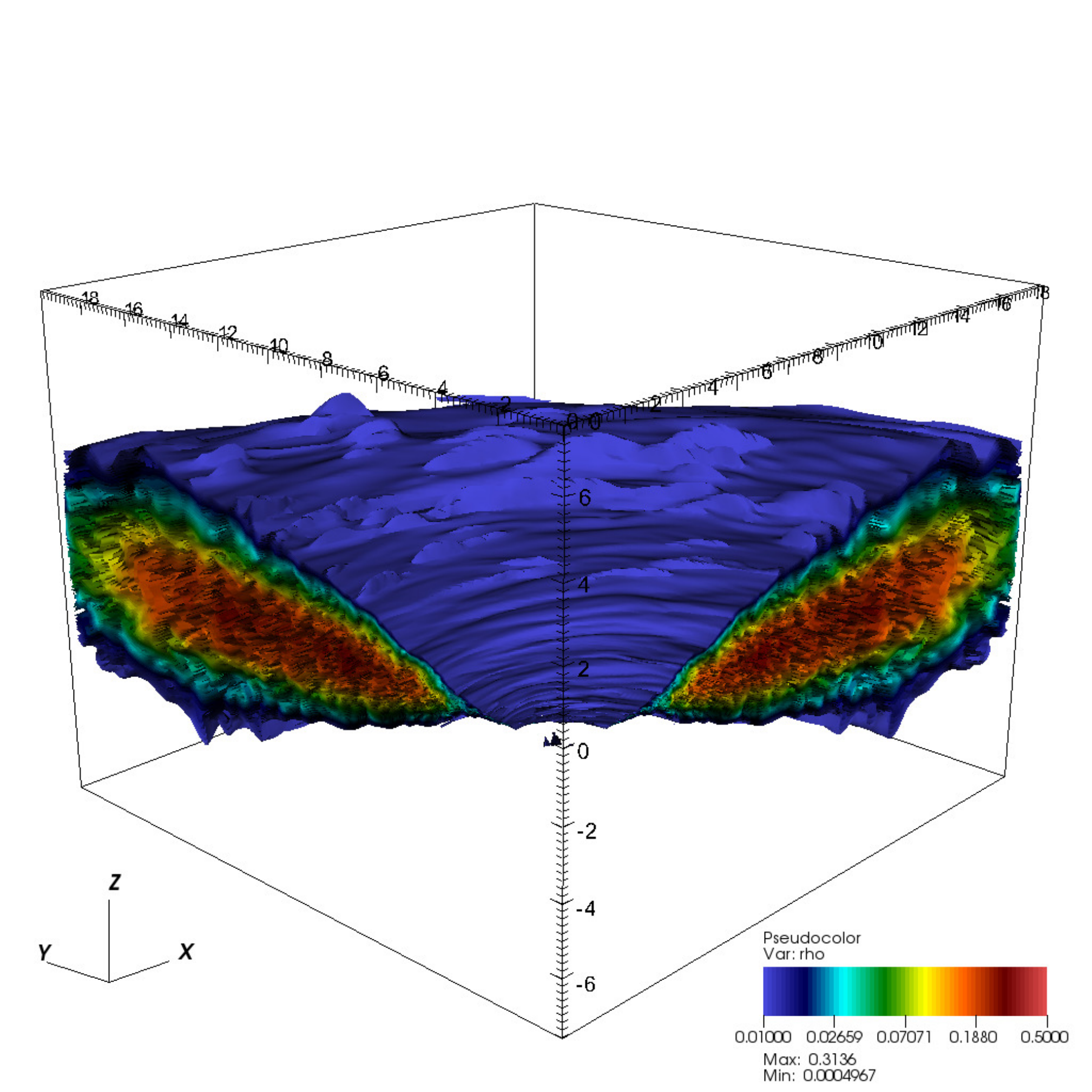}
\includegraphics[width=0.45\textwidth]{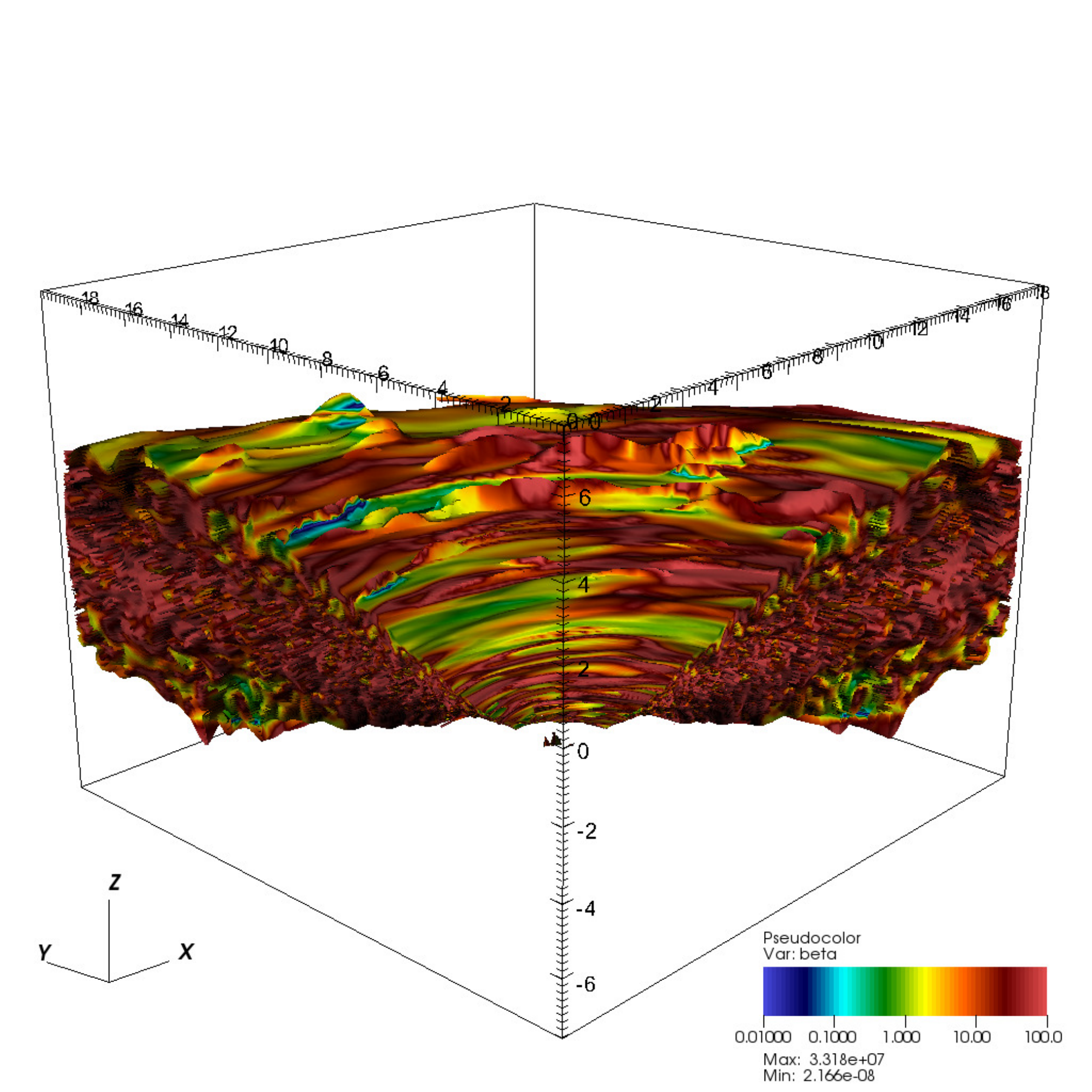}
\end{center}
\caption[]{Volume renderings of gas density (top panel)
and gas $\beta$ parameter (bottom panel) at $t=7000$. }
\label{fluid_state} 
\end{figure}

\begin{figure}
%\leavevmode
\begin{center}
\includegraphics[width=0.45\textwidth]{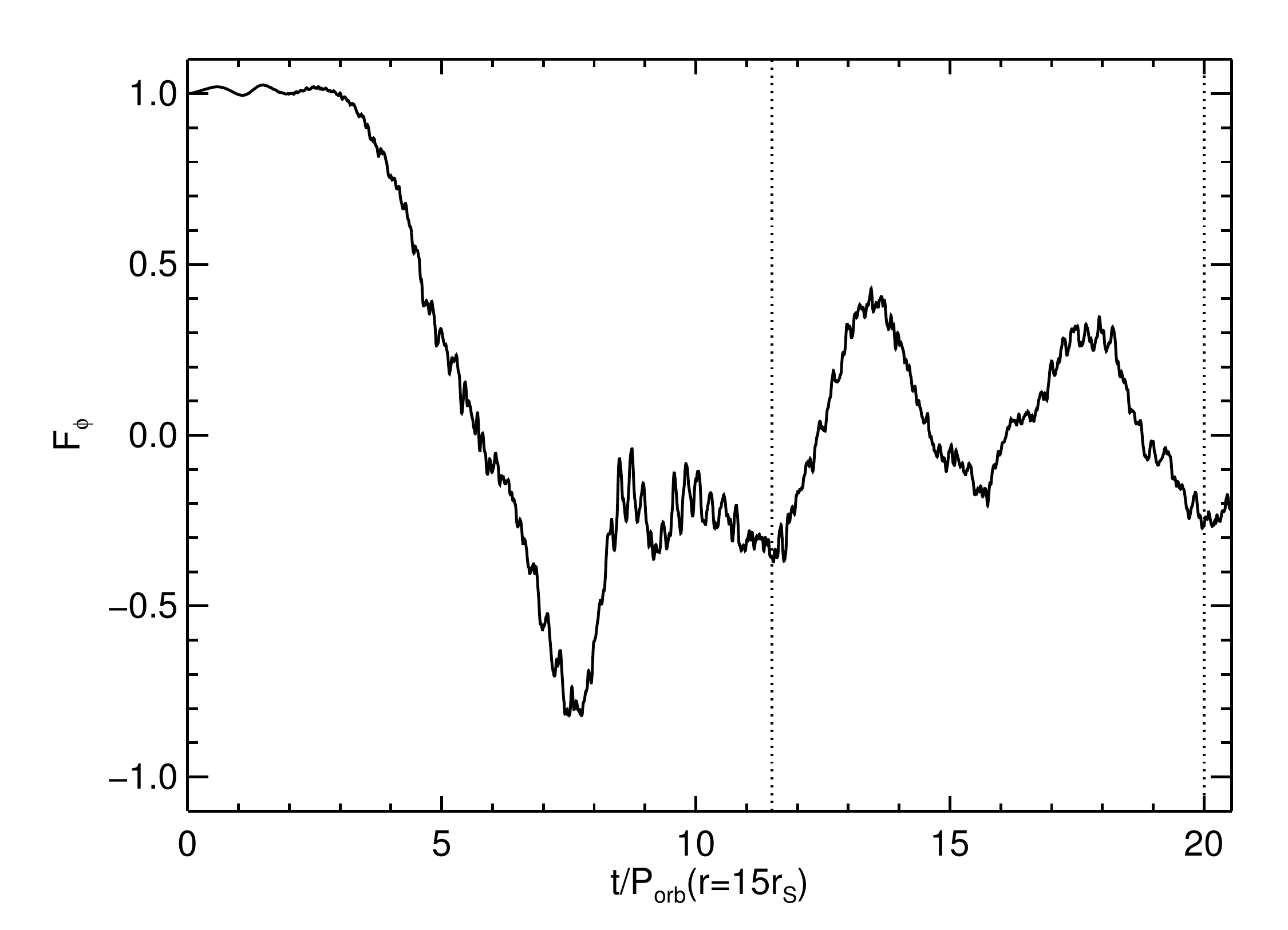}
\end{center}
\caption[]{Time history of the toroidal magnetic \emph{flux}, $F_\phi$, computed from the toroidal magnetic field evaluated at $\phi=\pi/4$ and  integrated over the radial region $5\le r \le 15$ and the entire vertical domain of the simulation, $|Z|\le5H$.}
\label{fphi} 
\end{figure}

\begin{figure}
%\leavevmode
\begin{center}
$
\begin{array}{cc}
\includegraphics[width=0.48\columnwidth, viewport=30 10 345 325,clip]
{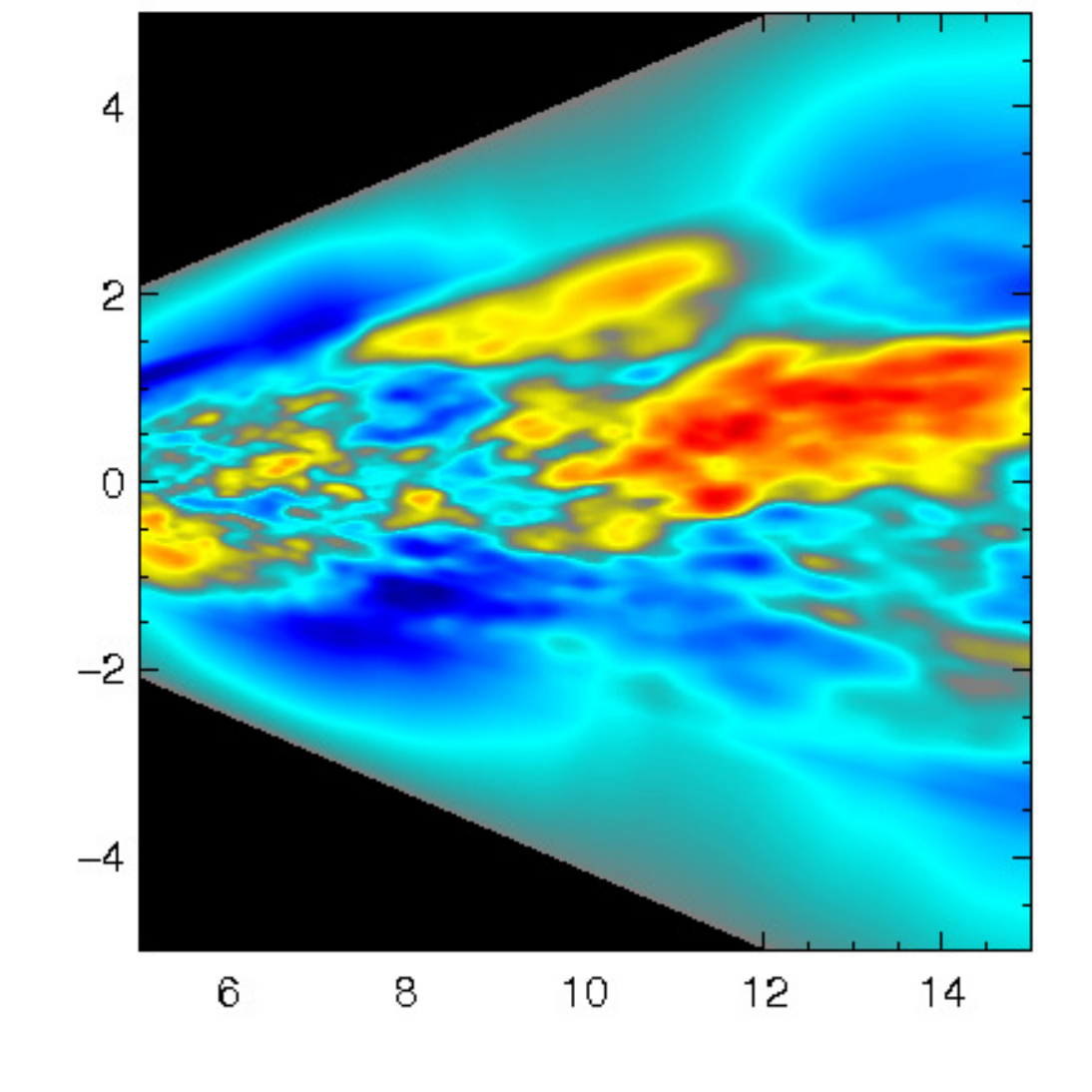} &
\includegraphics[width=0.48\columnwidth, viewport=30 10 345 325,clip]
{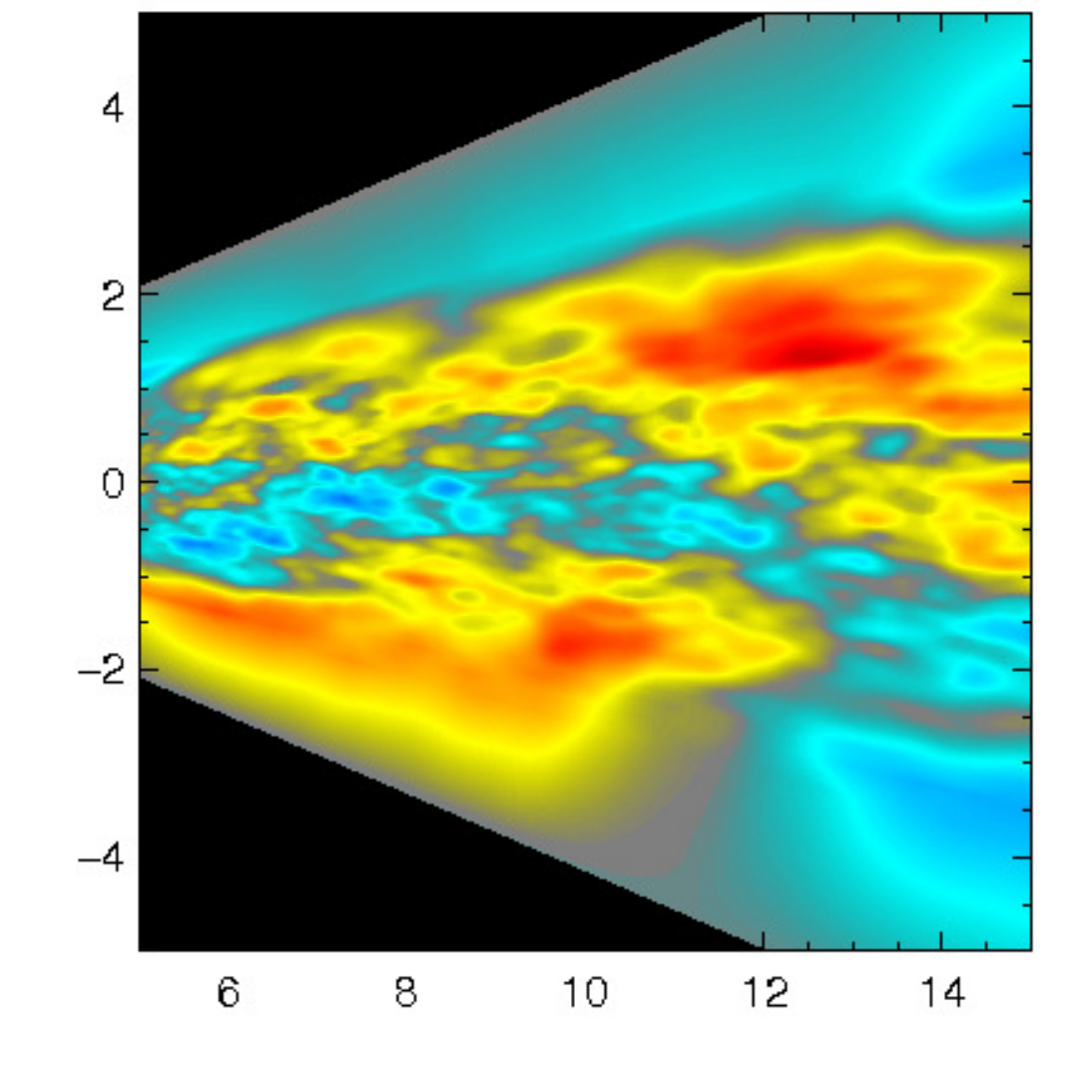} \\
\includegraphics[width=0.48\columnwidth, viewport=30 10 345 325,clip]
{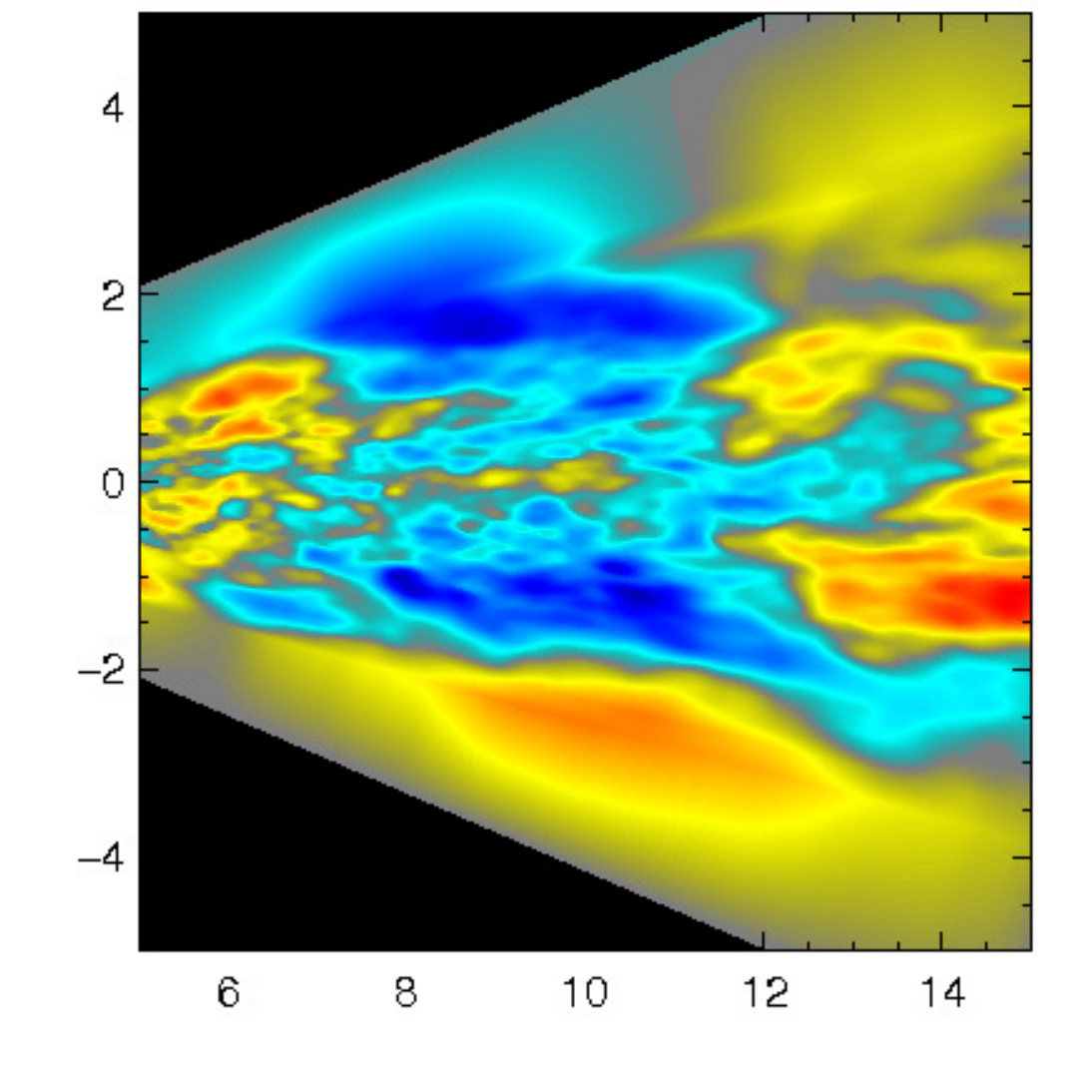} &
\includegraphics[width=0.48\columnwidth, viewport=30 10 345 325,clip]
{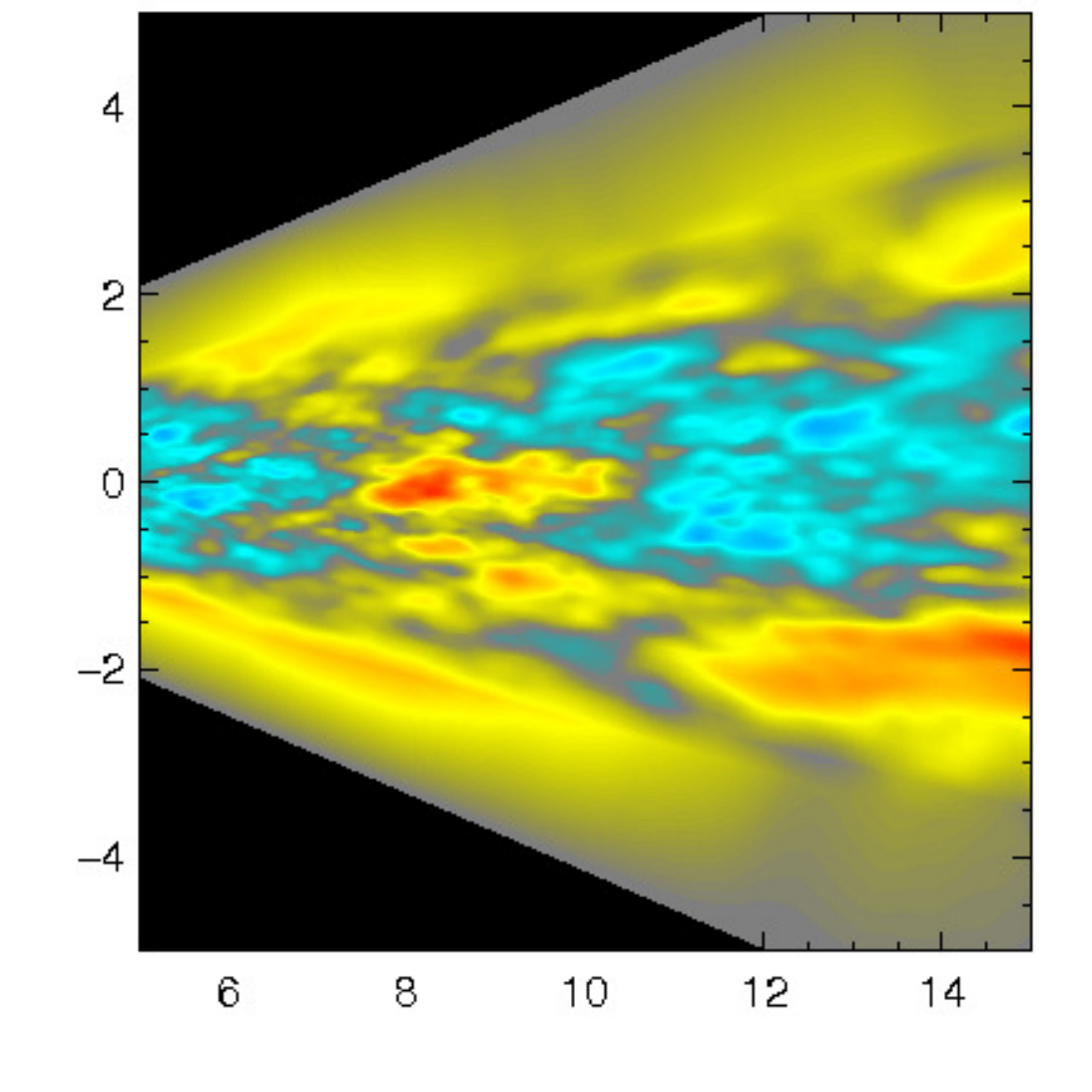} \\
\multicolumn{2}{c}
{\includegraphics[width=0.48\textwidth]{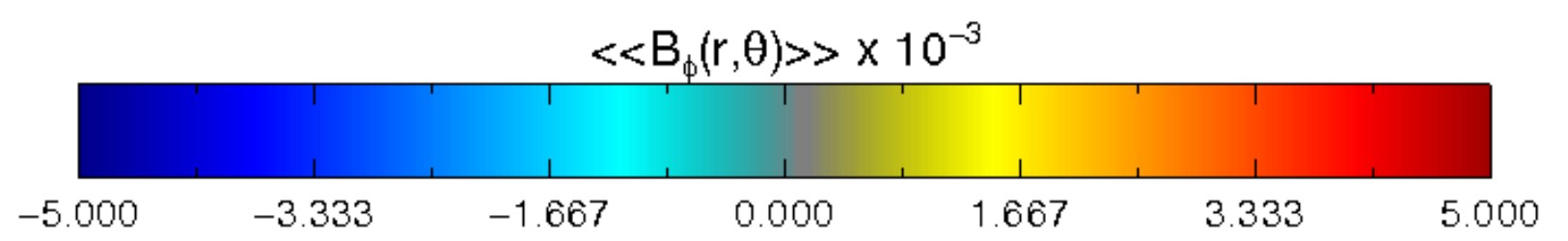}} \\
\end{array}
$
\end{center}
\caption[]{Evolution of the azimuthal average of the toroidal magnetic field, $\left< \left< B_\phi (r,\theta) \right> \right>$ on the poloidal plane, averaged between orbits $10-12$ (top left), $12-14.5$ (top right), $14.5-16$ (bottom left) and $16-19$ (bottom right) where the orbital period is measured at $r=15r_S$.}
\label{bphi_xz} 
\end{figure}

The evolution of the disk proceeds in a fashion consistent with prior toroidal field models, e.g. \cite{Hawley:2000,Hawley:2002,Beckwith:2008a}. Modes with high poloidal and high toroidal wave numbers grow first \cite[as described in][see top left panel of Figure \ref{bphi_evolve}]{Hawley:1995}. These modes gradually assemble into structures characterized by smaller toroidal and poloidal wave numbers, as can be seen in the top right panel of Figure \ref{bphi_evolve}. This process continues for a further $2.5$ orbits (measured at $r=15r_S$, until after about three orbits at this radius the amplitude of the turbulent fluctuations have become sufficient to drive accretion into the central object (bottom left panel of Figure \ref{bphi_evolve}). By five orbits at $r=15r_S$, turbulent fluctuations in the magnetic field have expanded to fill the entire radial domain of the simulation and the disk has entered the quasi-stationary state (bottom right panel of Figure \ref{bphi_evolve}). The simulation is evolved for a total of $20$ orbits at $r=15r_S$; the data of Figure \ref{fluid_state} shows the state of the disk at this time. Note the large amplitude non-axisymmetric surface density fluctuations and equipartition-strength magnetic fields evident in these plots. \cite{Beckwith:2008a} found that models initialized with a net toroidal field possessed corona with magnetic fields with volume-averaged strengths a factor of approximately three below equipartition. The data presented in Figure \ref{fluid_state} suggest that whilst that conclusion is approximately correct in a volume-averaged sense; it is possible to form equipartition-strength magnetic fields in non-axisymmetric structures, reminiscent of those suggested by \cite{Spruit:2005}.

\cite{Fromang:2006} demonstrate that net toroidal flux present in the initial state of their models is rapidly expelled from the disk into the coronal region, such that sustained MHD turbulence within the disk  is dependent on the small scale dynamo exhibited in stratified shearing boxes \cite[see e.g.][]{Davis:2010,Gressel:2010,Simon:2010}. Figure \ref{fphi} plots the time history of the toroidal flux, $F_\phi(t)$ in the simulation presented here, where:
\begin{equation}
F_\phi(t)= \int^{r=15r_S}_{r=5r_S} \int^{\theta=-5H}_{\theta=+5H} B_\phi (r,\theta,\phi=\pi/4,t) r dr d\theta
\end{equation}
The evolution of this quantity is similar to that reported by \cite{Fromang:2006};  the initial net toroidal flux distribution is expelled from the region of integration through the radial boundaries over the first five orbits in the outer disk (i.e. soon after the linear growth phase of the MRI) and is replaced with flux of opposite sign. After approximately eleven orbits at $r=15r_S$, a small scale dynamo reminiscent of that observed in vertically stratified shearing box simulations produces toroidal flux of opposite signs with a cycle of approximately five orbits, as previously reported by \cite{ONeill:2010}. The relatively rapid reversals of the toroidal flux suggest that the late-time analysis of the disk structure, presented below, ought to be largely independent of our choice of a net (rather than a zero) flux initial condition.
Figure \ref{bphi_xz} shows the structure of $\left< \left<B_\phi \right> \right>_\phi$ averaged over each period of the dynamo cycle shown in Figure \ref{fphi}, specifically orbits $10-12$, $12-14.5$ $14.5-16$ and $16-19$ where the orbital period is measured at $r=15r_S$. These periods are chosen to correspond to periodic minima and maxima in the dynamo cycle shown in Figure \ref{fphi}. The toroidal magnetic field is organized both radially and vertically over large spatial scales and the sense of this organization varies over the course of the dynamo cycle in a quasi-periodic fashion, reminiscent of the cycles seen in vertically stratified shearing box simulations \citep{Davis:2010,Gressel:2010,Simon:2010} and in previous global simulations at individual radii \citep{ONeill:2010}.

\subsection{Angular Momentum Transport and Resolved Turbulence}\label{amtrans}

\begin{figure}
%\leavevmode
\begin{center}
\includegraphics[width=0.45\textwidth]{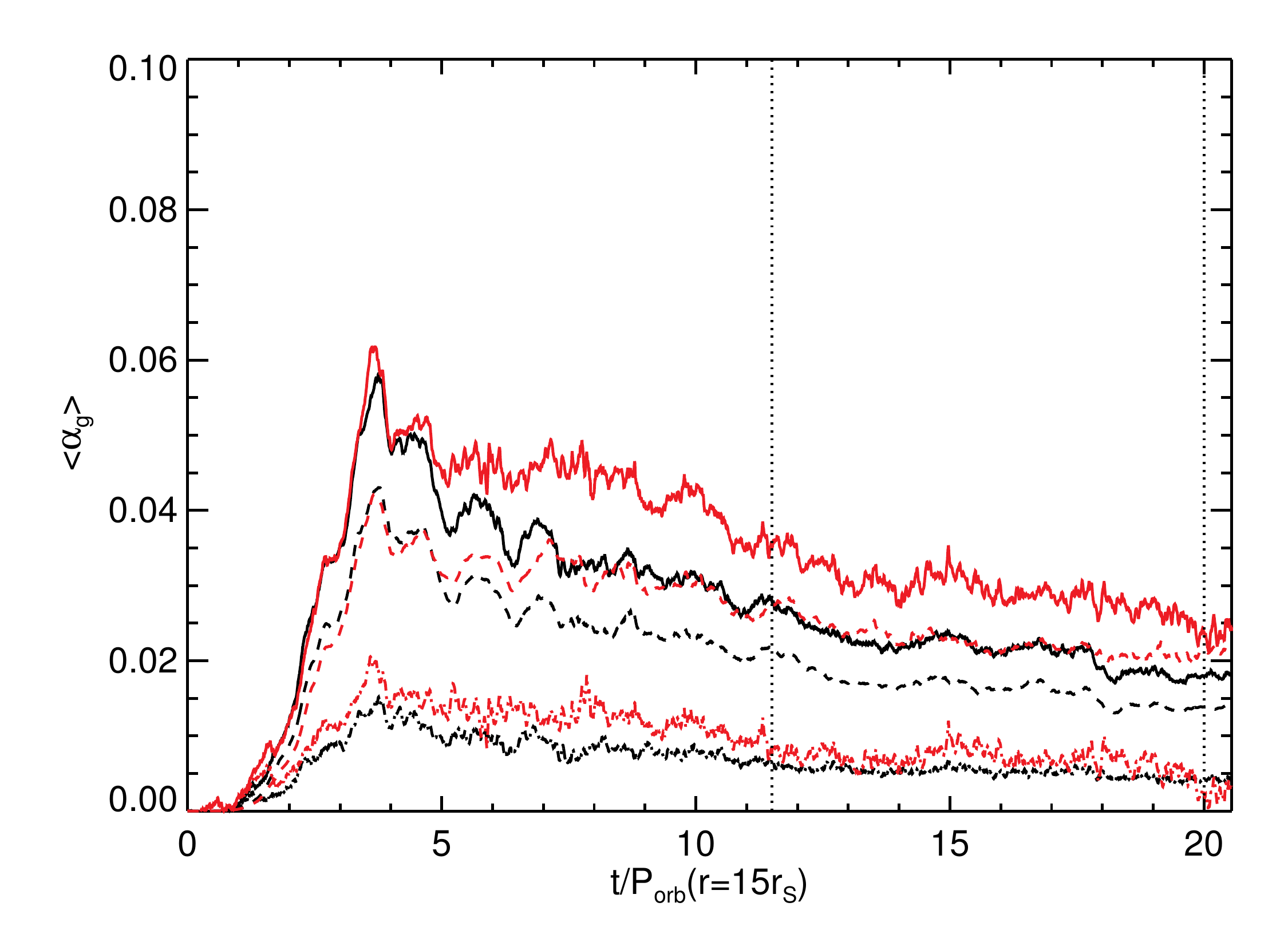}
\includegraphics[width=0.45\textwidth]{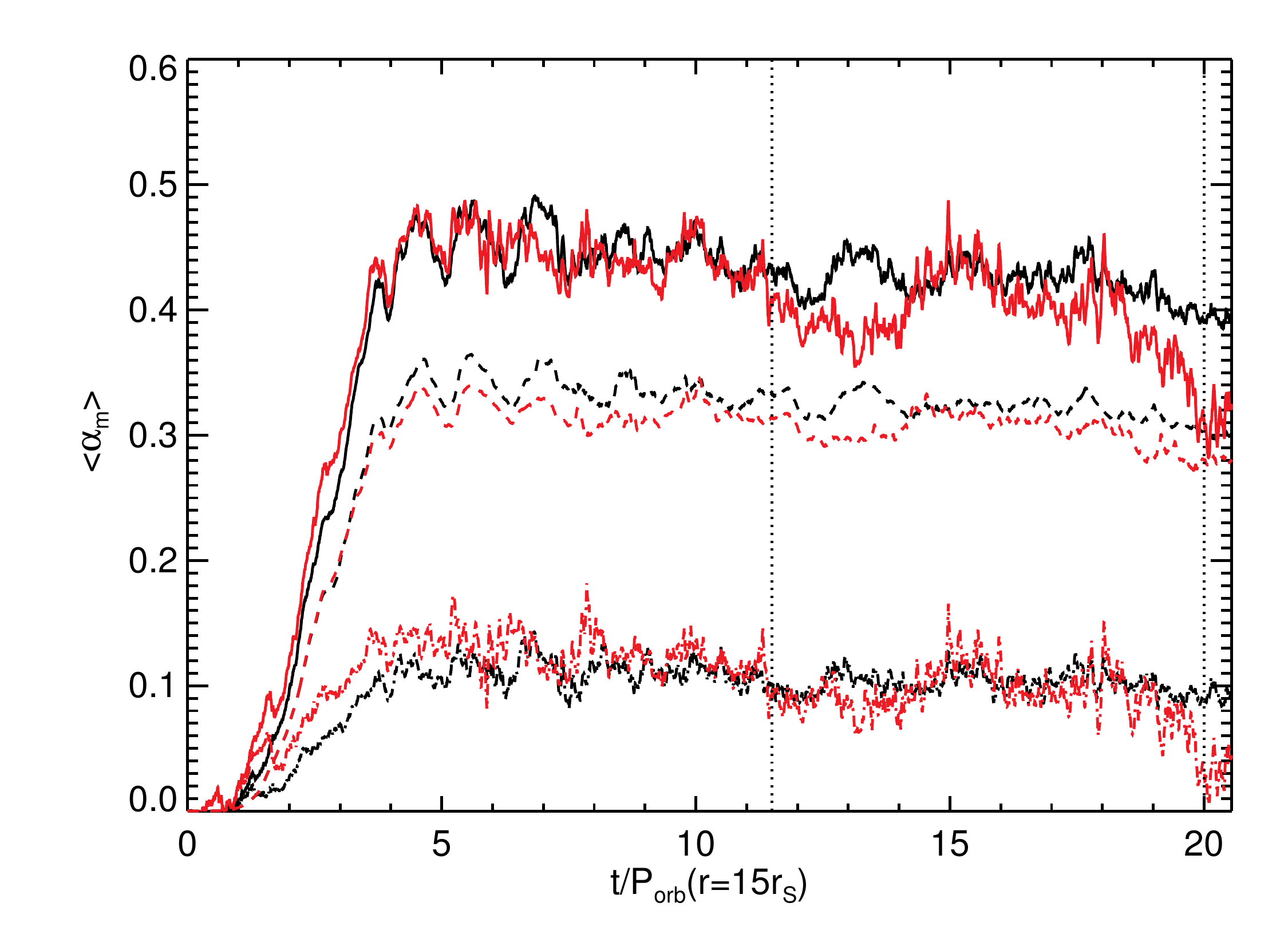}
\end{center}
\caption[]{Time history of accretion stress normalized to the gas pressure, $\left< \alpha^{t,f,m}_{g} \right> = \left< W^{t,f,m}_{r \phi} \right> / \left< P_g \right>$ (top panel) or the magnetic pressure, 
$\left< \alpha^{t,f,m}_{m} \right> = \left< W^{t,f,m}_{r \phi} \right> / \left< P_m \right>$ (bottom panel). All quantities were integrated inside the `disk-body' ($|Z| \le H$, black lines) or the `disk+corona' (red lines) over the radial region $5\le r \le 15$. In both panels, solid lines denote the total accretion stress, $ \left< W^{t}_{r \phi} \right>$, dashed lines the Maxwell stress,  $\left< W^{m}_{r \phi} \right>$ and dot-dash lines the Reynolds stress,  $\left< W^{f}_{r \phi} \right>$. Vertical dotted lines indicate the averaging interval. }
\label{alpha_vol} 
\end{figure}

A useful quantitative measure of the turbulence in these simulations is $\left< \alpha^{f,m,t}_{g,m} (t) \right> = \left< W^{f,m,t}_{r \phi} (t) \right> / \left<P_{g,m}(t) \right>$ where $W^{f}_{r \phi} = \rho \delta v_r \delta v_\phi$, $W^{m}_{r \phi} =  - B_r B_\phi$ are the Reynolds and Maxwell accretion stress respectively, $W^{t}_{r \phi} = W^{f}_{r \phi} + W^{m}_{r \phi}$ is the total stress and $P_{g,m}$ are the gas and magnetic pressures.
 The time-evolution of $\left< \alpha^{f,m,t}_{g} \right>$ is shown in the left-hand panel of Figure \ref{alpha_vol}, whilst that of $\left< \alpha^{f,m,t}_{m} \right>$ is shown in the right-hand panel of this figure. The data in these figures are computed by volume-averaging simulation data over the radial range $5 \le r/r_S \le 15 r_S$ such that we exclude regions of the disk likely to be influenced by the radial boundaries and either the entire vertical extent of the simulation (which we refer to as ``disk+corona'', denoted by red-lines in these figures) or $\pm H$ from the midplane (which we refer to as the ``disk body'', denoted by black-lines in these figures). The data of this figure are consistent with those of vertically stratified simulations computed at similar resolutions both with and without net flux configurations \cite[see e.g.][]{Guan:2009a,Simon:2010,Davis:2010}; specifically, we find $\left< \left< \alpha^{t}_{g} \right> \right> \sim 2.5\times10^{-2}$ and $\left< \left< \alpha^{m}_{m} \right> \right> \sim 0.3$ measured in the disk-body between $t=11.5-20 \; T_{orb}(r=15r_S)$ (the choice of averaging interval is such that it coincides with two dynamo cycles as detailed in the previous section). Note though that we find higher Reynolds stresses than reported by these authors, with $\left< \left< \alpha^{f}_{g} \right> \right> = 0.3-0.5 \left< \left< \alpha^{m}_{g} \right> \right>$ compared to $\left< \left< \alpha^{f}_{g} \right> \right> = 0.25 \left< \left< \alpha^{m}_{g} \right> \right>$. The magnitude of $\left< \left< \alpha^{t}_{g} \right> \right>$ is also significantly \emph{higher} than reported in previous \emph{global} calculations of magnetized thin accretion disks. \cite{Fromang:2006} reports $\alpha^{t}_{g} \sim 5.0\times10^{-3}$ for a toroidal field model computed at approximately twice the resolution of the calculation here, whilst \cite{Sorathia:2010} report for a model begun with a weak zero net poloidal flux configuration $6.0\times10^{-3} \lesssim \alpha^{t}_{g} \lesssim 9.0\times10^{-3}$ for a simulation of comparable resolution to that presented here\footnote{Note that \cite{Hawley:2002} reports $\alpha^{t}_{g} \sim 0.1$ for a disk with $H/R\sim0.15$.}.
  
 \begin{figure}
%\leavevmode
\begin{center}
\includegraphics[width=\columnwidth]
{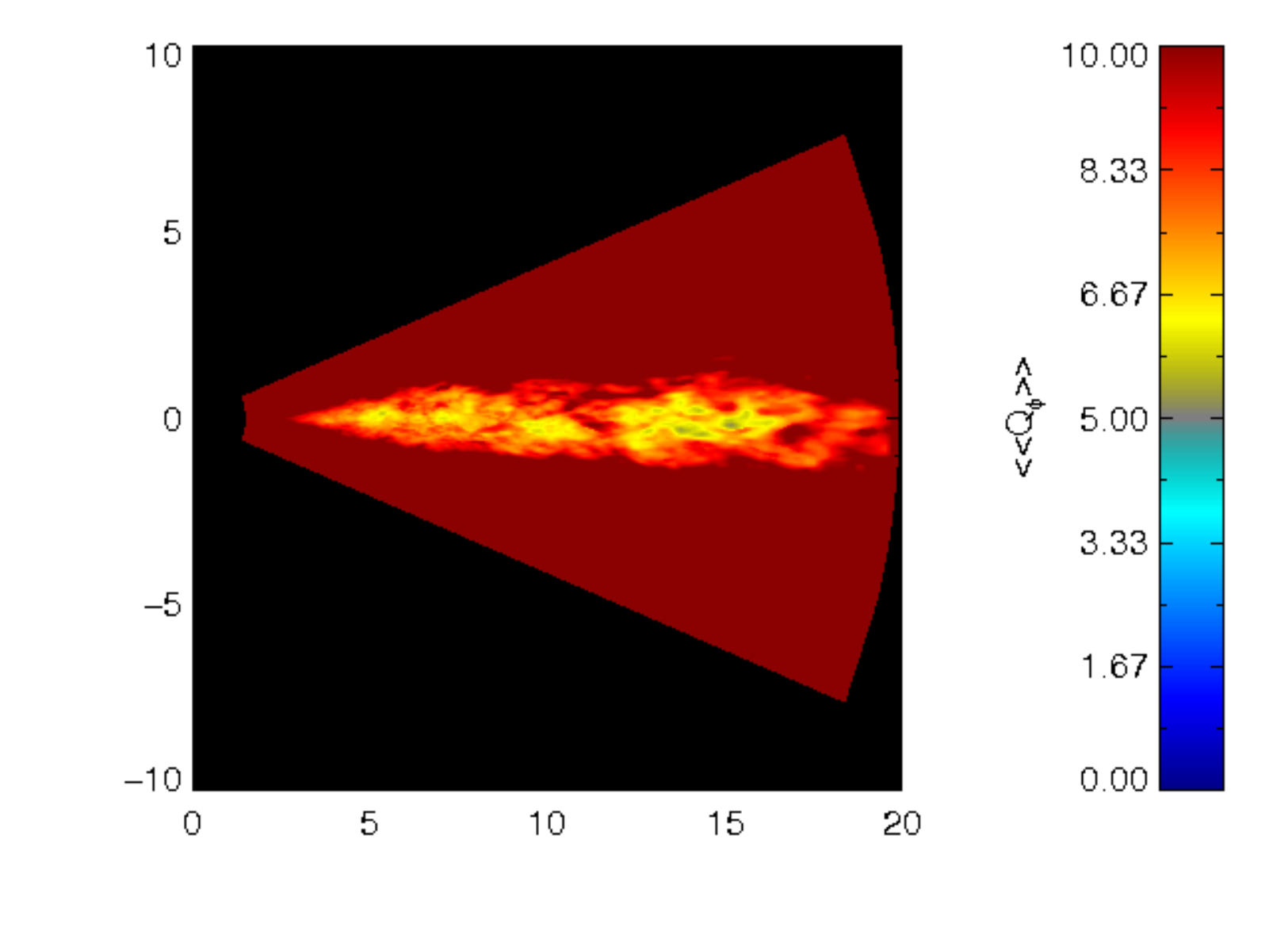}
\end{center}
\caption[]{Spatial dependence of number of zones per characteristic wavelength of the MRI, $\left< \left< Q_\phi \right> \right>$ on the $r-\theta$ plane. Simulation data was averaged in the toroidal direction and time-averaged over the period $\Delta T = 10-20 P_{orb} (r=15 r_S)$.}
\label{lmri_xz} 
\end{figure}

%FIGURE 5
\begin{figure}
%\leavevmode
\begin{center}
\includegraphics[width=0.45\textwidth]{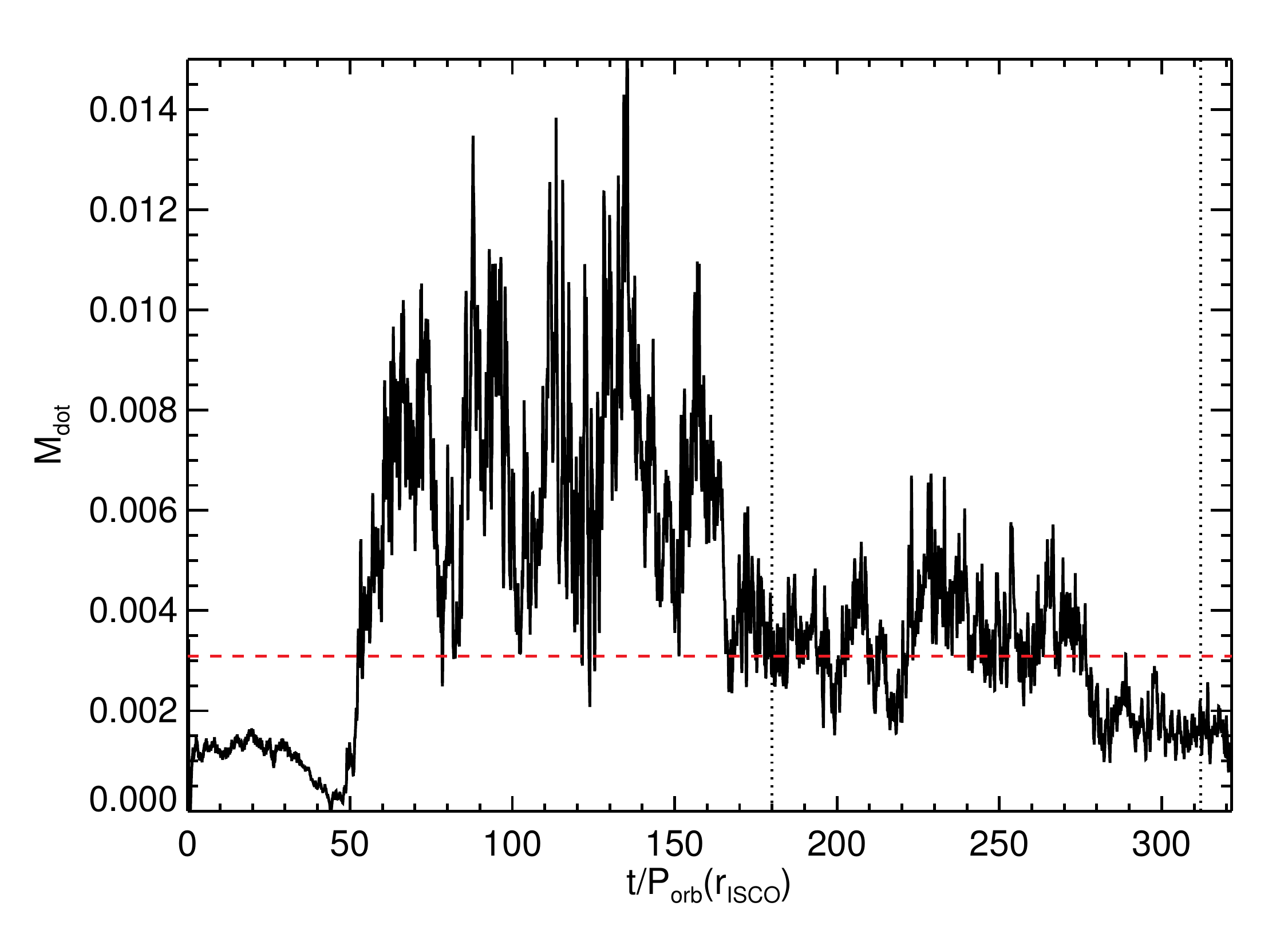}
\includegraphics[width=0.45\textwidth]{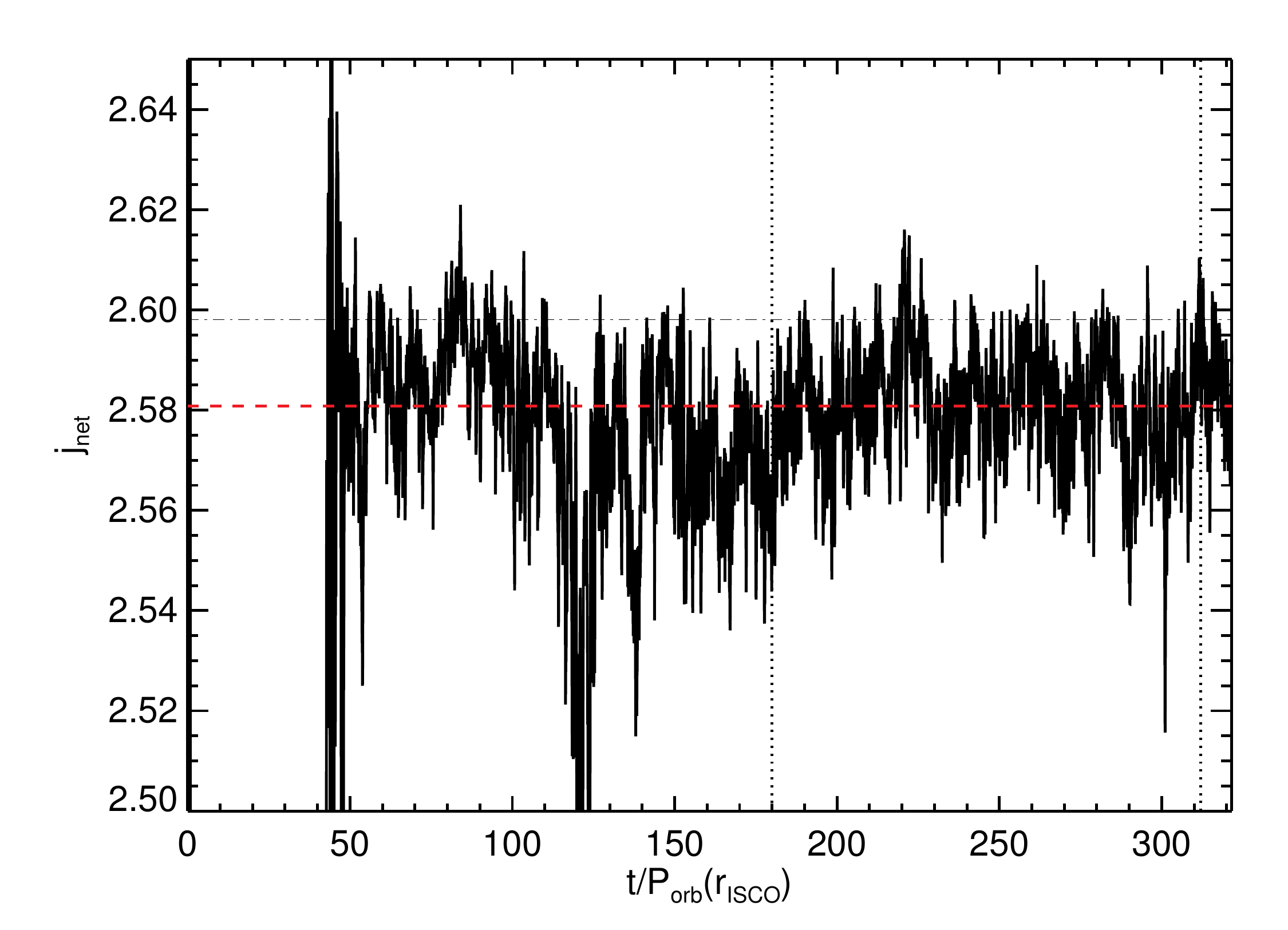}
\end{center}
\caption[]{Time histories of the mass-accretion rate, $\dot{M} (r_{ISCO},t)$ (top panel) and reduced specific angular momentum flux, $j_{net} (r_{ISCO},t)$ (bottom panel) evaluated at the ISCO. Both these quantities were evaluated from simulation data shell-integrated over the `disk-body', $|Z|<H$. In both panels, solid black lines denote simulation data, vertical dotted lines the averaging period, red dashed lines the mean value calculated over this period and in the right panel, black dot-dash lines the angular momentum of a circular orbit at the ISCO.}
\label{isco_flux} 
\end{figure}
 
One possible explanation of the substantially different levels of angular momentum transport measured in this work is that the simulation presented here has a significantly different resolution (measured in terms of the MRI) compared to those of (e.g.) \cite{Fromang:2006,Sorathia:2010}. \cite{Fromang:2006} find that for the MRI to be resolved in a global simulation begun with a net toroidal magnetic field, requires the fastest growing unstable mode of the toroidal field MRI to be resolved by at least $5$ zones. These authors demonstrate that if this criterion is not satisfied over a significant fraction of the disk volume in a time-averaged sense, then the MRI becomes under-resolved and angular momentum transport decays. This requirement can alternatively be expressed in terms of the toroidal ``quality factor'' described by \cite{Hawley:2011}:
\begin{equation}
Q_\phi = \frac{\lambda^{MRI}_\phi}{r d\phi} > 5
 ; \;\;
\lambda^{MRI}_\phi = \frac{2 \pi |B_\phi|}{\Omega \sqrt{\rho}}
\end{equation}
where $\lambda^{MRI}_\phi$ is the ``characteristic'' wavelength of the MRI \cite[closely related, but not precisely equal to the wavelength of the fastest unstable mode, see][]{Hawley:2011}. Figure \ref{lmri_xz} shows  $\left<  \left< Q_\phi \right> \right>$ calculated over the interval $\Delta T=11.5-20 \; P_{orb}(r=15r_S)$. The data of this Figure demonstrate that the MRI is well resolved, i.e. $Q_\phi \ge 5$ throughout the magnetized region of the disk, at least in terms of the criteria specified by \cite{Fromang:2006}. We also note that $Q_\phi > 10$ within the coronal region. The MRI is well-resolved here because the density decreases due to vertical stratification. Magnetic fields rise buoyantly from the disk-body into the corona at constant (if not increasing) field
strength (see e.g. Figure \ref{bphi_xz}). As a result, the Alfven velocity in
the corona is (significantly) greater than in the disk body and hence $\lambda^{MRI}_\phi$ increases. Beyond this, buoyant magnetic structures tend to become dominated by power at large scales \cite[e.g.][]{Suzuki:2009,Blackman:2009}.
Vertically averaging inside the disk body, we find $Q_\phi \sim 10$, corresponding to a characteristic mode with toroidal wave number $m=40$. These results demonstrate that the simulation presented here is of approximately the same effective resolution as the well-resolved models described in \cite{Fromang:2006} which exhibited sustained accretion stresses over many hundreds of inner disk orbits. That is, the different levels of angular momentum transport measured in this work are not the result of substantial differences in the effective resolution of the simulations. We will return to the origin of this discrepancy in \S\ref{stresses}.

That $Q_\phi \ge 5$ is a necessary, but not sufficient demonstration that the MRI is well-resolved in the simulation presented here.
The data of Figure \ref{alpha_vol} shows an apparent secular decrease of $\left< \alpha^m_g(t) \right>$, which is not observed in $ \left< \alpha^m_{m}(t) \right>$. Inspection of simulation data reveals that both that the volume integrated gas and magnetic pressure both decline after $t\sim 5 P_{orb} (r=15r_S)$, i.e. after the linear growth phase of the MRI has completed. The decline of these quantities does not occur in ``lock-step'' however, magnetic pressures decrease more rapidly than gas pressure and $\beta = P_g / P_m$ decreases from $\beta = 10$ initially to $\beta = 25$ at the end of the simulation, behavior consistent with that reported by \cite{Beckwith:2008a}. The data of Figure \ref{alpha_vol} show that as this process occurs, $\left< \alpha^m_g (t) \right>$ remains constant, suggesting that the MRI remains well-resolved throughout this process \cite[see e.g.][]{Hawley:2011}. We therefore conclude that the secular decrease in $\left< \alpha^m_g(t) \right>$ is due to evolution of the accretion flow itself, rather than the MRI being under-resolved here.

%\newpage

\begin{figure}
%\leavevmode
\begin{center}
\includegraphics[width=0.45\textwidth]{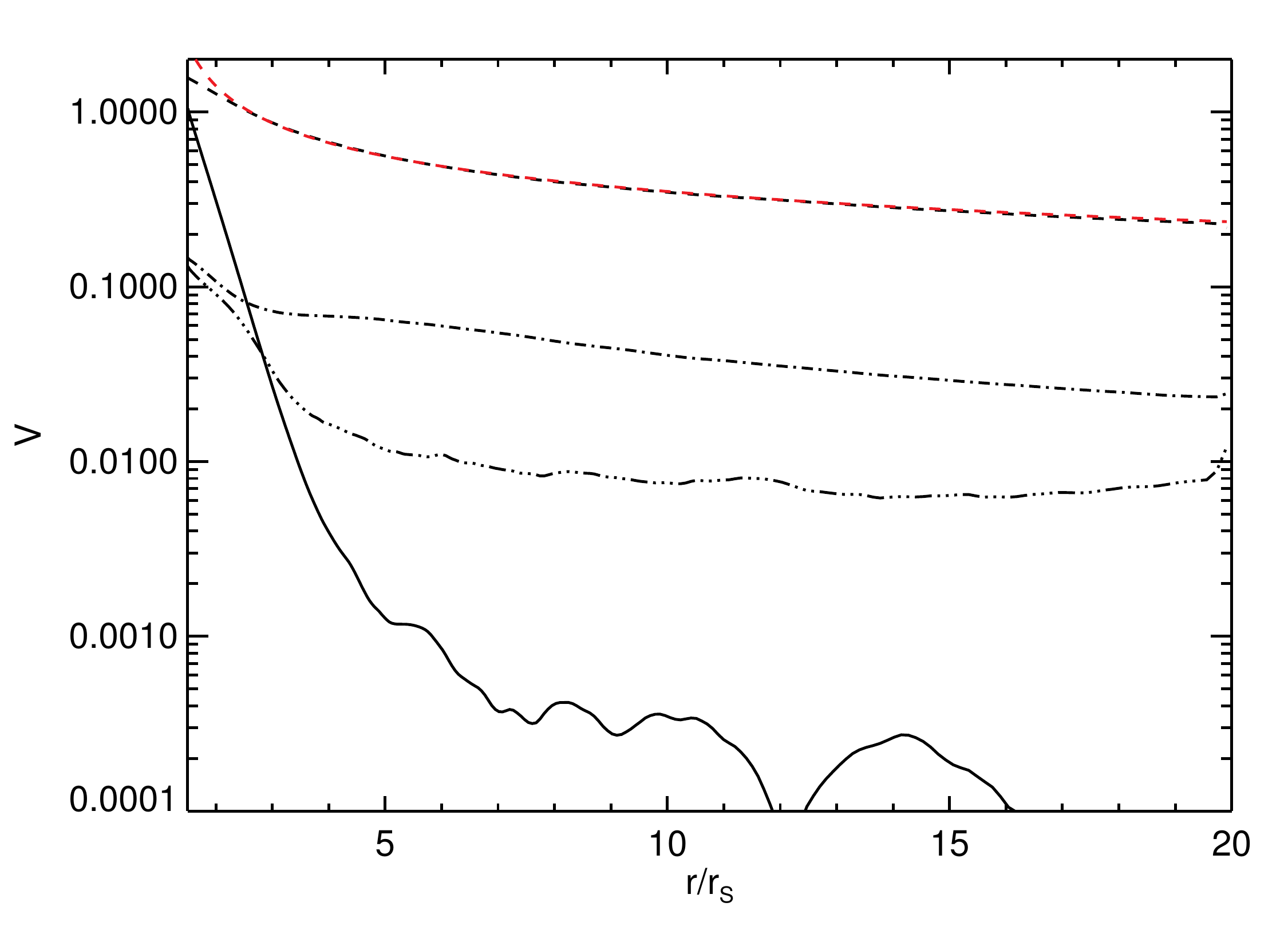}
\end{center}
\caption[]{Radial profiles of characteristic velocities within the accretion flow, $\left< \left< |V| (r) \right> \right>$. All quantities are time-averaged over the period $\Delta T = 10-20 P_{orb} (r=15r_S)$ and shell-averaged over the `disk-body' , $|Z|<H$. Solid lines denote the accretion velocity, $ \left<\left< v_{acc} (r) \right> \right>$, dashed lines the orbital velocity, $\left<\left< v_{\phi} (r) \right> \right>$, dot-dash lines the sound speed, $\left<\left< c_{s} (r) \right> \right>$ and en-dash lines the Alfven speed, $\left<\left< v_{A} (r) \right> \right>$. The red dashed lines show the orbital velocity for a Keplerian angular momentum profile (included for reference). Note that the flow becomes both super-Alfvenic and super-sonic before reaching the inner radial boundary.}
\label{vfield_ravg} 
\end{figure}

\begin{figure}
%\leavevmode
\begin{center}
\includegraphics[width=0.45\textwidth]{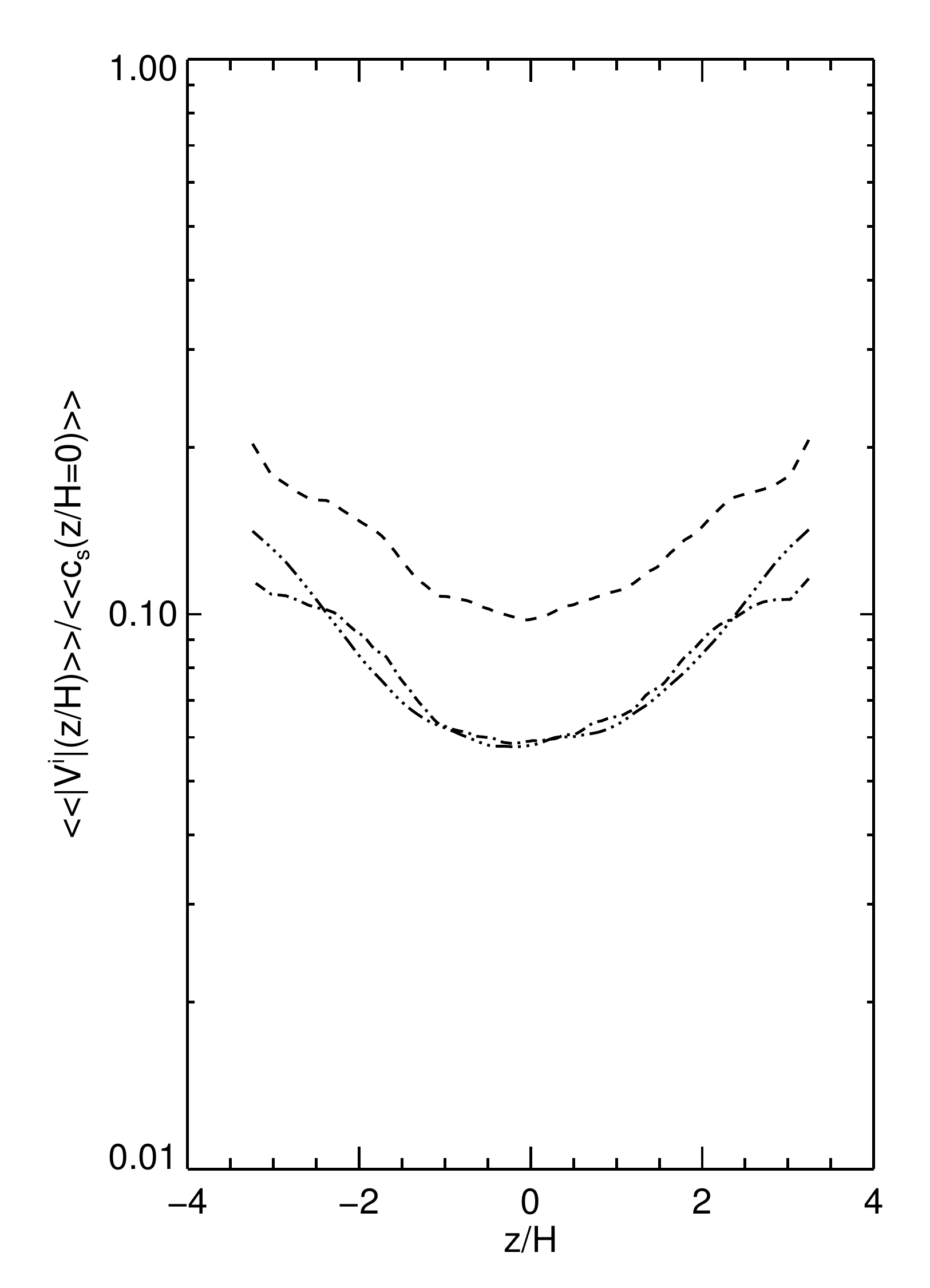}
\end{center}
\caption[]{Vertical profile of turbulent velocity fluctuations, $\left< \left< |\mathbf{V}^i| (z/H) \right> \right>$, in units of the gas sound speed evaluated in the midplane, $\left< \left< c_s (z/H=0) \right> \right>$. Both quantities are time-averaged over the period $\Delta T = 10-20 P_{orb} (r=15r_S)$ and radially-averaged over the region $5\le r/r_S \le 15$. Dashed lines denote $\left< \left< |\mathbf{V}^r| (z/H) \right> \right>$, dot-dash lines $\left< \left< |\mathbf{V}^\phi| (z/H) \right> \right>$ and en-dash lines $\left< \left< |\mathbf{V}^\theta| (z/H) \right> \right>$.}
\label{vel_vert} 
\end{figure}

A further test as to whether the MRI remains well resolved is to examine fluxes of mass and angular momentum through the ISCO. \cite{Noble:2010} demonstrated that if the MRI becomes under-resolved during the evolution, then the ratio of these two quantities shows a secular increase. Figure \ref{isco_flux} shows $\dot{M} (r_{ISCO, t}) =-\left< \rho v_r (r_{ISCO}, t) \right >$ and the net specific angular momentum carried through this surface, $j_{net} (r_{ISCO}, t) = \dot{L}(r_{ISCO}, t) / \dot{M}(r_{ISCO}, t)$, where
\begin{equation}
\dot{L}(r_{ISCO}, t) = r_{ISCO}
\left( \left< \rho v_r v_\phi (r_{ISCO}, t) \right> +
\left< W^t_{r \phi} (r_{ISCO}, t) \right> \right)
\end{equation}
in analogy to the discussion of \cite{McKinney:2002}. These quantities are integrated within the ``disk-body'' (as defined above); extending the integral outside of this region includes contributions from low density, low angular momentum and rapidly inflowing fluid present in the corona. The purpose of this diagnostic is to demonstrate that turbulent fluctuations within the disk body remain well-resolved and so we exclude the coronal material here. The data of Figure \ref{isco_flux} demonstrate that over the duration of the simulation, there are no long term trends in $j_{net}$ evaluated at the ISCO; we note that this quantity time-averaged over $\Delta T =11.5-20 \; P_{orb}(r=15r_S)$ is reduced by approximately $0.5\%$ compared to the angular momentum of a Keplerian orbit at the ISCO, a result consistent with the work of \cite{Beckwith:2008a,Beckwith:2008b} who found significant reductions in electromagnetic stresses at and inside the ISCO for models begun with toroidal as compared to poloidal magnetic field distributions and hence net angular momentum fluxes consistent with that of a Keplerian orbit at the ISCO. The mass accretion rate through this surface displays more complex behavior with at least two different states evident; a high state prior to $150$ ISCO orbits (corresponding to $11.5$ orbits at $r=15r_S$) and a subsequent low state. Fluctuations about the mean in the low state occur within $\sim \pm 1$ standard deviation; suggesting that disk has entered a quasi-stationary state during this period. This is again consistent with the results of \cite{Beckwith:2008a} who found that it took approximately $12$ outer disk orbits for toroidal field models to establish a quasi-stationary state in terms of the mass accretion rate.
 
\section{Structure of the Disk}\label{disk}

Having demonstrated that the development of non-linear magnetohydrodynamic turbulence within the disk conforms to the expectations of previous studies and that the MRI remains well-resolved throughout the simulation, we now characterize the radial and vertical structure of the disk. This is accomplished through the shell- and time-averaged radial profiles, where the limits of integration are restricted to the ``disk-body'' (i.e. $|Z|<H$) and disk area and time-averaged vertical profiles averaged over the radial range $5 \le r/r_S \le 15$. Both of these diagnostics are calculated as described in \S\ref{reduce} and time-averaged over the period $\Delta T=11.5-20 \; P_{orb}(r=15r_S)$.

\subsection{Velocities}\label{diskvel}

\begin{figure}
%\leavevmode
\begin{center}
\includegraphics[width=0.45\textwidth]{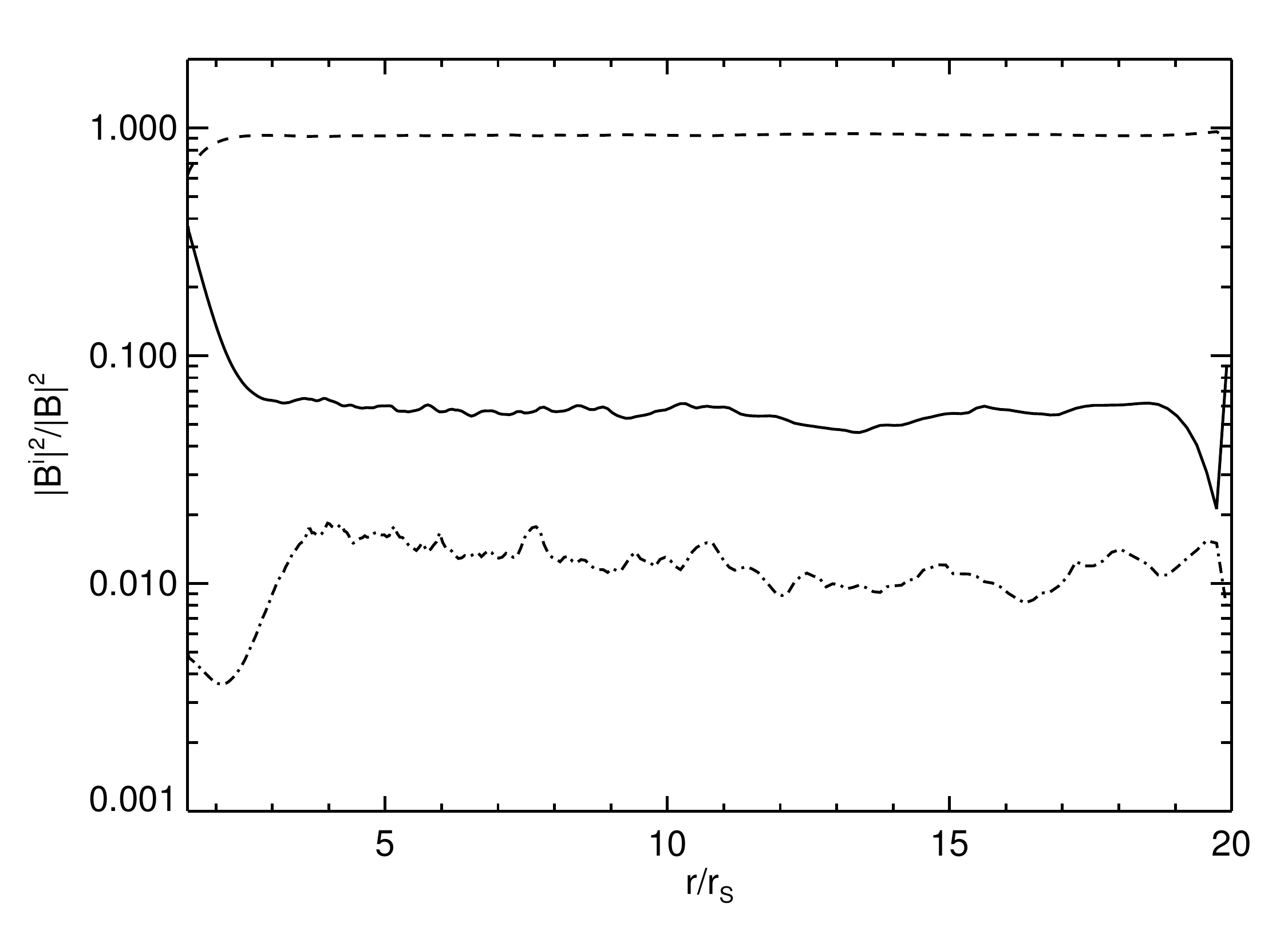}
\end{center}
\caption[]{Radial profiles of components of the magnetic field, time-averaged and vertically integrated within the disk-body, $\left< \left< |\mathbf{B}^i|^2 (r/r_S) \right> \right>$, normalized to magnetic field strength, $\left< \left< |\mathbf{B}|^2 (r/r_S) \right> \right>$. Solid lines denote $\left< \left< |\mathbf{B}^r|^2 (r/r_S) \right> \right>$, dashed lines $\left< \left< |\mathbf{B}^\phi|^2 (r/r_S) \right> \right>$ and dot-dash lines $\left< \left< |\mathbf{B}^\theta|^2 (r/r_S) \right> \right>$.}
\label{bfield_ravg} 
\end{figure}

We begin by considering radial profiles of several characteristic velocities within the accretion disk body, including the Alfven speed $\left<\left< v_{A} (r) \right> \right>$, sound speed $\left<\left< c_{s} (r) \right> \right>$, accretion velocity,
\begin{equation}
\left<\left< v_{acc} (r) \right> \right> =-\left<\left< \rho v_r (r) \right> \right> / \left<\left< \rho(r) \right> \right>
\end{equation}
and orbital velocity
\begin{equation}
\left<\left< v_{\phi} (r) \right> \right> = r^{-1} \left<\left< \rho \ell (r) \right> \right> /  \left<\left< \rho (r) \right> \right>
\end{equation}
Radial profiles of these quantities are shown in Figure \ref{vfield_ravg}. The data of this figure show that, well outside the ISCO ($r\ge5r_S$ which we term the ``outer'' disk) the disk is characterized by highly supersonic orbital motion, subsonic Alfven velocities (corresponding to $\beta = P_g / P_m \sim 25$) and slow inward radial drift characterized by $\left<\left< v_{acc} (r) \right> \right> / \left<\left< v_{\phi} (r) \right> \right> \lesssim 10^{-3}$. As we move inwards through the disk towards the ISCO, the accretion velocity increases, eventually exceeding the Alfven speed at approximately the radius of the ISCO and the sound speed just inside this point,  i.e. the flow outside of the ISCO is upstream of both a sonic and Alfven point, such that the inner boundary should not influence the structure of the outer accretion disk \cite[as described by][see \S\ref{numerics}]{McKinney:2002}. We also note that both the ordering of velocities within the outer disk and their radial dependence is consistent with those reported by \cite{Hawley:2000}, with two caveats. Firstly, the accretion velocity in the outer disk is approximately an order of magnitude less and furthermore, the Alfven point is closer to the ISCO by $\Delta R \sim 1$ than reported by Hawley. The simulations reported by Hawley are, however, moderately thick accretion torii. A better point of comparison then is the data of \cite{Fromang:2006,Reynolds:2008} who examine the accretion velocity within thin magnetized accretion disks. \cite{Fromang:2006} find radial velocities consistent with those demanded by the equations of standard thin accretion disk theory \citep{Pringle:1981}, where the inflow timescale is much longer than the orbital timescale. Clearly, our results are consistent with this requirement. \cite{Reynolds:2008} demonstrate that the form of the accretion velocity within the disk is consistent with slow inward radial drift well outside of the ISCO transitioning to ballistic infall close to the ISCO. Additionally, \cite{Reynolds:2008} find that the Alfven point is located a distance $\sim H$ inside the ISCO, again consistent with the results presented here.

\begin{figure}
%\leavevmode
\begin{center}
\includegraphics[width=0.45\columnwidth, viewport=10 10 350 620,clip]{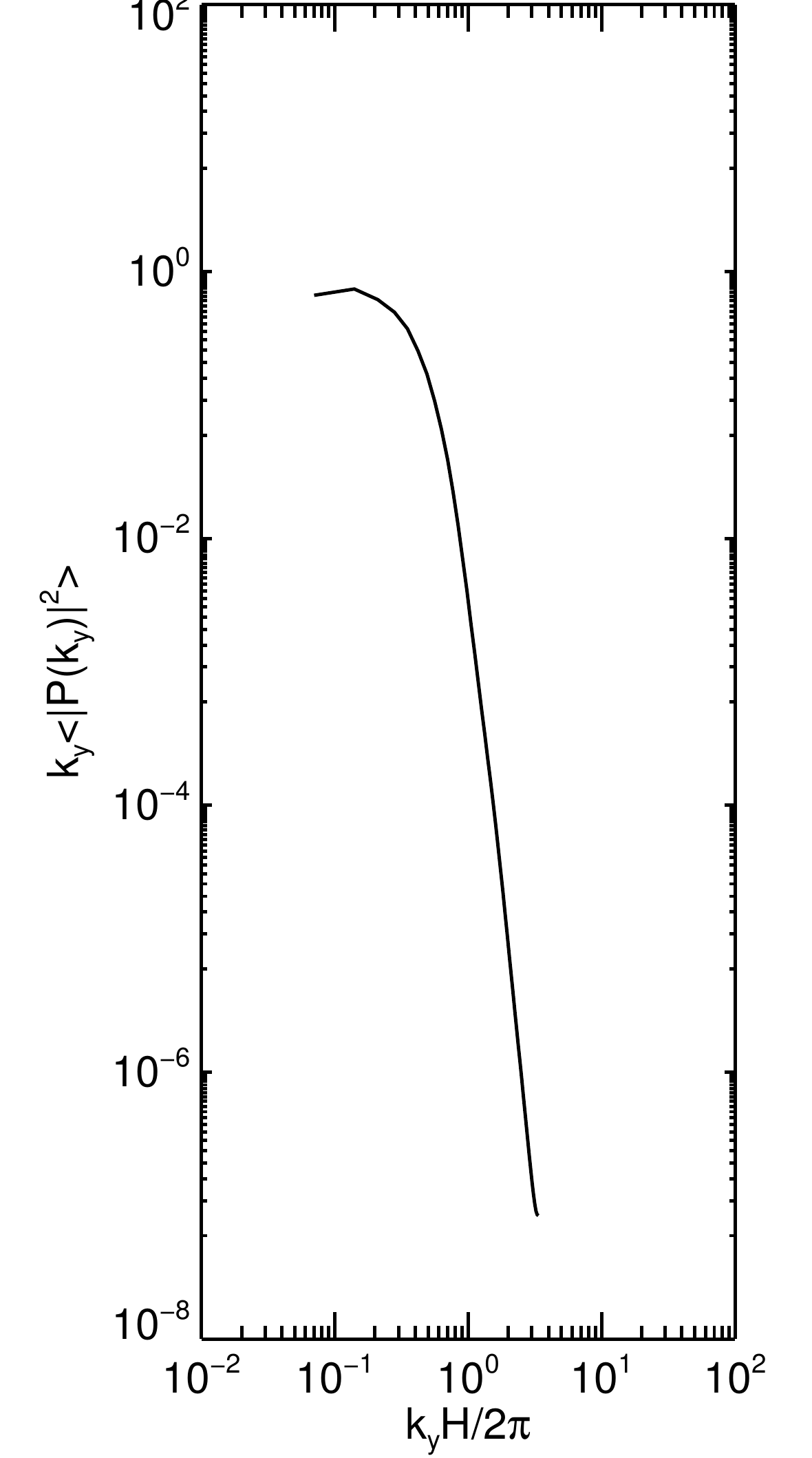}
\end{center}
\caption[]{Toroidal power spectrum of magnetic field strength, $\left< \left< \left| \left< P_{|\mathbf{B}|^2}(k_y) \right> \right|^2 \right> \right>$, plotted as $k_y \left< \left< \left| \left< P_{|\mathbf{B}|^2}(k_y) \right> \right|^2 \right> \right>$ such that the $y$-axis is proportional to the fractional contribution to the  total power per logarithmic interval in $k_y$. In calculating this quantity, simulation data was first integrated vertically inside the `disk-body' , $|Z|<H$, to yield $\left< |B|^2 (r/r_S,\phi,t) \right>$. This quantity was then Fourier transformed along the $\phi$-axis to obtain $\left| \left< P_{|\mathbf{B}|^2}(k_y) \right> \right|^2$. Finally, $\left| \left< P_{|\mathbf{B}|^2}(k_y) \right> \right|^2$ was time-averaged over the period $\Delta T = 11.5-19 P_{orb} (r=15r_S)$ and radially-averaged over the region $5\le r/r_S \le 15$.}
\label{bsq_y_fft} 
\end{figure}

\begin{figure}
%\leavevmode
\begin{center}
\includegraphics[width=0.45\textwidth]{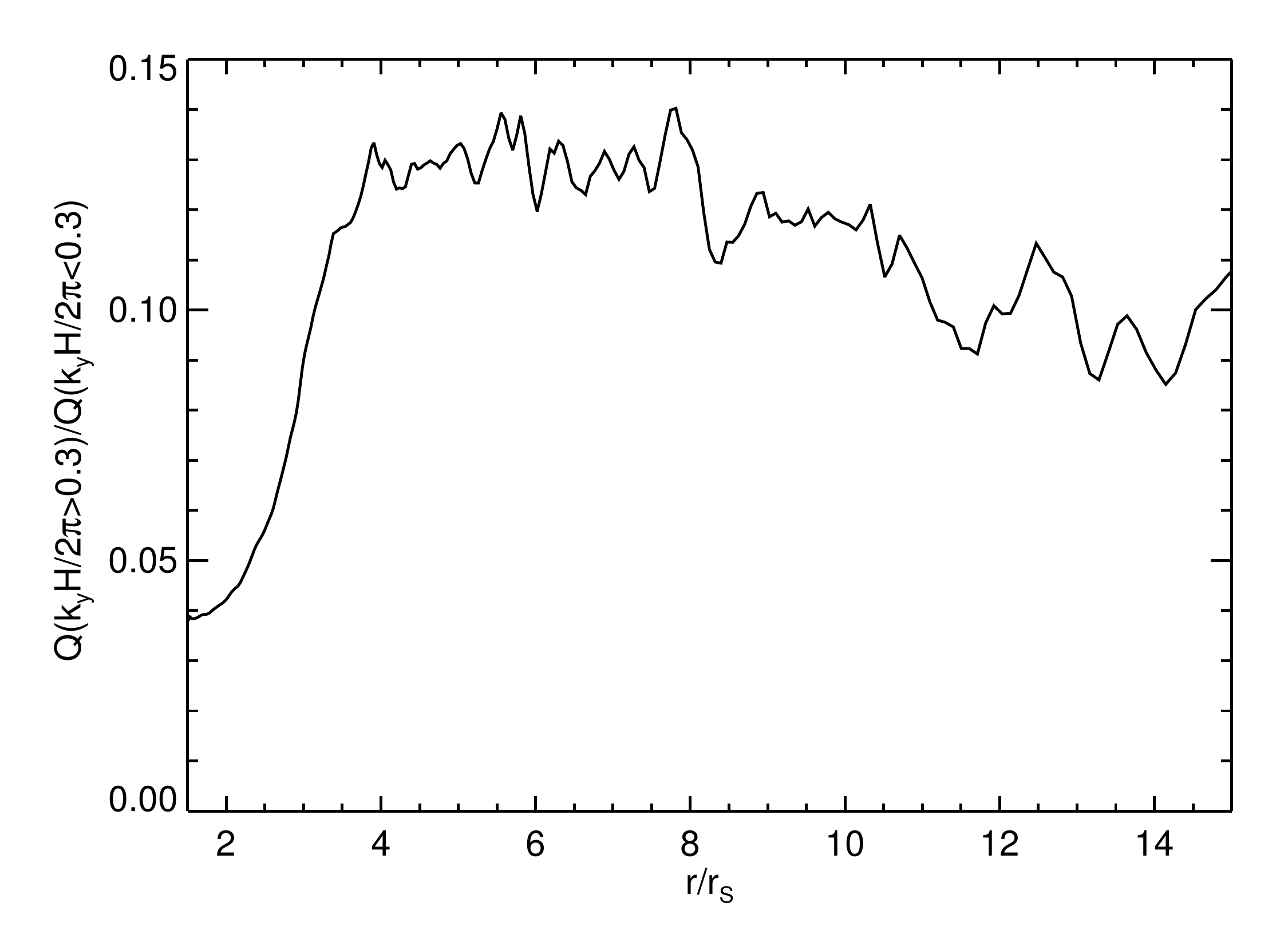}
\end{center}
\caption[]{Ratio of power in the magnetic field strengthon small toroidal scales ($k_yH/2\pi\ge0.3$, corresponding to $\sim3H$) to power in this same quantity on large toroidal scales ($k_yH/2\pi<0.3$). In calculating this quantity, simulation data was integrated vertically inside the `disk-body' , $|Z|<H$, to yield $\left< |B|^2 (r/r_S,\phi,t) \right>$. This quantity was filtered in Fourier space to include contributions from either $k_yH/2\pi\le0.2$ or $k_yH/2\pi<0.2$ and then transformed back into real space. Finally, the resulting filtered data was averaged in the toroidal direction and time-averaged over the period $\Delta T = 11.5-19 P_{orb} (r=15r_S)$.}
\label{bsq_fluct} 
\end{figure}

Figure \ref{vel_vert} shows the vertical structure of turbulent velocity fluctuations, $\left< \left< |\delta V^i| (z/H) \right> \right>$ in units of the gas sound speed evaluated at the midplane, $\left< \left< c_s (z/H=0) \right> \right>$. In calculating the data shown in this figure, we averaged simulation data over the disk surface area between $5 \le r/r_S \le 15$ over all $\phi$ and time-averaged over the period $\Delta T = 10-20 P_{orb} (r=15 r_S)$. The data of this figure are directly comparable with that of Figure 11 of \cite{Fromang:2006}. We find a similar structure to the turbulent velocity fluctuations within the disk to that described by these authors; fluctuations are dominated by the radial component of the velocity, which are approximately a factor of two large at the midplane than fluctuations in the $\phi-$ or $\theta-$ components of the velocity. \cite{Fromang:2006} associate this enhanced fluctuations in the turbulent radial velocities with radially propagating spiral density waves, which are evident in the simulation presented here (as can be seen from Figure \ref{fluid_state}). At the midplane, the radial velocity fluctuations are approximately $10\%$ that of the sound speed here, rising to approximately $20\%$ of the midplane sound speed at $|Z|=3H$, a profile closely followed by turbulent fluctuations in the $\phi-$ and $\theta-$ components of the velocity, but lower in amplitude at all heights by factors $\sim2$.

\subsection{Magnetic Fields}\label{diskfield}

Figure \ref{bfield_ravg} shows the radial profile of the shell- and time-averaged magnetic field, $\left<\left< |\mathbf{B}^i|^2(r) \right> \right>_S$ ($i=r,\theta,\phi$), normalized to the sum of these quantities at each radius.  The magnetic field is predominately toroidal outside of the ISCO; approximately $10\%$ of the total energy in the field is radial while vertical field accounts for only $\sim2\%$ of the total field energy. Inside the ISCO, the balance in field components changes as the fluid plunges into the black hole; here the relative importance of the radial field increases with decreasing radius and there is a corresponding decrease in the relative importance of toroidal and vertical magnetic fields. This result, taken in combination with the data of the Figure \ref{vfield_ravg} suggests that the behavior of the magnetic field inside of the ISCO is determined by flux-freezing, rather than turbulence. To confirm this suggestion, consider the one-dimensional toroidal power spectrum of the magnetic field strength, $\left< \left< \left| \left< P_{|\mathbf{B}|^2}(k_y) \right> \right|^2 \right> \right>$, shown in Figure \ref{bsq_y_fft}, calculated as described in \S\ref{power_spectra} where the radial average was computed over $5 \le r/r_S \le 15$. Note the location of the break in the power spectrum at $k_y H / 2\pi = 0.3$, corresponding to a spatial scale of $3H$ (the significance of which will be discussed in \S\ref{turbulence}). Figure \ref{bsq_fluct} shows the radial profile of the ratio of power in the magnetic field strength on toroidal scales smaller than this break ($k_yH/2\pi\ge0.3$) to power in this same quantity on toroidal scales \emph{larger} than this break ($k_y H / 2\pi<0.3$) (calculated as defined in eqn. \ref{power_ratio} with $k_1 H / 2\pi = 0.3$, see \S\ref{power_spectra}). In the region $4\le r/r_S \le 8$, we find that the ratio of these two measures is $\sim 0.13$, while inside of $4r_S$, power on toroidal scales smaller than the break drops rapidly compared to that on toroidal scales larger than the break. That is, as one approaches the ISCO from larger radii, we measure decreasing power in small scale turbulent fluctuations of the magnetic field in the azimuthal direction, with a corresponding increasing in power in large scale modes, exactly as we would expect if the dynamics of the magnetic field were controlled by flux-freezing, rather than small scale turbulent fluctuations.

\begin{figure}
%\leavevmode
\begin{center}
\includegraphics[width=0.45\columnwidth]{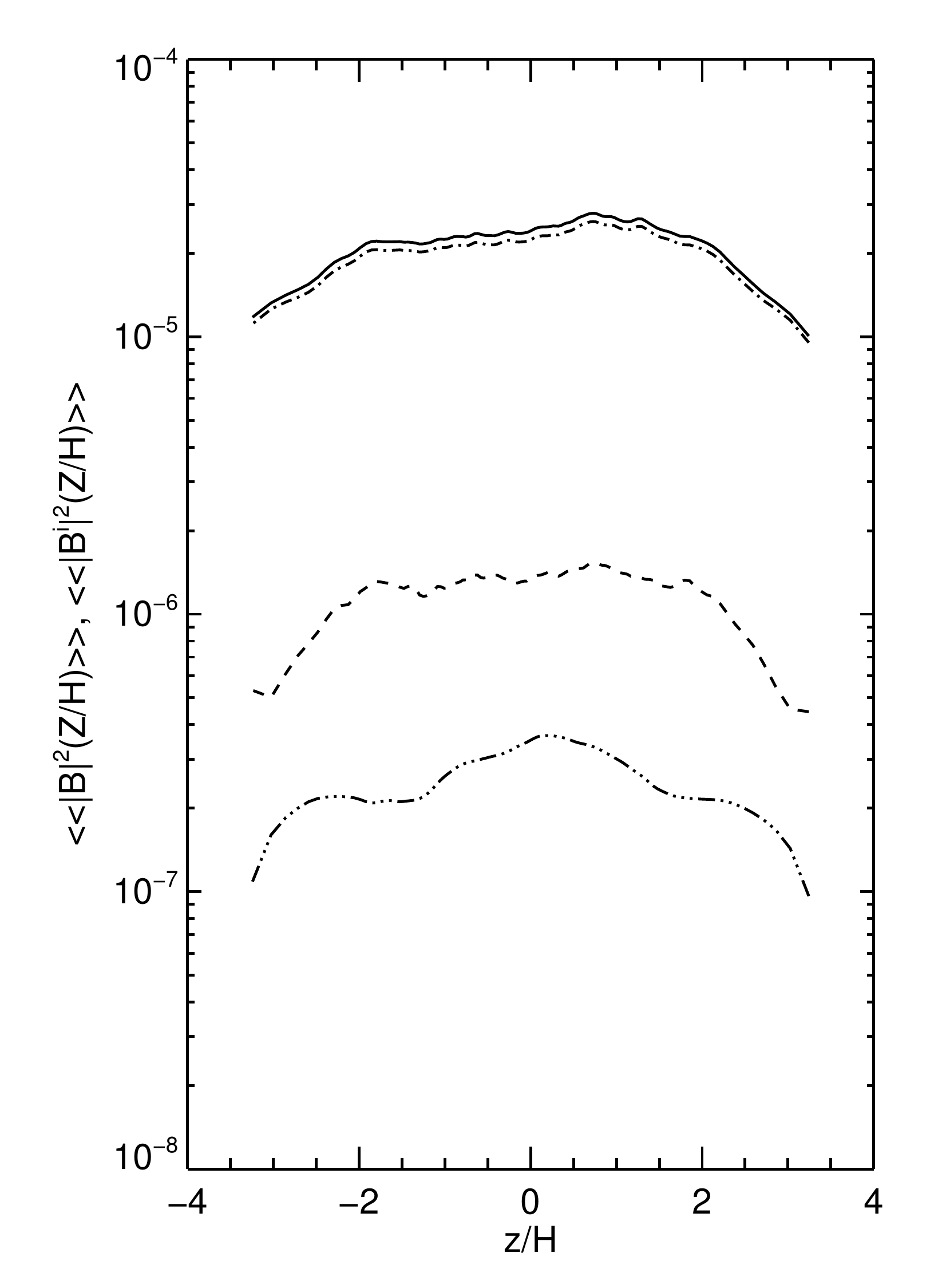}
\includegraphics[width=0.45\columnwidth]{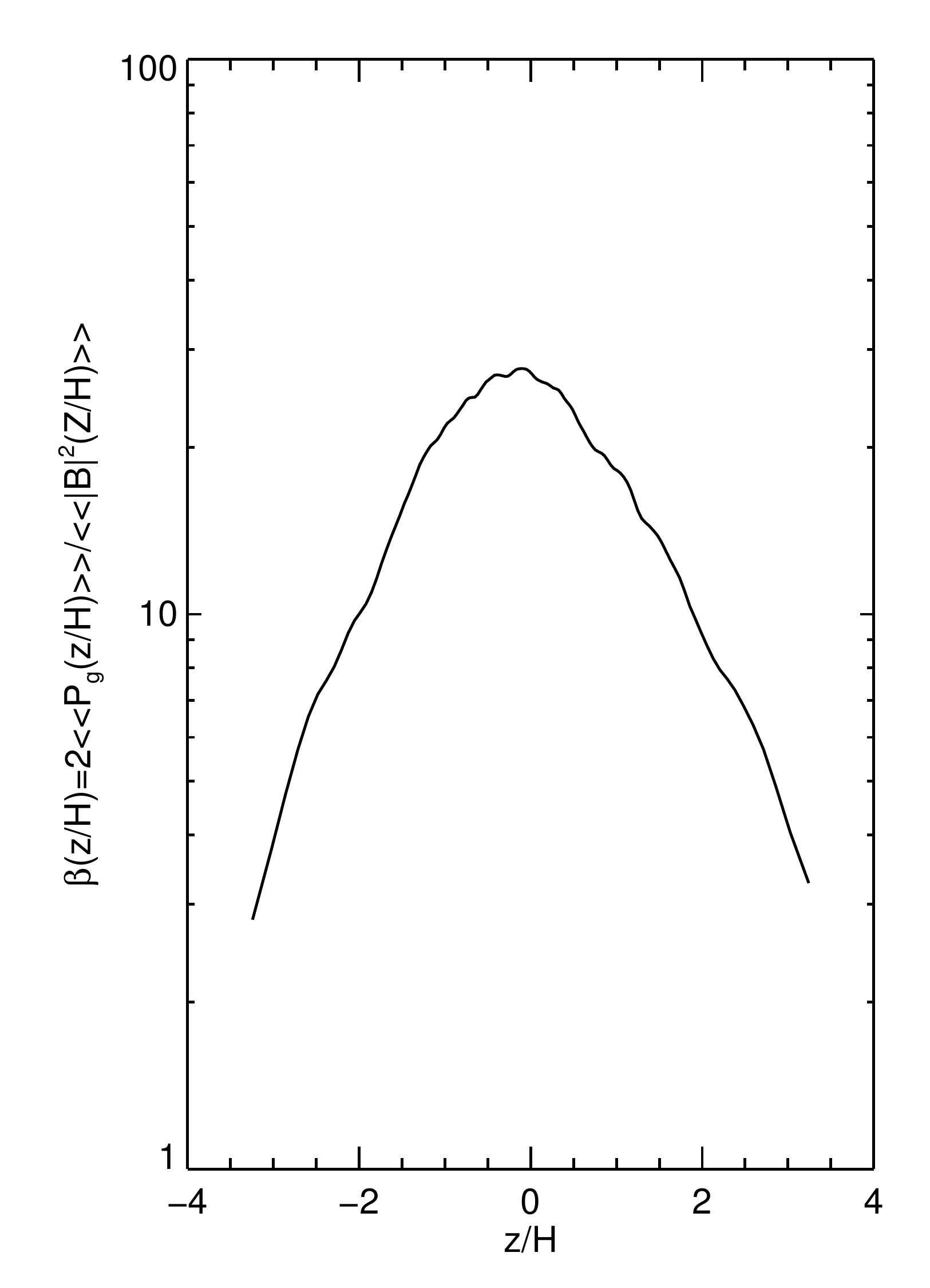}
\end{center}
\caption[]{Vertical profiles of magnetic field strength, $\left< \left< |\mathbf{B}|^2 (z/H) \right> \right>$ and $\left< \left< |\mathbf{B}^i|^2 (z/H) \right> \right>$ (left panel) and $\beta = 2\left< \left< P_g (z/H) \right> \right>/\left< \left< |\mathbf{B}|^2 (z/H) \right> \right>$ (right panel). All quantities are time-averaged over the period $\Delta T = 11.5-19 P_{orb} (r=15r_S)$ and radially-averaged over the region $5\le r/r_S \le 15$. In the left-hand panel, solid lines denote $\left< \left< |\mathbf{B}|^2 (z/H) \right> \right>$, dashed lines denote $\left< \left< |\mathbf{B}^r|^2 (z/H) \right> \right>$, dot-dash lines $\left< \left< |\mathbf{B}^\phi|^2 (z/H) \right> \right>$ and en-dash lines $\left< \left< |\mathbf{B}^\theta|^2 (z/H) \right> \right>$.}
\label{bfield_vert} 
\end{figure}

Figure \ref{bfield_vert} shows the vertical structure of the magnetic field strength, $\left< \left< |\mathbf{B}|^2 (z/H) \right> \right>$ and the magnetic field components, $\left< \left< |\mathbf{B}^i|^2 (z/H) \right> \right>$, along with that of the gas $\beta$ parameter, where 
$$\beta = 2\left< \left< P_g (z/H) \right> \right>/\left< \left< |B|^2 (z/H) \right> \right>$$
As with the vertical profile of turbulent velocity fluctuations discussed previously, to compute the data shown in this figure, we averaged simulation data over the disk surface area between $5 \le r/r_S \le 15$ over all $\phi$ and time-averaged over the period $\Delta T = 10-20 P_{orb} (r=15 r_S)$.  The vertical profile of each of the magnetic field components, along with that of the magnetic field strength,  show a similar vertical structure; approximately constant inside of $|Z|\le2 H$ and the falling by a factor $\sim2-3$ at higher latitudes. The relative strengths of the magnetic field components is as expected from the radial profile shown in Figure \ref{bfield_ravg}, the field is predominantly toroidal, with radial contributions at roughly the $10\%$ level and small contributions from vertical fields. Unlike in \cite{Fromang:2006}, there is no hint that the magnetic field topology changes as one moves from the `disk-body' to higher latitudes, perhaps due to our utilization of periodic vertical boundary conditions. The vertical profile of the gas $\beta$ parameter is shown in the right-hand panel of Figure \ref{bfield_vert}. We find that this quantity varies by roughly an order of magnitude over the vertical extent of the disk, from $\beta \sim 30$ in the disk midplane, to $\beta = 3$ at $|Z|=3H$. On average, therefore, the corona is only moderately magnetized in this simulation. Recall, however, that there are patches of the corona where $\beta \lesssim 1$, as evident in Figure \ref{fluid_state}, emphasizing the importance of non-axisymmetric structures. The vertical dependence of $\beta$ the disk magnetization intermediate between the results of \cite{Fromang:2006} and \cite{Beckwith:2008a}. The former set of authors report stronger magnetizations in the disk midplane ($\beta \sim 10$) and slightly stronger magnetizations in the corona ($\beta \sim 1$), whilst the latter set of authors report magnetizations weaker in the midplane ($\beta \sim 100$) and slightly stronger in the corona ($\beta \sim 1$). The range in magnetization found here is therefore similar to that found in \cite{Fromang:2006}, but smaller than that of \cite{Beckwith:2008a}, this latter contrast perhaps due to the use of full GRMHD by \cite{Beckwith:2008a}. We also note here that even though the magnetic field strengths found in this simulation are approximately a factor of three weaker than those reported by \cite{Fromang:2006}, the volume averaged Maxwell stress levels found here are nearly an order of magnitude greater than those reported by these authors.

\subsection{Accretion Stresses}\label{stresses}

\begin{figure}
%\leavevmode
\begin{center}
\includegraphics[width=0.45\textwidth]{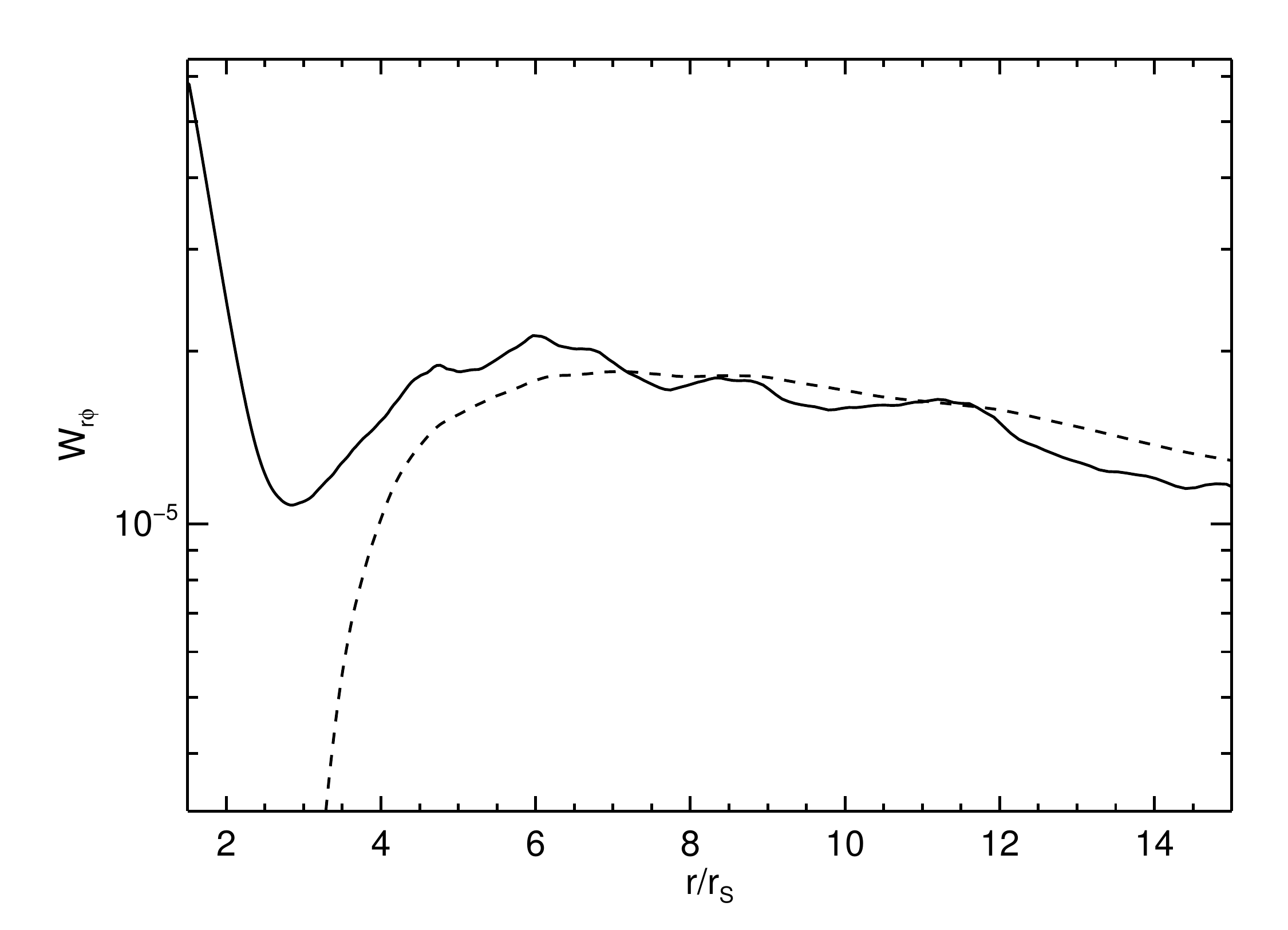}
\end{center}
\caption[]{Radial profile of accretion stress, $\left< \left < W^{t}_{r\phi} (r)\right> \right>$ (solid line) compared to prediction of a model that assumes zero ISCO stress (dashed line). Simulation data were time-averaged over the period $\Delta T = 11.5-19 P_{orb} (r=15r_S)$ and shell-averaged over the `disk-body' , $|Z|<H$.}
\label{stress_ravg} 
\end{figure}

Our next probe of the radial structure of the disk is the accretion stress. As we noted in the previous section, the magnetic field strength in the simulation presented here is a factor of three weaker than that reported by \cite{Fromang:2006}, whilst the volume averaged accretion stress is approximately an order of magnitude greater than reported by these authors. Taken together, this result suggests that at fixed magnetic field strength, the simulation reported here exhibits accretion stresses a factor \emph{thirty} larger than those reported by \cite{Fromang:2006}. Since we have already found that the effective resolution of the simulation presented here is the same as that found by \cite{Fromang:2006} and that the MRI is well-resolved, the origin of this discrepancy must be related to the physical properties of the turbulence in this simulation. We will therefore analyze the properties of the accretion stress directly in order to understand the discrepancy.

\begin{figure}
%\leavevmode
\begin{center}
\includegraphics[width=0.45\columnwidth, viewport=10 10 350 620,clip]
{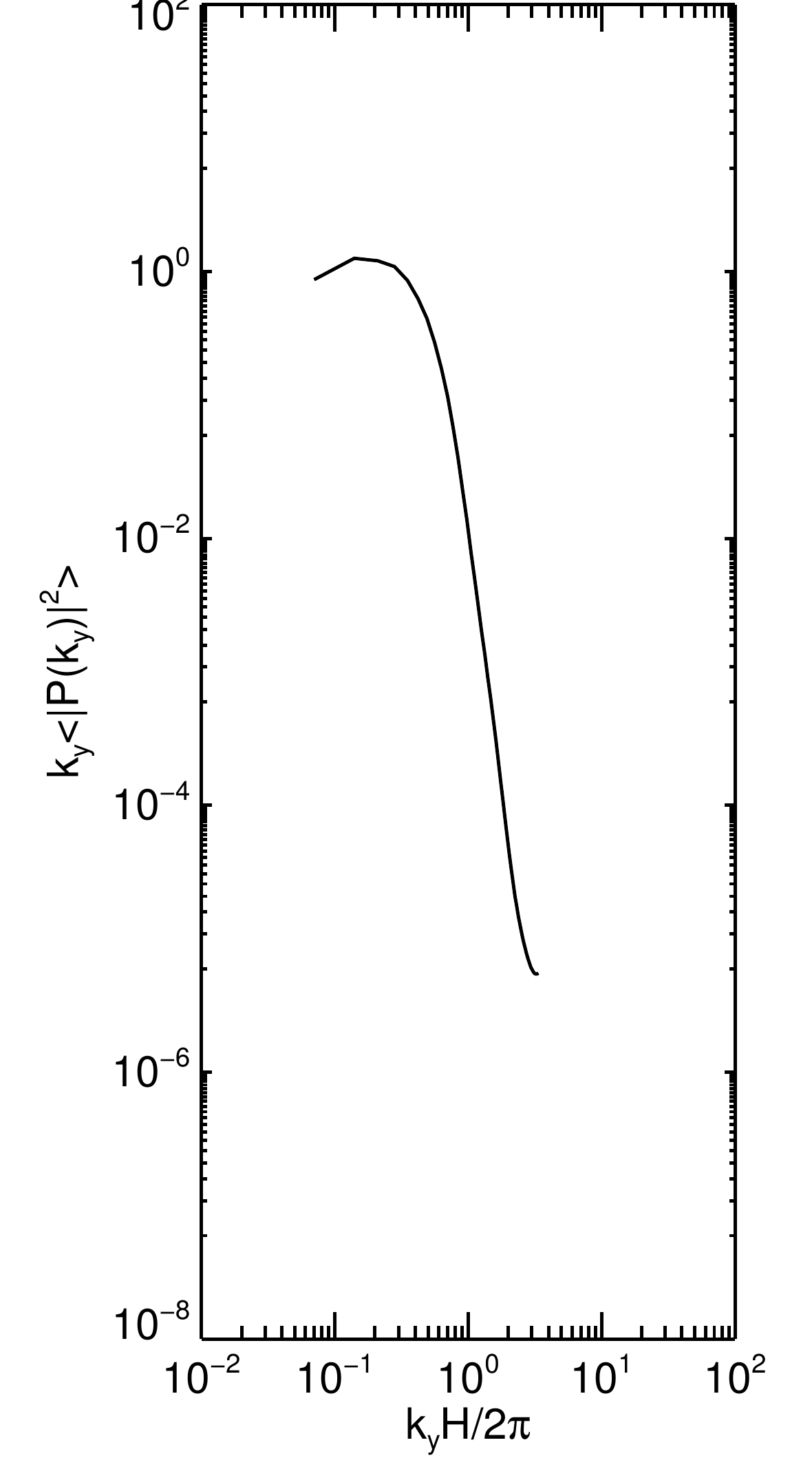}
\end{center}
\caption[]{Toroidal power spectrum of the Maxwell stress, $\left< \left< \left| \left< P_{W^m_{r\phi}}(k_y) \right> \right|^2 \right> \right>$, calculated as described for the magnetic field strength in Figure \ref{bsq_y_fft}.}
\label{stress_y_fft} 
\end{figure}

\begin{figure}
%\leavevmode
\begin{center}
\includegraphics[width=0.45\textwidth]{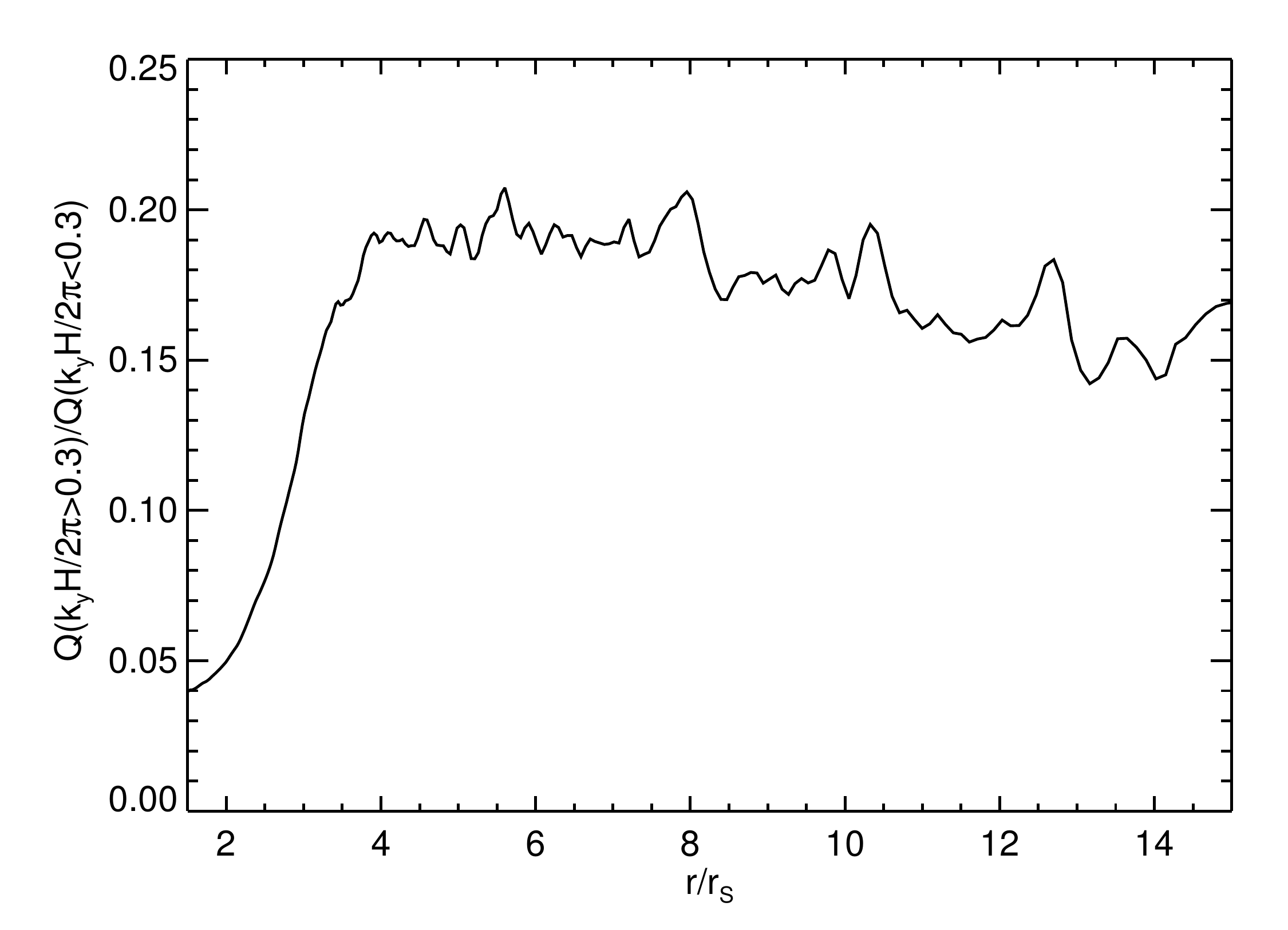}
\end{center}
\caption[]{Ratio of power in the total accretion stress, $\left< \left< W^t_{r \phi} (r) \right> \right>$ on small toroidal scales ($k_yH/2\pi\le0.3$, corresponding to $\sim3H$) to power in this same quantity on large toroidal scales ($k_yH/2\pi<0.3$), calculated as described for the magnetic field strength in Figure \ref{bsq_fluct}.}
\label{stress_fluct} 
\end{figure}

The data of Figure \ref{stress_ravg} shows the time-averaged radial profile of the total accretion stress, $\left<\left< W^{t}_{r \phi} (r) \right> \right>$ compared to the expectation from a combination of arguments regarding angular momentum conservation and the assumption that the ISCO is stress free \cite[see e.g.][for a detailed description]{Hawley:2002,Beckwith:2008b}:
\begin{equation}
W_{r\phi} (R) = \frac{\Omega(R) \dot{M}_{ISCO}}{\pi/2} \left(1 - \frac{\ell_{ISCO}}{\ell(R)} \right)
\end{equation}
Here, $\dot{M}_{ISCO}$ is the time-averaged mass accretion rate through the ISCO, $\Omega(r)$ is the time-averaged orbital angular velocity, $\ell_{ISCO}$ is the time-averaged specific angular momentum of the fluid at the ISCO and $\ell(R) = R^2 \Omega(R)$ is the time-averaged specific angular momentum of the fluid. The data of Figure \ref{stress_ravg} makes clear that angular momentum conservation combined with a stress-free boundary condition applied at the ISCO provides an excellent description of the turbulent stresses within the disk body for $r\ge4r_S$. As one moves in towards the ISCO from $r = 4r_S$, the radial form of the total stress deviates from the prediction of the stress-free ISCO assumption and well inside of the ISCO, the stress profile becomes singular as the inner boundary as approached from above. The preceding discussion suggests that departures from the stress-free inner boundary condition originate from the increasing importance of flux freezing effects as the ISCO is approached from above. We demonstrate this using the same approach outlined above for the magnetic field strength. Figure \ref{stress_y_fft} shows the toroidal power spectrum of the Maxwell stress, $\left< \left< \left| \left< P_{W^m_{r\phi}}(k_y) \right> \right|^2 \right> \right>$, calculated as described in \S\ref{power_spectra} where the radial average was computed over $5 \le r/r_S \le 15$. As was the case for the toroidal power spectrum of the magnetic field strength, a break in this power spectrum is evident on scales $k_y H / 2\pi = 0.3$. Figure \ref{stress_fluct} shows the radial profile of the ratio of power in the total accretion stress on toroidal scales smaller than this break ($k_yH/2\pi\ge0.3$) to power in this same quantity on toroidal scales \emph{larger} than this break ($k_yH/2\pi<0.3$) (calculated as defined in eqn. \ref{power_ratio} with $k_1 H / 2\pi = 0.3$, see \S\ref{power_spectra}). The data of these panels clearly illustrate that contributions from small scale fluctuations to the accretion stress decrease dramatically in importance inside of $4r_S$, confirming the suggestion that inside of $4r_S$, accretion stresses are controlled by flux freezing, rather than turbulence.

\begin{figure}
%\leavevmode
\begin{center}
\includegraphics[width=0.45\textwidth]{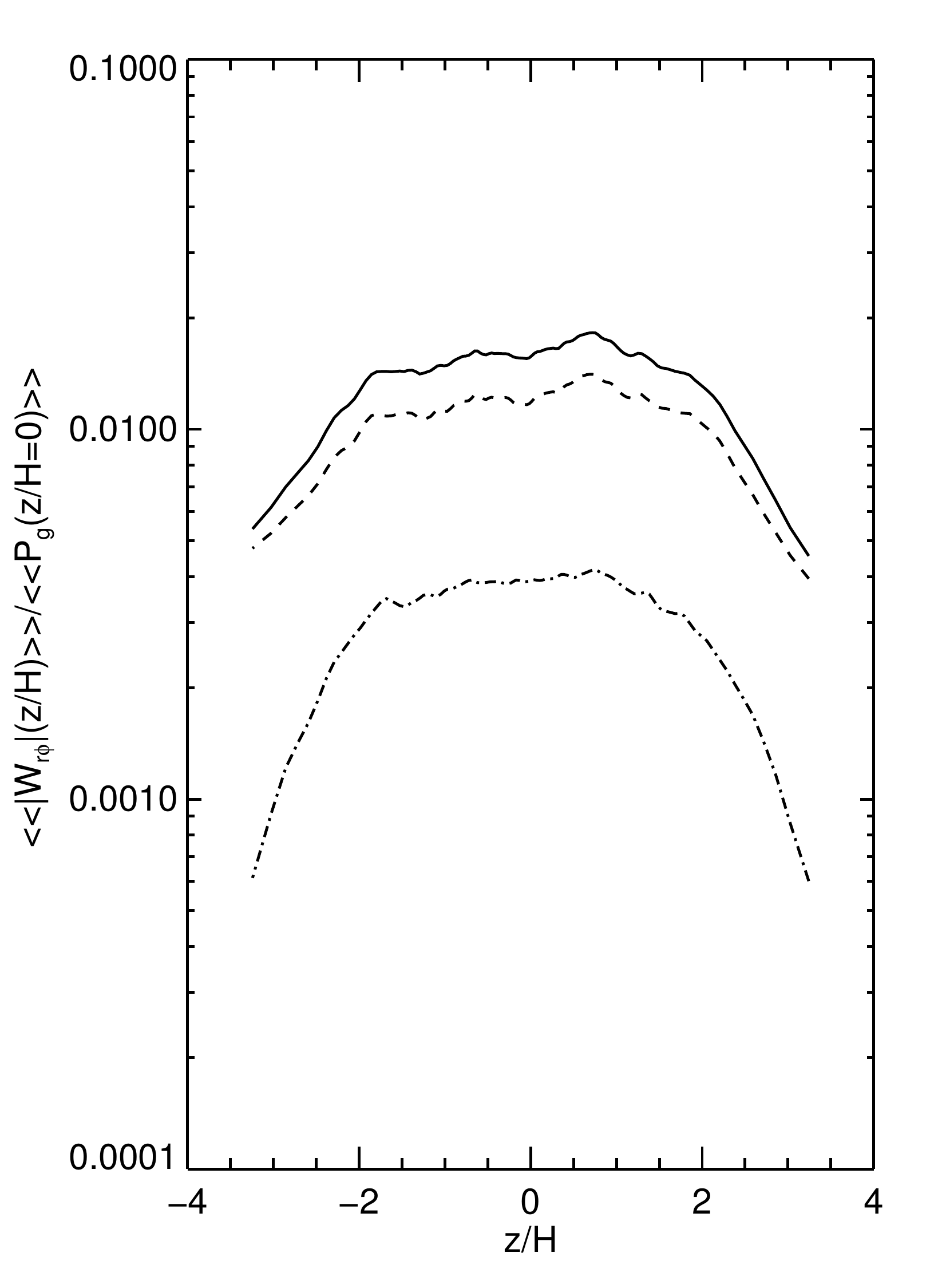}
\end{center}
\caption[]{Vertical profile of accretion stress, $\left< \left< W_{r \phi} (z/H) \right> \right>$, in units of the gas pressure evaluated in the midplane, $\left< \left< P_g (z/H=0) \right> \right>$. Both quantities are time-averaged over the period $\Delta T = 11.5-19 P_{orb} (r=15r_S)$ and radially-averaged over the region $5\le r/r_S \le 15$. Solid lines denote the total stress, $\left< \left< W^t_{r \phi} (z/H) \right> \right>$, dashed lines denote the Maxwell stress, $\left< \left< W^m_{r \phi} (z/H) \right> \right>$ and dot-dash lines the Reynolds stress, $\left< \left< W^f_{r \phi} (z/H) \right> \right>$.}
\label{stress_vert} 
\end{figure}

Next, we consider the vertical structure of the accretion stress, $\left<\left< W^{t,f,m}_{r \phi} (z/H) \right> \right>$ normalized by the gas pressure evaluated at the midplane, $\left<\left< P_g (z/H=0) \right> \right>$. These data are shown in Figure \ref{stress_vert} and were obtained from averaged simulation data over the disk surface area between $5 \le r/r_S \le 15$ over all $\phi$ and time-averaged over the period $\Delta T = 10-20 P_{orb} (r=15 r_S)$. Notably, the vertical profiles of the total, Maxwell and Reynolds stresses are all approximately constant for $|Z|\le2H$, a result consistent with those of \cite{Simon:2010} for vertically stratified shearing box simulations performed at resolutions of $32$ zones per scale height in the ``high'' state. Note that both \cite{Fromang:2006} and \cite{Sorathia:2010} found a distinct `double-peaked' form to the vertical accretion stress profile, reminiscent of the `low' state reported by \cite{Simon:2010}, where resistive effects had resulted in a disk that had little-to-no angular momentum transport and low levels of magnetohydrodynamic turbulence. Intermittent ``double peaked'' structures in the vertical accretion stress profile are also evident in the spacetime diagram of the Maxwell stress presented in \cite{Davis:2010}  Such double peaks are also evident in the data of \cite{Simon:2010} for the ``high'' state. However, when these data are time-averaged over many local orbital periods, these double peaked structures are removed in the ``high'' state, but not for the ``low'' state. The data of both \cite{Fromang:2006,Sorathia:2010} was time-averaged over many orbital periods to obtain the vertical stress profile, suggesting that both of these simulations were in a state equivalent to the ``low'' state described by \cite{Simon:2010}.

A possible explanation as to why the simulations of \cite{Fromang:2006,Sorathia:2010} were in a ``low'' state can be found in the data of Figure \ref{stress_fluct}. This data of this figure suggest that approximately $80\%$ of the total accretion stress is found in modes on scales with $k_y H / 2\pi < 0.3$, that is in modes with toroidal length scales greater than $\sim 3H$, corresponding to an approximate angular extent of $\sim40^\circ$. The simulations described by \cite{Fromang:2006,Sorathia:2010} utilized toroidal domains with angular extents $45^\circ$ and $60^\circ$ respectively. Our results suggest that in the turbulent disk $80\%$ of the accretion stress is found on scales \emph{greater} than $40^\circ$ and as such the simulations described by  \cite{Fromang:2006,Sorathia:2010} \emph{must} significantly underestimate the total accretion stress. The observation that the simulations presented by these authors exhibit similar vertical stress profiles to the ``low'' state described by \cite{Simon:2010} emphasizes the non-linear nature of accretion disk turbulence. Choices regarding toroidal domain size can have a profound impact on the properties of the turbulence itself. An interesting question is therefore whether extending the toroidal domain from $\pi/2$ to $2\pi$ will again influence the turbulence. Previous studies suggest that this is not the case \citep{Hawley:2001}, however, further investigation of this issue is necessary. These calculations are in progress and will be presented in future work.

\section{Structure of the Turbulence}\label{turbulence}

The results of the preceding section, taken in combination of those of \cite{Fromang:2006,Sorathia:2010}, suggest that the extent of the toroidal domain in global simulations plays a key role in determining both the saturation level and the vertical profile of accretion stress within these simulations. This result is somewhat surprising as the toroidal domain size utilized by these authors ($\pi/4$ and $\pi/3$ respectively) is sufficient to capture many disk scale heights, which should be sufficient since results from shearing box calculations suggest that correlations in the turbulence are \emph{local} \citep{Guan:2009a}. This result points to a more fundamental question about the nature of MRI-driven MHD turbulence within accretion disks, namely, is the nature of the turbulence \emph{local} (correlations on size scales smaller than the disk scale height) or \emph{global} (correlations on size scales many times the disk scale height)?

Beyond considerations of the influence of azimuthal domain size, there remain questions regarding the behavior of the periodic variations in the magnetic field described in \S\ref{dynamo}, including: How are these fluctuations arranged spatially? Is there an energy injection scale? What temporal variability patterns characterize the dynamo? Is there a distinction between global and local behavior in this system? In this section, we address the question of whether MRI-driven MHD turbulence within accretion disks is \emph{local} or \emph{global} in nature and investigate the properties of the turbulence on both small and large scales. Our primary diagnostic in performing this analysis are Fourier transforms of scalar (rather than vector) quantities, calculated as described in \S\ref{diagnostics}. We utilize scalar quantities for reasons of both simplicity and computational tractability (this approach reduces the required number of Fourier transforms by a factor three). We have found that while this approximation results in some quantitative changes to the outcome of these calculations, the qualitative conclusions that we draw remain the same. We note that other limitations of our calculations, e.g. finite grid resolution and restricted vertical domain will play at least as important a role in determining precise quantitative outcomes as our use of scalar rather than vector quantities and so, to this extent, we argue that this approximation is justified. The remainder of the section examines the power spectrum of the magnetic field and density fluctuations on the poloidal plane, temporal fluctuations of these quantities at small and large spatial scales, fluctuations in the magnetic  field and density within the disk body and finally the autocorrelation functions of these quantities, again within the disk body.

\subsection{Characteristics of Fluctuations in the Magnetic and Kinetic Energies}\label{fluct}

\begin{figure}
%\leavevmode
\begin{center}
\includegraphics[width=0.45\columnwidth, viewport=10 10 350 625,clip]
{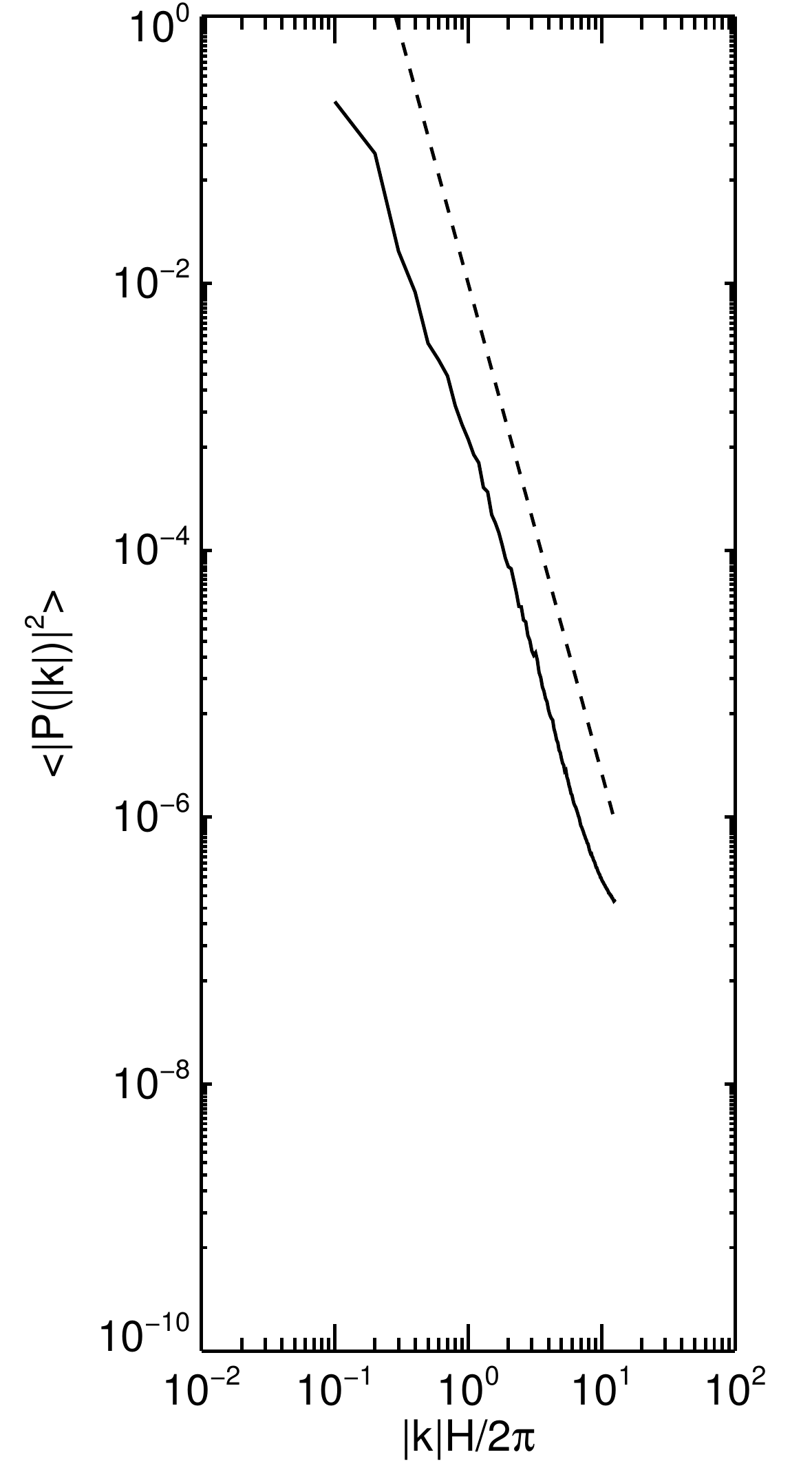}
\includegraphics[width=0.45\columnwidth, viewport=10 10 350 625,clip]
{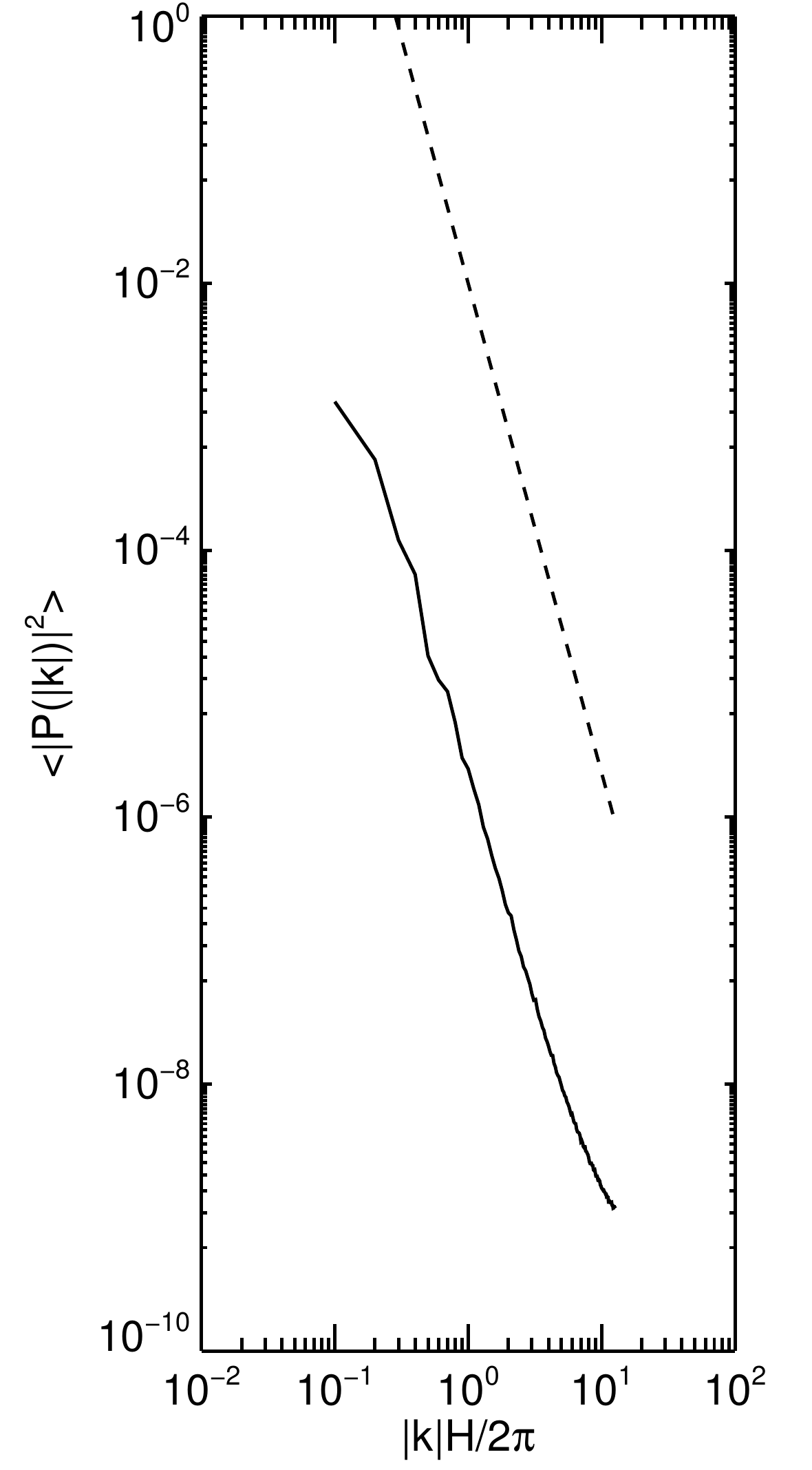}
\end{center}
\caption[]{Two-dimensional power spectrum of magnetic field strength (left panel) and kinetic energy (right panel). Each panel shows $\left< \left| \left< P_{{\cal Q}}(|k|) \right> \right|^2 \right>$ (solid lines) compared to the power spectrum expected for isotropic incompressible turbulence, $|k|^{-11/3}$ (dashed lines) where $|k| = \sqrt{k^2_x + k^2_y}$. The power spectra are calculated as described in \S\ref{diagnostics} and time-averaged over the period $\Delta T = 11.5-19 P_{orb} (r=15r_S)$.}
\label{xz_2d_fft} 
\end{figure}

\begin{figure}
%\leavevmode
\begin{center}
\includegraphics[width=0.45\columnwidth, viewport=10 10 380 625,clip]
{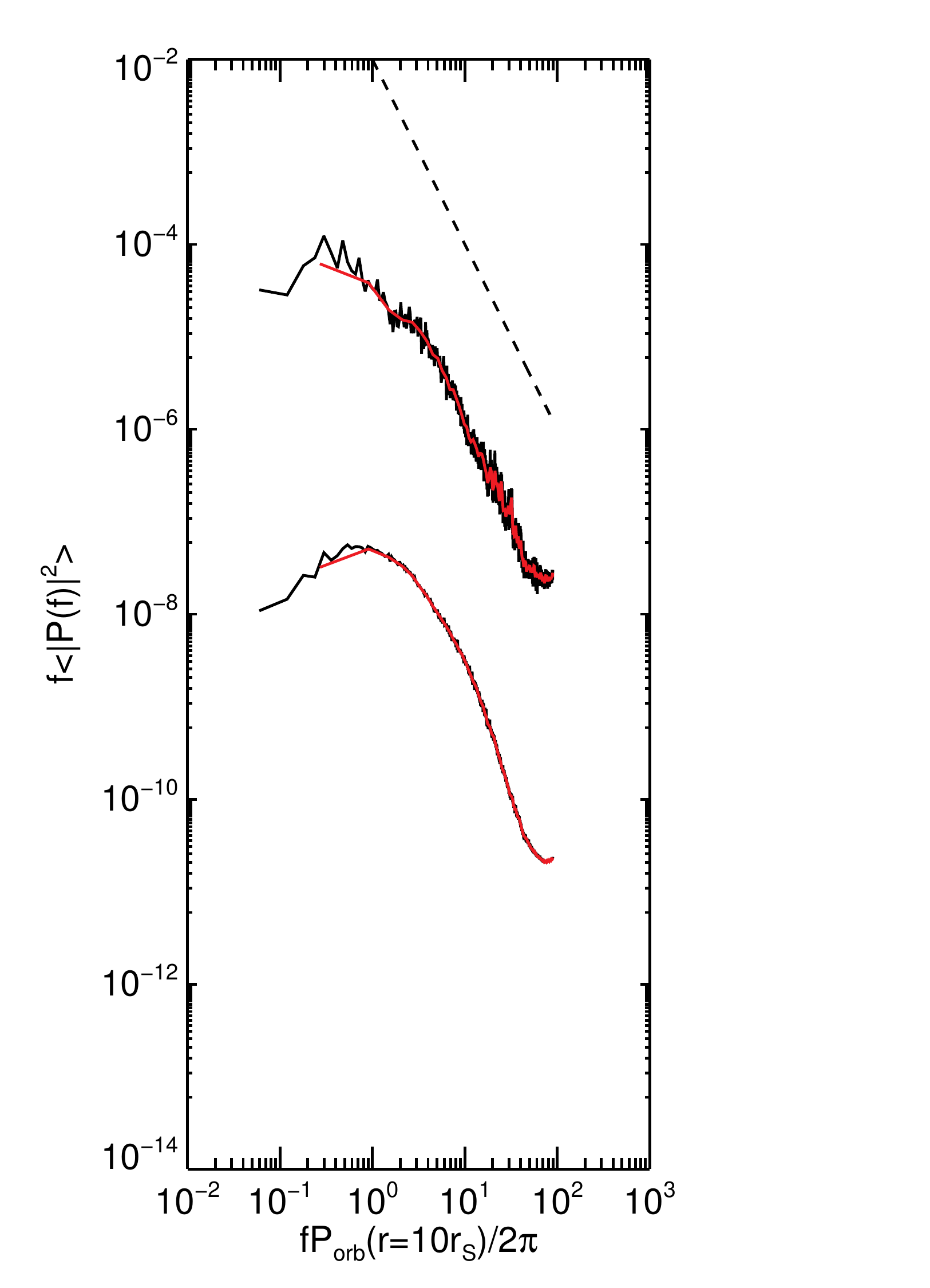}
\includegraphics[width=0.45\columnwidth, viewport=10 10 380 625,clip]
{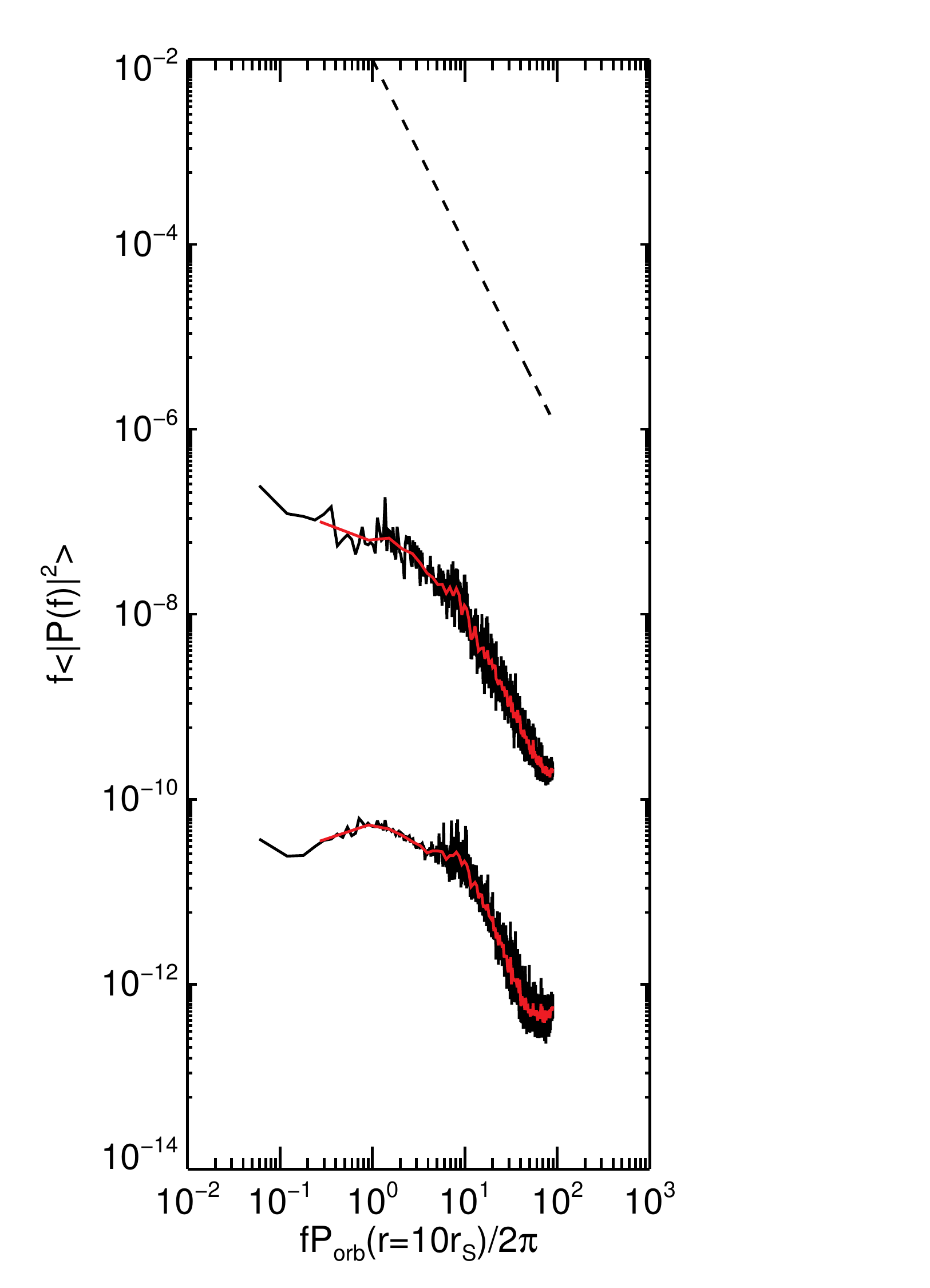}
\end{center}
\caption[]{Power spectra of temporal fluctuations in magnetic field strength (left panel) and kinetic energy (right panel). In both panels, solid black curves denote results from simulation data, the upper curve fluctuations on spatial scales larger than a disk scale height, the lower curve fluctuations on spatial scales smaller than a disk scale height. The solid red curves represent the result of rebinning the raw power spectra onto a grid a factor of $10$ coarser in frequency space. The dashed line denotes the scaling $f^{-2}$. The power spectra are calculated as described in \S\ref{diagnostics}.}
\label{xzt_3d_fft} 
\end{figure}

As a first step in understanding the behavior of MRI-driven MHD turbulence within the disk, we further investigate the periodic variations in the poloidal structure of the magnetic field discussed in \S\ref{dynamo}. The first step in this discussion is to understand how structures in the magnetic and kinetic energies are arranged spatially. We probe this aspect of the simulations via the shell- and time-averaged two-dimensional Fourier transform of toroidally-averaged, spatially normalized simulation data, $\left< \left| \left< P_{|\mathbf{B}|^2}(|k|) \right> \right|^2 \right>$ and $\left< \left| \left< P_{\rho}(|k|) \right> \right|^2 \right>$ calculated as described in \S\ref{power_spectra}. These data, time-averaged over $\Delta T = 10-20 P_{orb} (r=15 r_S)$ using $3000$ frames and calculated from simulation data in the region $5\le r/r_S \le 15$, $|Z| \le 5$ are shown in Figure \ref{xz_2d_fft}. One point of contrast between the two power spectra is that the total power in kinetic energy fluctuations  is approximately an order of magnitude less than that in magnetic energy fluctuations. Apart from this, the power spectra are remarkably similar. The scaling of both power spectra with $|k|$ is well described by the power law $|k|^{-11/3}$, consistent with the Kolmogorov spectrum for homogeneous incompressible turbulence \citep{Hawley:1995}. In both cases, the power spectrum is rather featureless, with no evidence of a break at large scales, a behavior reminiscent of that reported by \cite{Hawley:1995,Simon:2010} for unstratified shearing boxes. These results suggest  that energy is injected at the largest scales within the turbulence and is transferred to progressively smaller scales via a direct cascade, a result consistent with those of \cite{Lesur:2010}. That the power spectrum of the magnetic and kinetic energies is approximately incompressible at small scales is perhaps unsurprising, as velocity fluctuations and magnetic field strengths are largely subthermal in these simulations close to the disk midplane, $|Z|<H$. The extension of this same power law to large scales therefore implies that turbulent fluctuations within the disk remain incompressible  at large scales, despite the increase in magnetic field strength and turbulent velocity fluctuations as one moves to large scales within the disk (see the discussion of \S\ref{diskvel} and \S\ref{diskfield}).

\begin{figure*}
%\leavevmode
\begin{center}
\includegraphics[width=0.32\textwidth, viewport=30 15 475 445,clip]
{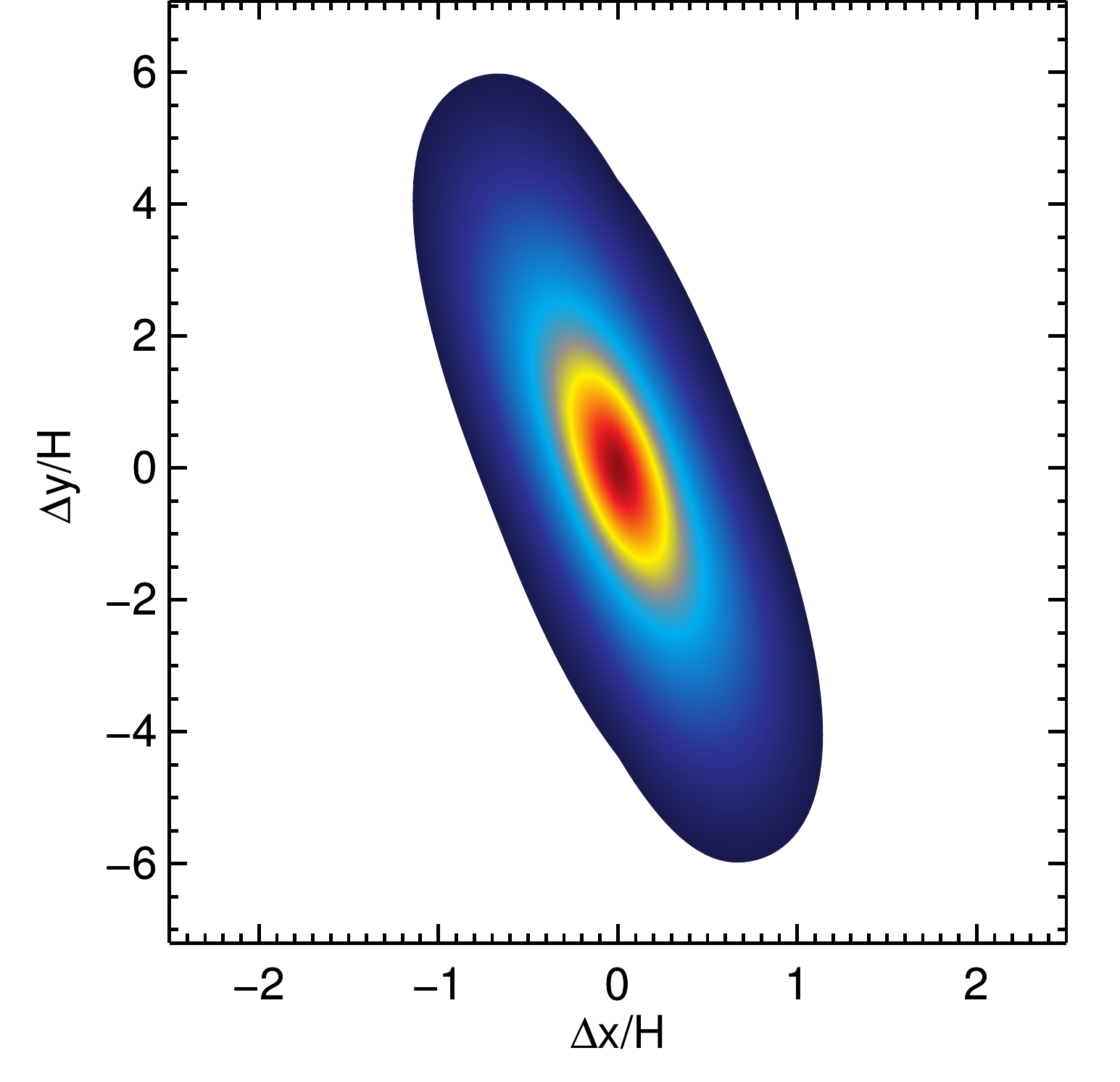}
\includegraphics[width=0.32\textwidth, viewport=30 15 475 445,clip]
{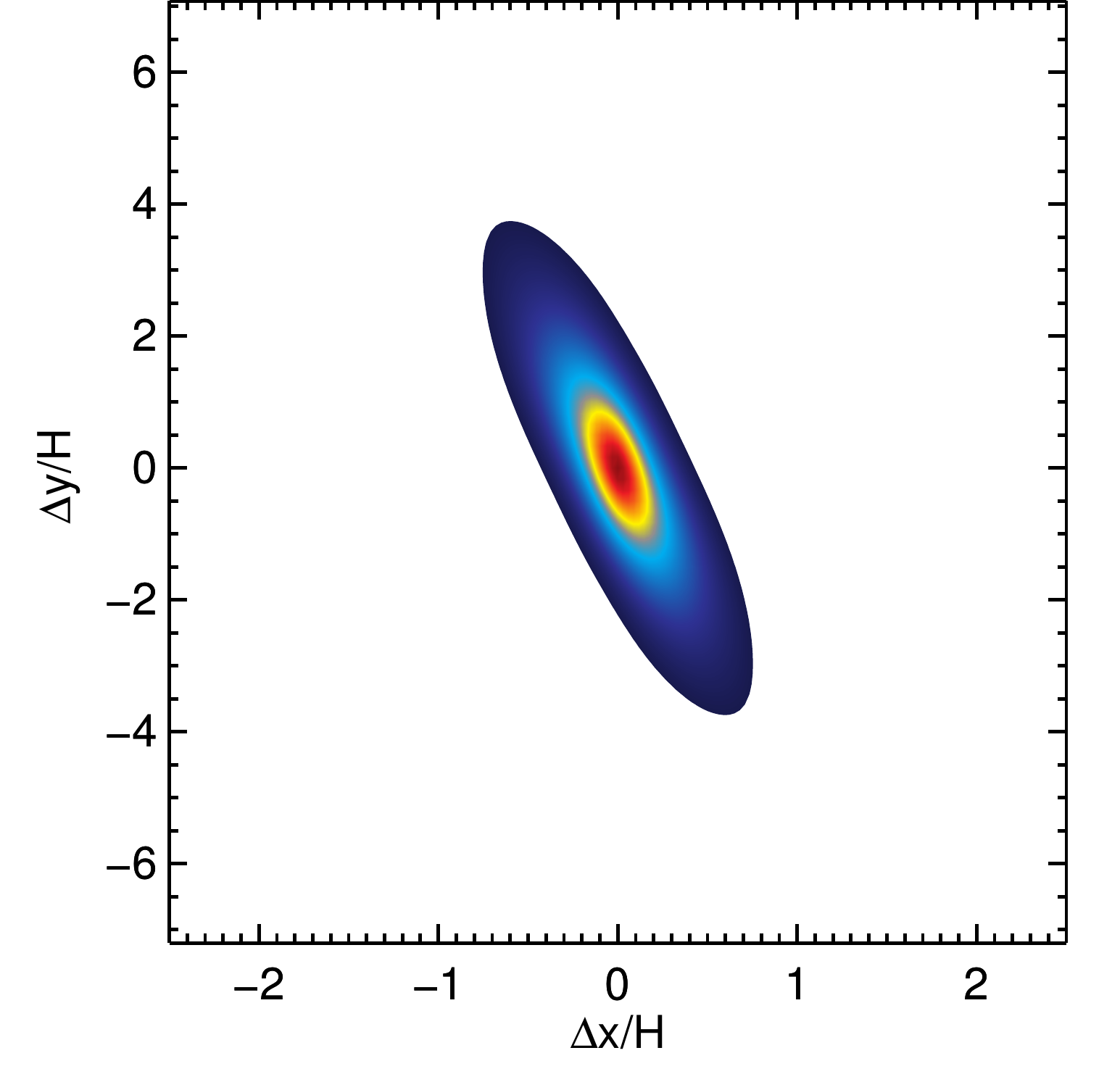}
\includegraphics[width=0.32\textwidth, viewport=30 15 475 445,clip]
{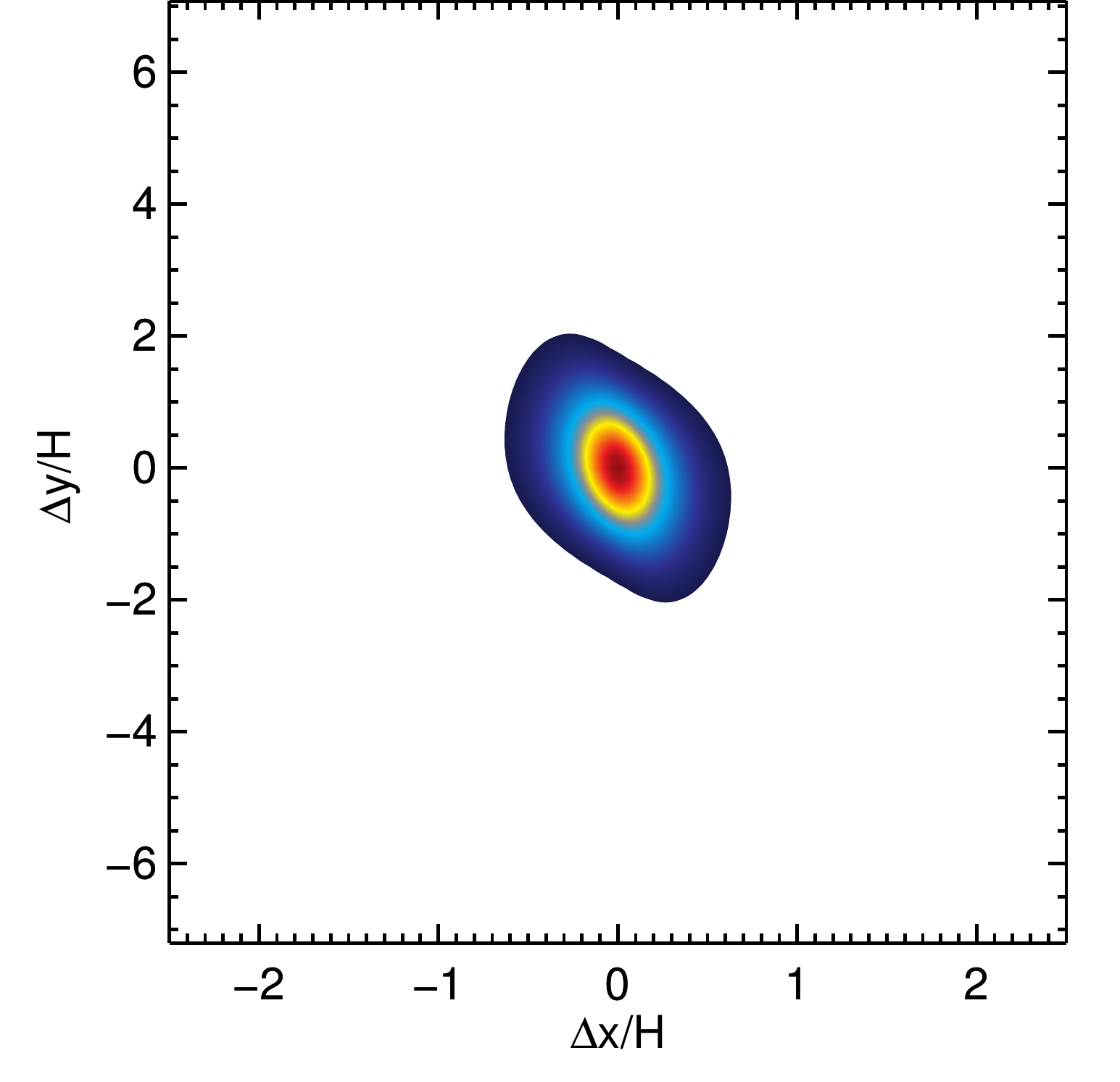}
{\includegraphics[width=0.48\textwidth]{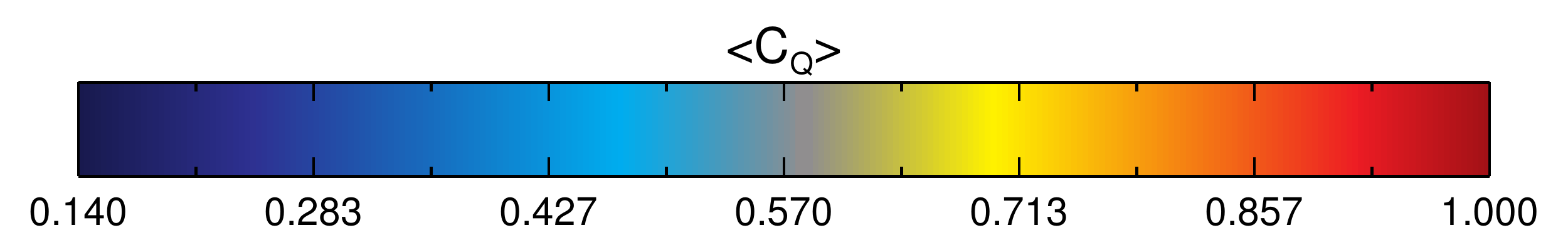}} \\
\end{center}
\caption[]{Structure of $\left< C_{\rho}(\Delta \mathbf{x}) \right>$ (left panel), $\left< C_{|\mathbf{B}|^2}(\Delta \mathbf{x}) \right>$ (center panel)  and $\left< C_{\rho |\delta \mathbf{V}|^2}(\Delta \mathbf{x}) \right>$ (right panel) where $\Delta \mathbf{x} = ( \Delta x, \Delta y, \Delta z)$ along the plane $\Delta z=0$. The correlation functions are calculated over the region $5 \le r/r_S \le 11$, $|Z|\le\pm H$) and $0\le\phi\le\pi/2$, time-averaged over $\Delta T = 16-19P_{orb}(r=15r_S)$ and normalized to their value at $\Delta \mathbf{x} = 0$.}
\label{xy_corr} 
\end{figure*}

One of the underlying assumptions of the Kolmogorov model for homogeneous, incompressible turbulence is that the fluctuations are self-similar; that is, fluctuations on different scales are statistically indistinguishable. One way to probe whether this is the case is to examine the character of temporal fluctuations in the magnetic and kinetic energies on large ($|k|H/2\pi < 1$) versus small ($|k|H/2\pi \ge 1$) spatial scales. This is accomplished by taking the three-dimensional Fourier transform of toroidally-averaged, spatially normalized simulation data, $\left| \left< P_{|\mathbf{B}|^2}(|k|,f) \right> \right|^2$ and $\left| \left< P_{\rho}(|k|,f) \right> \right|^2$ as described in \S\ref{power_spectra}. We perform a one-dimensional average over $|k|H/2\pi < 1$ and $|k|H/2\pi \ge 1$ to yield the temporal power spectrum on large and small spatial scales respectively (see \S\ref{power_spectra}). In performing these calculations, we utilize simulation data from the same domain as above, $5\le r/r_S \le 15$, $|Z| \le 5$ and include contributions from frames in the range $\Delta T = 10-20 P_{orb} (r=15 r_S)$. These data are shown in Figure \ref{xzt_3d_fft}. Whilst both sets of power spectra are dominated by contributions from large scales (a consequence of the $|k|^{-11/3}$ scaling described above), there are now clear contrasts between the temporal power spectra for these two measures of the turbulence. Firstly, the magnetic energies are characterized by broadband power extending over the entire frequency range and characterized by the power law, $\propto f^{-1/2}$. Kinetic energies are instead characterized by a broken power law, $\propto f^{-2}$ at frequencies higher than $fP_{orb} (r=10r_S)/2\pi > 10$ and $\propto f^{-0.3}$ at frequencies lower than this.
In addition, we find evidence for a break at $f \sim P_{orb}(r=10r_S)$ in the power spectra for the magnetic energy at large spatial scales, a feature that is absent for the magnetic energy at small spatial scales and also in the kinetic energy at all scales. These results suggest that while the temporal fluctuations in the kinetic energy are similar at different scales within the turbulence, the same is not true for the magnetic energy. That is, the temporal behaviour of the magnetic field at large versus small scales is \emph{not} consistent with the underlying Kolmogorov model for homogeneous, incompressible turbulence, namely that fluctuations on different scales are statistically indistinguishable. In the next section, we examine properties of the temporal correlation function in these quantities to confirm this suggestion.

\subsection{Correlation Functions}\label{corr}

\begin{figure*}
%\leavevmode
\begin{center}
\includegraphics[width=0.32\textwidth]{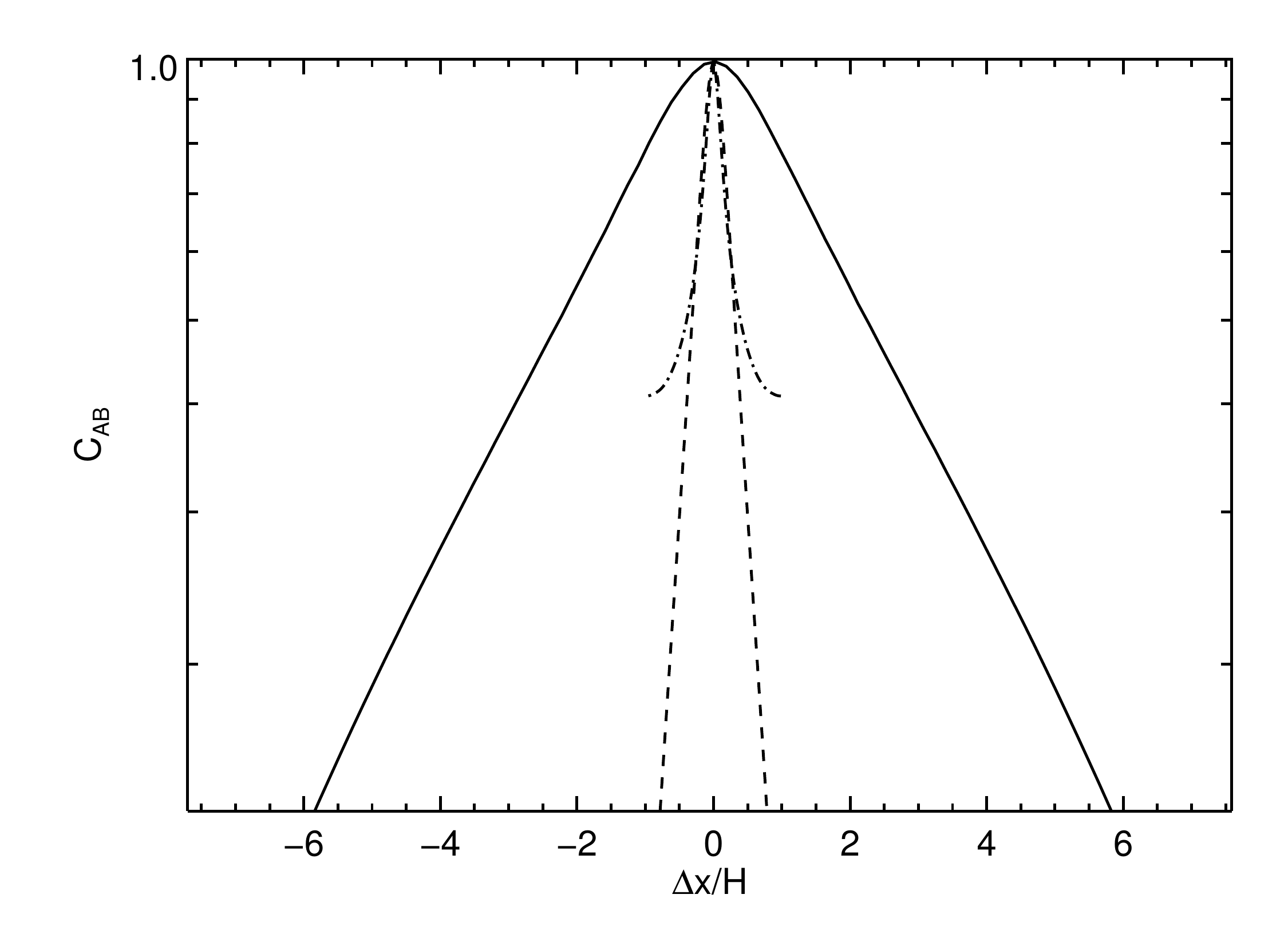}
\includegraphics[width=0.32\textwidth]{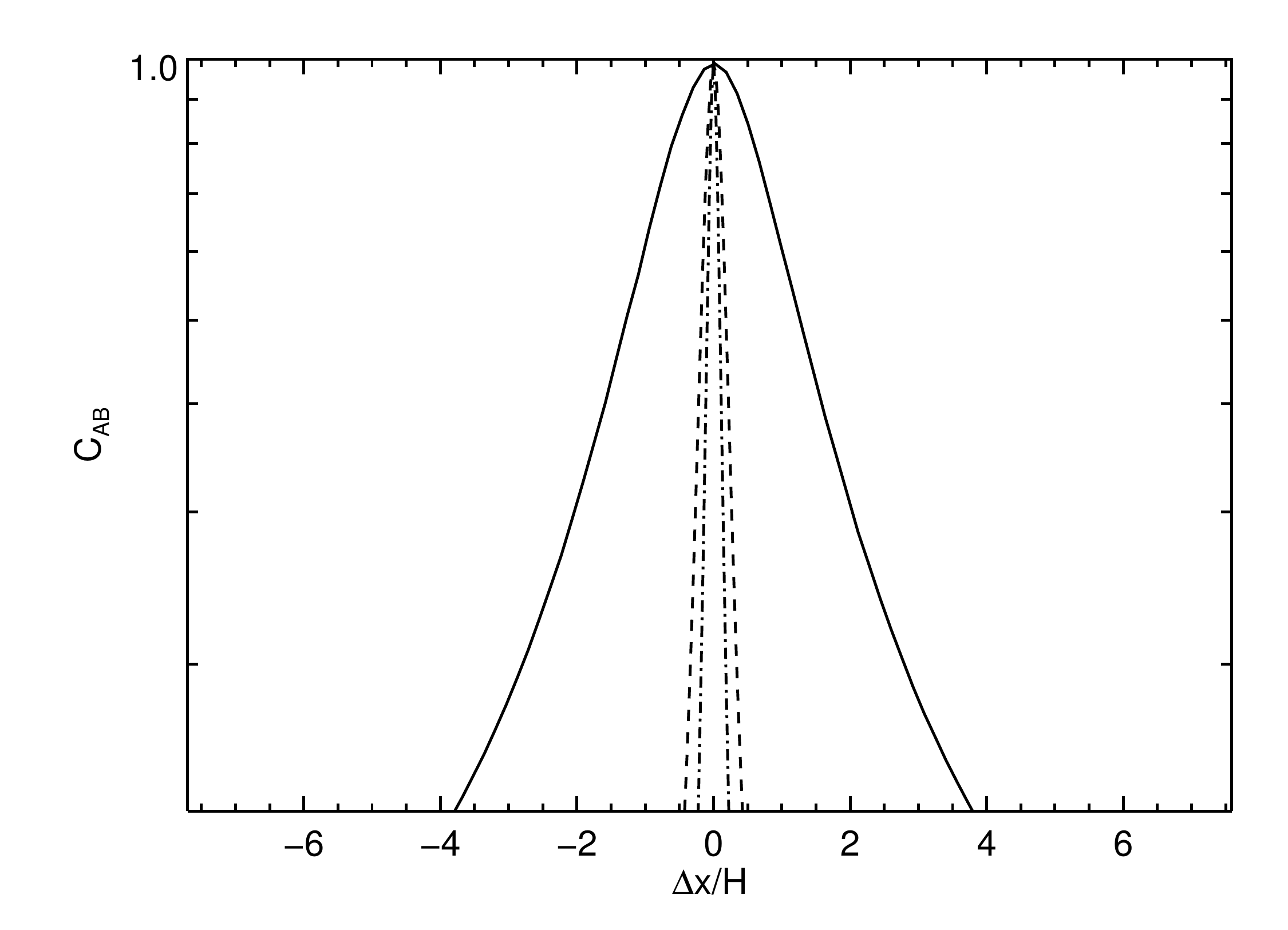}
\includegraphics[width=0.32\textwidth]{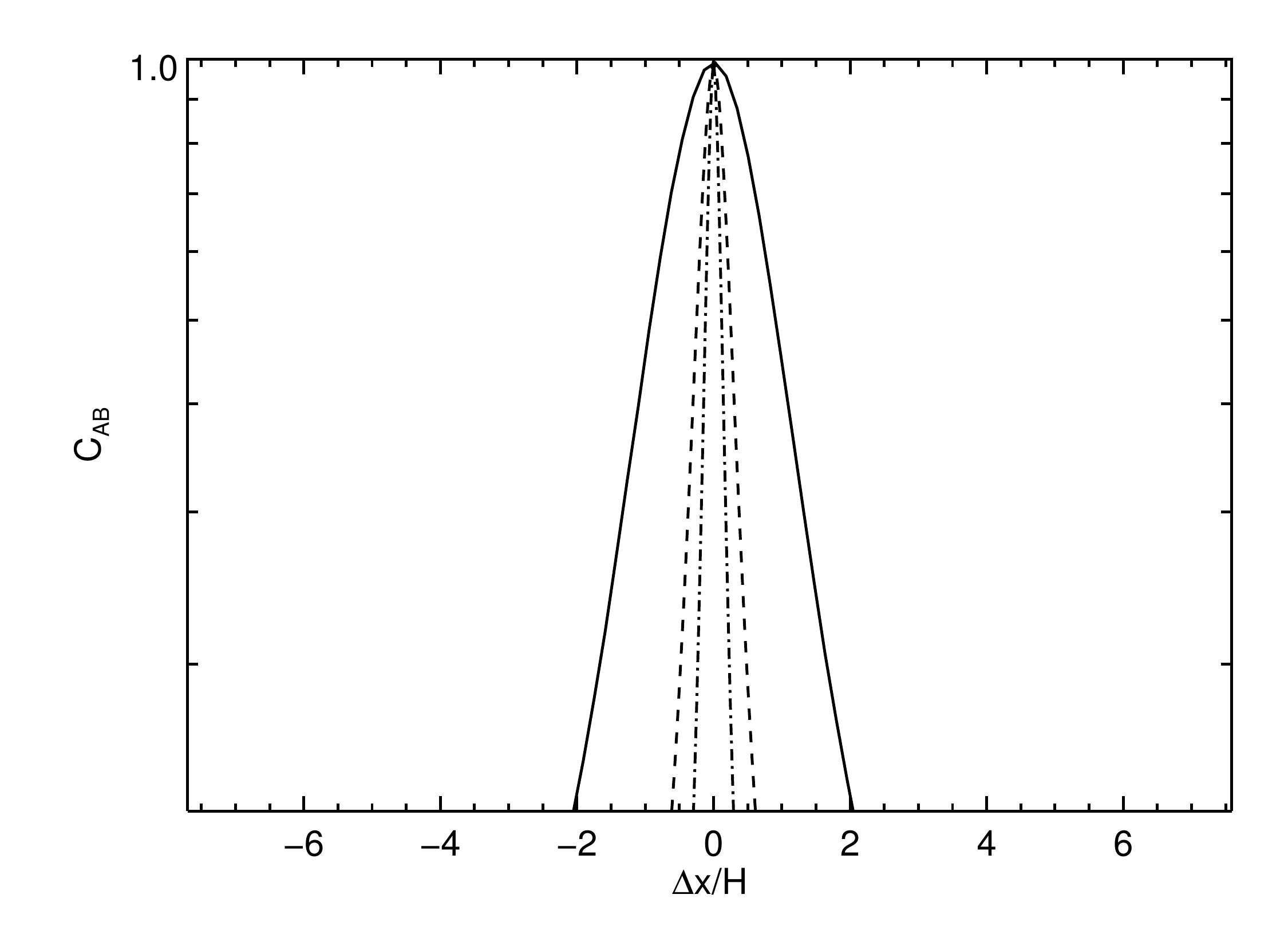}
\end{center}
\caption[]{Slices through the major (solid lines), minor (dashed lines) and vertical (dot-dash lines) axes of $\left<C_{\rho}\right>$ (left panel), $\left<C_{|B|^2}\right>$ (center panel) and $\left<C_{\rho |\delta V|^2}\right>$ (right panel), calculated as described in Fig. \ref{xy_corr}}
\label{corr_slice} 
\end{figure*}

\begin{figure*}
%\leavevmode
\begin{center}
\includegraphics[width=0.32\textwidth]{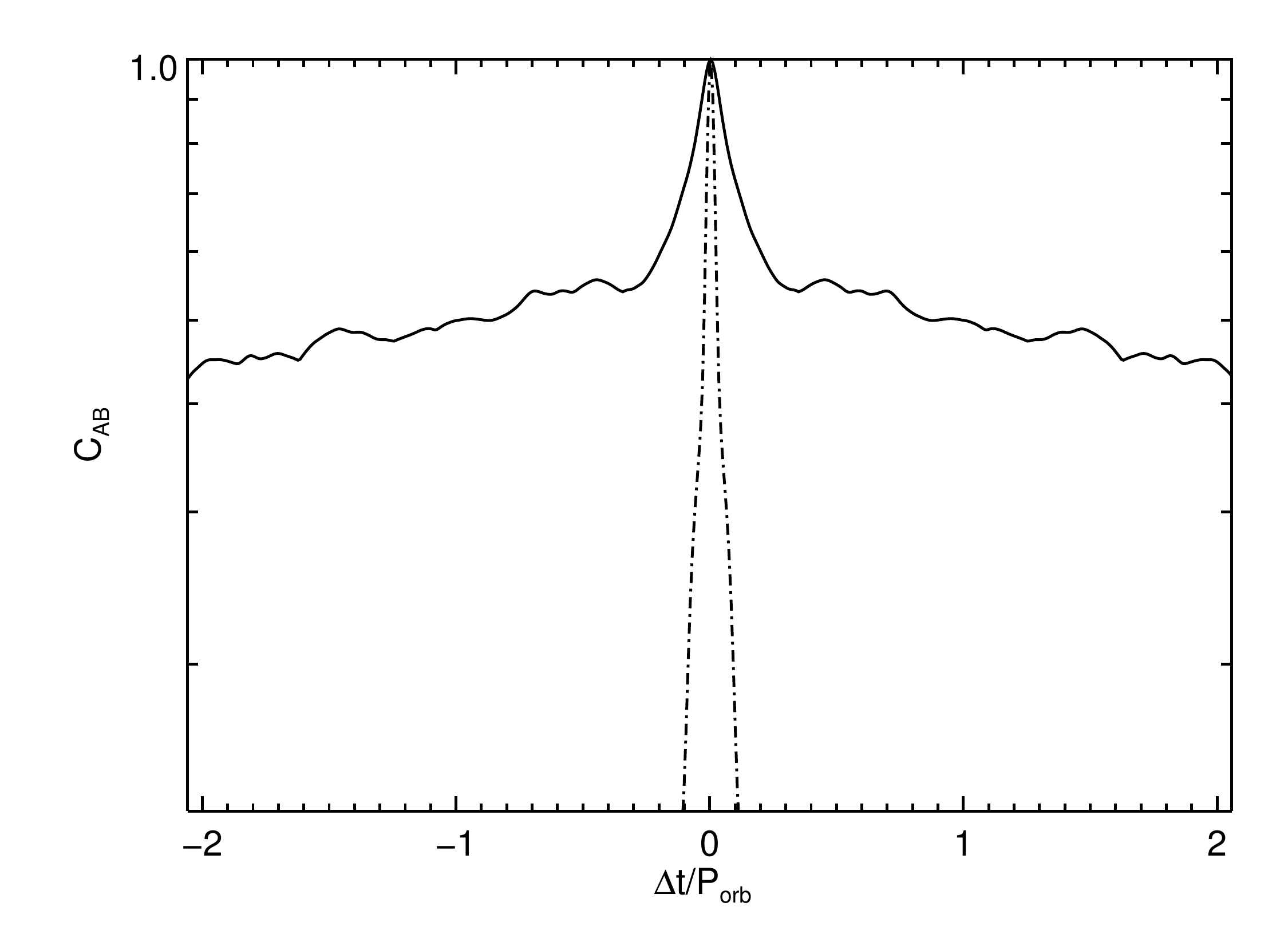}
\includegraphics[width=0.32\textwidth]{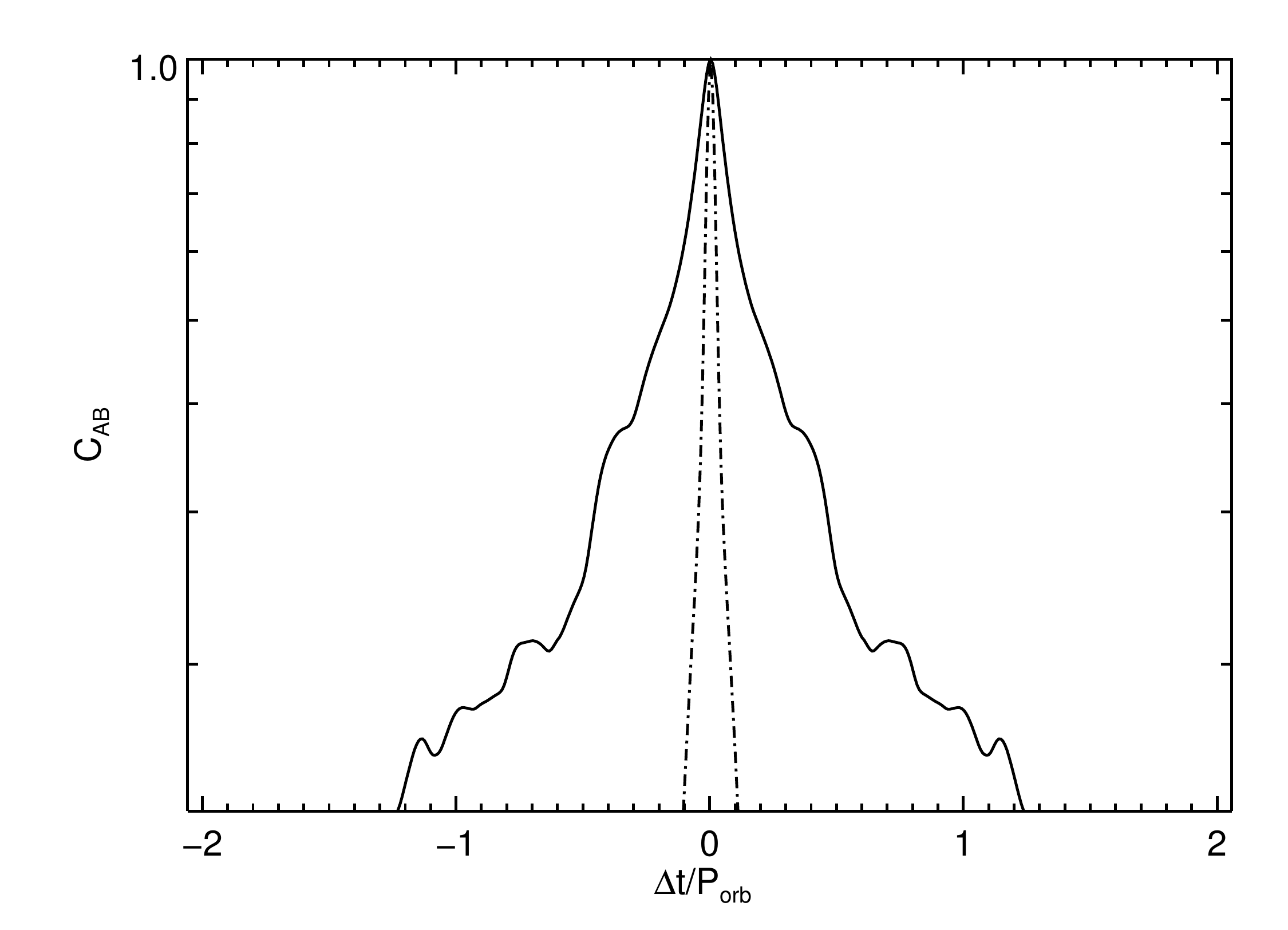}
\includegraphics[width=0.32\textwidth]{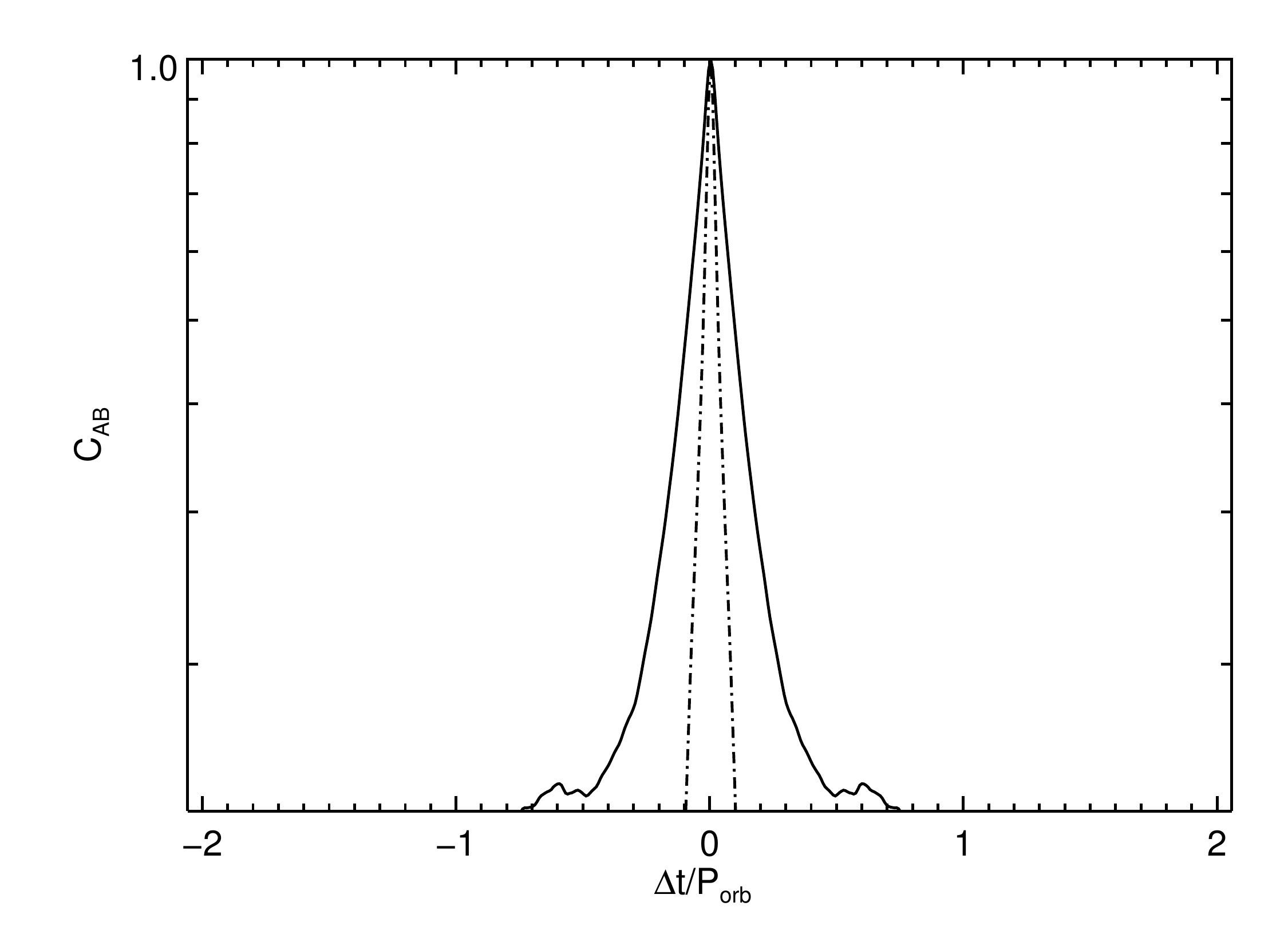}
\end{center}
\caption[]{Lifetimes of modes with $\Delta \mathbf{x} > H$ (solid lines) compared to modes with $\Delta \mathbf{x} \le H$ (dashed lines) for $\left< \left< C_{\rho}(\Delta{x},\Delta{y},\Delta{t}) \right> \right>$ (left panel), $\left< \left< C_{|\mathbf{B}|^2}(\Delta{x},\Delta{y},\Delta{t}) \right> \right>$ (center panel) and $\left< \left< C_{\rho |\delta \mathbf{V}|^2}(\Delta{x},\Delta{y},\Delta{t}) \right> \right>$. The correlation functions are calculated from vertically integrated data over the region $5 \le r/r_S \le 11$, $0\le\phi\le\pi/2$ where the vertical integral is performed over $|Z|\le\pm H$ and over the time-interval $\Delta t = \pm2P_{orb}(r=8r_S)$. The resulting correlation functions are then time-averaged over $\Delta T = 16-19P_{orb}(r=15r_S)$ and normalized to their value at $\Delta{x},\Delta{y},\Delta t = 0$.}
\label{t_corr} 
\end{figure*}

\cite{Guan:2009a} use unstratified local shearing box calculations to demonstrate that the magnetic autocorrelation function is localized on the $x-y$ ($r-\phi$) plane with a correlation length, $\lambda$, along the major axis of the correlation function of $\lambda \sim 0.3H$ for net toroidal magnetic field geometries, suggesting that the turbulence is \emph{local}.
The autocorrelation function is useful in addressing these questions as it provides an improved statistical measure of properties of the turbulence at large spatial scales \citep{Guan:2009a}. \cite{Davis:2010} examine the shape of the magnetic autocorrelation function on the $x-y$ plane for large radial domain, vertically stratified shearing boxes. These authors obtain results approximately consistent with those of \cite{Guan:2009a} for zero net flux configurations at small scales, but also find the existence of large scale correlations which enforce uniformity in the Maxwell stress and magnetic energies over many scale heights, suggesting that \emph{global} correlations exist within the turbulence. That global correlations do exist within the turbulence was further demonstrated by \cite{Nelson:2010}. These authors compare the amplitude of density fluctuations obtained in vertically stratified shearing boxes with those obtained in global simulations, finding that shearing boxes of dimensions $4H\times16H\times2H$ in $x,y,z$ are necessary to provide correlations in the density which are consistent with those obtained from global simulations performed using a $\pi/2$ toroidal domain. This is because at least six scale heights are necessary in the azimuthal domain to correctly capture the excitation of spiral density waves \cite{Heinemann:2009a,Heinemann:2009b}. As suggested in \S\ref{diskvel}, these spiral density waves are intimately tied to the strength of velocity fluctuations within the disk and hence to the turbulence itself.

If an azimuthal domain size of \emph{sixteen} scale heights is necessary to accurately reproduce the results of $\pi/2$ toroidal domain global simulations within shearing boxes, it is worth asking whether $\pi/2$ toroidal domain global simulations accurately reproduce the results of full $2\pi$ toroidal domain global simulations? \cite{Hawley:2001} performs an explicit comparison of cylindrical global simulations simulations computed using a full $2\pi$ domain in the toroidal dimension versus those using a restricted $\pi/2$ domain, finding that, whilst the linear growth stage of the toroidal field MRI can be influenced by use of a restricted domain, properties of the quasi-steady state are approximately similar between the two domains. In particular, the one-dimensional toroidal power spectra of the density and magnetic field components were found to be approximately independent of the increase in toroidal domain size from $\pi/2$ to $2\pi$. Taken together, these results suggest that it is necessary to consider toroidal domains at least a factor $16$ greater than the disk scale height in order to obtain accretion stresses independent of the domain size.

To further investigate these ideas, we have calculated three-point auto-correlation functions of the gas density,  $\left< C_{\rho}(\Delta \mathbf{x}) \right>$, the magnetic energy density, $\left< C_{|\mathbf{B}|^2}(\Delta \mathbf{x}) \right>$ and the kinetic energy density, $\left< C_{\rho |\delta \mathbf{V}|^2}(\Delta \mathbf{x}) \right>$ (where $\Delta \mathbf{x} = ( \Delta x, \Delta y, \Delta z)$) with the goal of determining correlation lengths within the turbulence. Each of these auto-correlation functions are calculated over the region $5 \le r/r_S \le 11$ inside the disk body (i.e. $|Z|\le\pm H$) and over the entire toroidal domain (i.e. $0\le\phi\le\pi/2$). The extent of the radial domain is chosen so that the correlation functions pass through zero before interacting with the radial boundaries. The extent of the domain in the vertical dimension is chosen both for the reasons described in \cite{Davis:2010}, namely so that large scale correlations from the regions close to the vertical boundaries do not influence the shape of the correlation function and so that we are confident that the included region corresponds to incompressible turbulence. The correlation functions are calculated as described in \S\ref{power_spectra} over $\Delta T = 16 - 19 P_{orb} (r=15r_S)$ (corresponding to $\sim10$ orbits at $8 r_S$) from $204$ full three-dimensional data dumps (a resolution of $\sim20$ dumps per orbital period at $8r_S$).

\begin{table}
\begin{center}
\caption{Correlation Lengths}
\label{length}
\begin{tabular}{@{}rcccc}
\hline
Autocorrelation function          &
$\lambda_{maj}/H$ &
$\lambda_{min}/H$ &
$\lambda_{vert}/H$ \\
\hline
$\left< C_\rho \right>$ & 3 & 0.38 & 1 \\
$\left< C_{|B|^2} \right>$ & 2 & 0.25 & 0.13 \\
$\left< C_{\rho|\delta V|^2} \right>$ & 1 & 0.25 & 0.13 \\
\hline
\end{tabular}
\end{center}
\end{table}

Figure \ref{xy_corr} shows each correlation function on the $x-y$ plane at zero offset from the vertical axis, $\Delta Z = 0$. Before discussing the properties of these functions in detail, we recall that the ratio of the Maxwell stress to the magnetic pressure defines a characteristic tilt angle of the magnetic field with respect to the toroidal direction \citep{Guan:2009a}:
\begin{equation}
|\left< \left< \alpha^{m}_{m} \right> \right>| =
\frac{2 |\left< \left< B_{r} B_\phi \right> \right>|}{|B|^2}
= 2 \sin \theta_t \cos \theta_t = \sin 2 \theta_t
\end{equation}
where $\theta_t$ is the `tilt' angle with respect to the $\phi$-axis. For the simulation presented here, $|\left< \left< \alpha^{m}_{m} \right> \right>| \sim 0.3$ (see Figure \ref{alpha_vol}) within the disk body ($|Z| < H$), where we are confident that the turbulence is incompressible, implying that $\theta_t \sim 9^\circ$, somewhat smaller than $\theta_t = 12^\circ$ reported by \cite{Guan:2009a} for net toroidal field unstratified shearing boxes computed at $32$ zones per scale height. The data of Figure \ref{xy_corr} show that both the auto-correlation function in the density and the magnetic field are both characterized by ellipses, with the major axis tilted with respect to the $y$-axis at an angle consistent with $\theta_t \sim 9^\circ$. The auto-correlation function of the kinetic energy density is more circular than these two previous measures and as such it is harder to assess the alignment of the major axis. What evidence there is does suggest that the tilt angle of this auto-correlation function is also consistent with  $\theta_t \sim 9^\circ$. To determine the correlation lengths along the major, $\lambda_{maj}$, minor, $\lambda_{min}$ and vertical axes, $\lambda_{vert}$, we rotate each correlation function through $\theta_t \sim 9^\circ$ on the $x-y$ plane and measure the distance from the axis where the correlation function falls to $e^{-2}$ of its maximum value. Assuming that the correlation functions along each axis take an exponential profile, $\left< C (\Delta x) \right> = \left< C (0) \right> e^{-\Delta x / \lambda}$ \cite[as in][]{Guan:2009a}, we can then determine the correlation lengths, $\lambda$ along each axis. Plots showing the shape of the correlation function along the major, minor and vertical axes are shown in Figure \ref{corr_slice} and the correlation lengths are given in Table \ref{length}. As these data make clear, there is a hierarchy of scales within the turbulence. The longest correlation lengths exist in the density, correlations in the magnetic field exist inside these and finally the kinetic energy has the smallest correlations lengths. All of these correlation lengths are significantly longer (in units of the disk scale height) than those reported by \cite{Guan:2009a} for a net toroidal field model. We note, however, that the correlation length along the major axis in the density is identical to that found by \cite{Nelson:2010}, suggesting that these extended correlation lengths are due to the excitement of spiral density waves in the disk \citep{Heinemann:2009a,Heinemann:2009b}. Finally, the correlation length in the density coincides exactly with the break in the toroidal field power spectrum found in both the magnetic field and the accretion stress in \S\ref{diskfield} and \S\ref{stresses}, suggesting an intimate link between the the formation of spiral density waves and fluctuations in the magnetic field and accretion stresses. We leave detailed investigation of this possibility to future work.

The discussion of \S\ref{fluct} suggests that fluctuations within the turbulence can display different temporal variability patterns at different spatial scales. If correct, this would indicate that the turbulence is not self-similar at different spatial scales. We can further investigate this suggestion by examining the lifetimes of modes at large versus small spatial scales by using the scale filtered space-time auto-correlation functions described in \S\ref{power_spectra}. Figure \ref{t_corr} shows the lifetimes of modes in the density, magnetic field and perturbed kinetic and large and small spatial scales. For each of these quantities, the lifetimes of modes on small spatial scales $\tau \sim 0.05P_{orb}$ (estimated using $\left< C (\Delta t) \right> = \left< C (0) \right> e^{-\Delta t / \tau}$) where $\tau$ is the mode lifetime. The lifetimes of modes on large spatial scales display different behavior. Fluctuations in the density on large spatial scales do not de-correlate on the timescales considered here, indicating that they are long-lived compared to the local orbital time. Modes on large spatial scales in the magnetic energy density have $\tau \sim 0.5 P_{orb}$, whilst those in the kinetic energy density have $\tau \sim 0.25 P_{orb}$. We also note that the form of the temporal correlation function in the magnetic energy on large scales has a different dependence on $\Delta t$ at $\Delta t > 0.5 P_{orb}$ to $\Delta t <0.5 P_{orb}$; this is the signature of the break in the temporal power spectrum in magnetic energy on large spatial scales at $f \sim P_{orb}$ discussed in \S\ref{fluct}. These data show that the lifetimes of modes on large scales are significantly longer than those on small scales, with typical lifetimes of $\sim P_{orb}$ in the former case, compared to lifetimes $\sim 0.1 P_{orb}$ in the latter, confirming the suggestion of \S\ref{fluct} that the properties of the turbulence are \emph{not} statistically indistinguishable at different spatial scales.

\section{A Local Flux-Stress Relationship}\label{flux_stress_relation}

In unstratified shearing box simulations, there is a direct relationship between the strength of the vertical magnetic
flux threading the domain and the time-averaged accretion stress arising from the MRI \citep{Hawley:1995}. In a
box of fixed physical size, the relation takes the simple form \citep{Pessah:2007},
\begin{equation}
\frac{W^{m}_{r \phi}}{P_g} \propto -B_z \propto -\lambda^{z}_{MRI},
\end{equation}
where $\lambda^{z}_{MRI}$ is the wavelength of the fastest unstable mode of the MRI in the vertical direction for
the specified net initial magnetic field $B_z$. The proportionality is limited in the high field limit by the requirement
that the most unstable wavelength must fit within the box, if instead $\lambda^z_{MRI} > L$, then the MRI freezes out and
the accretion stress drops to zero. This is a physical limit that would suppress (at least) the linear growth of the MRI in a real disk
threaded by a field with $\lambda^z_{MRI} \gg H$. The proportionality is also limited, purely numerically, in the low
field limit by the requirement that $\lambda^z_{MRI}$ must be resolved, i.e. that $\lambda^z_{MRI} > \Delta^z$, where
$\Delta^z$ is the vertical grid spacing. Because unstratified zero-net flux local simulations fail to converge in the absence
of physical dissipation \citep{Fromang:2007}, a consequence for {\em those simulations only} is that the mean accretion stress is entirely
determined by the boundary conditions for the magnetic field and the grid resolution \citep{Pessah:2007}.

Although the existence of a flux-stress relation has a straightforward anchor in the linear physics of the MRI, its
relevance to systems more realistic than the unstratified shearing box remains unclear. We first note that
stratified shearing boxes do not exhibit the same pathological convergence properties as unstratified simulations,
even when only numerical dissipation is present. Rather, such simulations converge to a non-zero accretion stress
even in the limit of zero net flux \citep{Davis:2010}. This suggests that for local, stratified disks, there ought to exist
a threshold net flux, below which the properties of the disk turbulence are universal. A flux-stress relation would then apply
only for larger net magnetic fields, up to the (physical) limit where MRI quenching occurs. The simulations necessary
to test such expectations in detail have not yet been done, though there is considerable evidence that vertical flux
continues to stimulate MRI-driven turbulence irrespective of the presence or absence of stratification.

\begin{figure*}
%\leavevmode
\begin{center}
\includegraphics[width=0.32\textwidth, viewport=0 0 480 445,clip]
{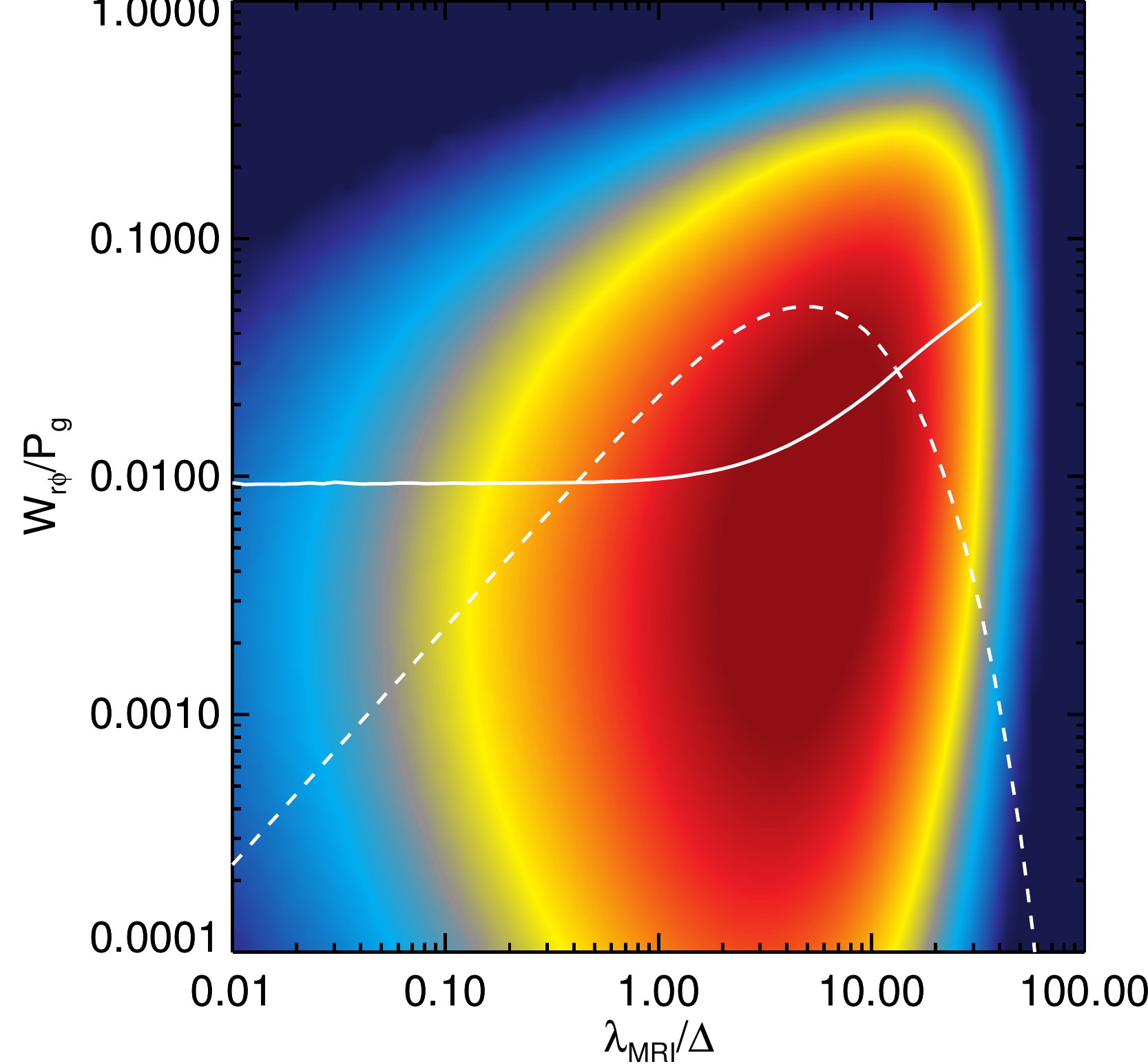}
\includegraphics[width=0.32\textwidth, viewport=0 0 480 445,clip]
{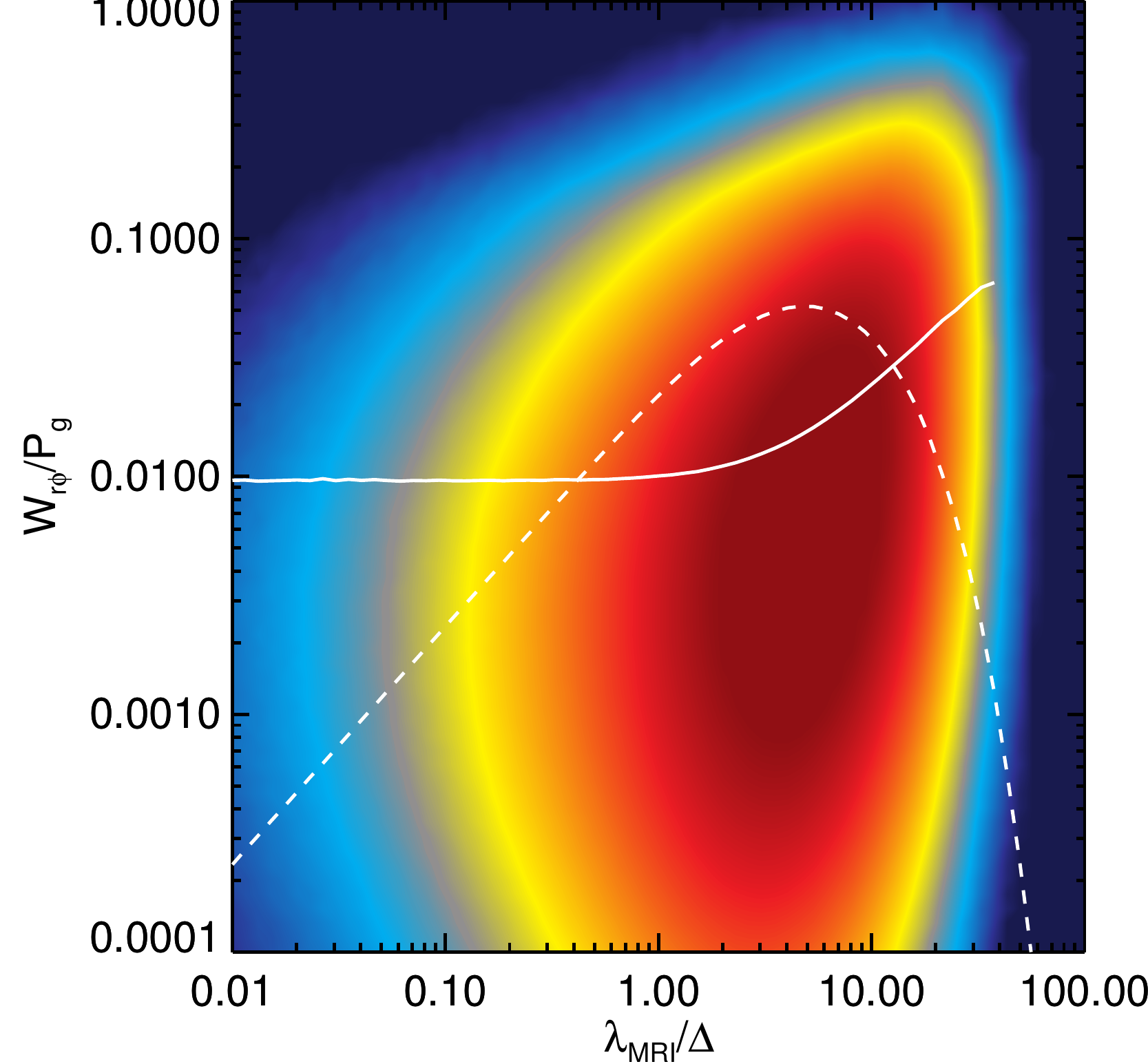}
\includegraphics[width=0.32\textwidth, viewport=0 0 480 445,clip]
{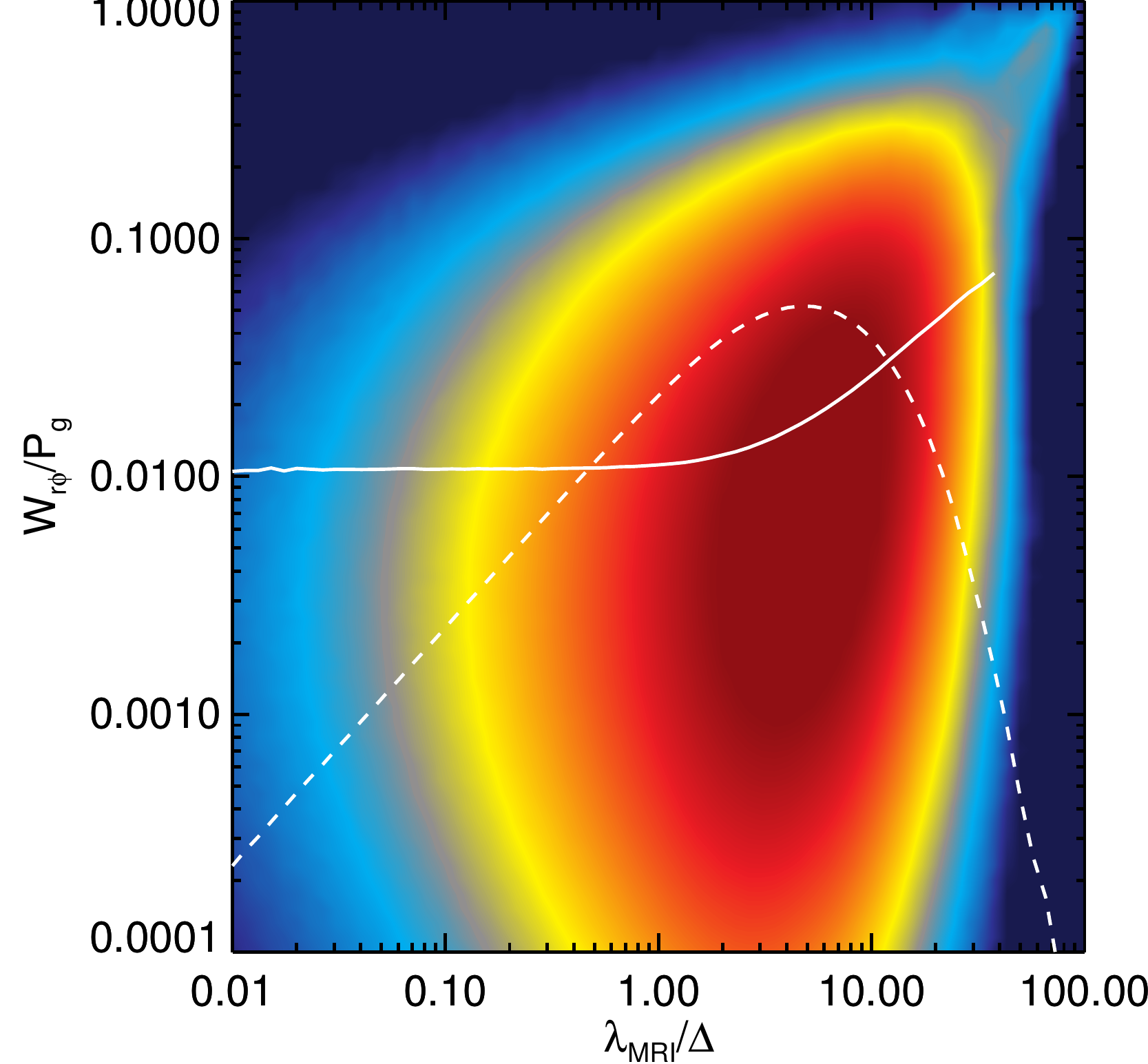}
\includegraphics[width=0.48\textwidth]{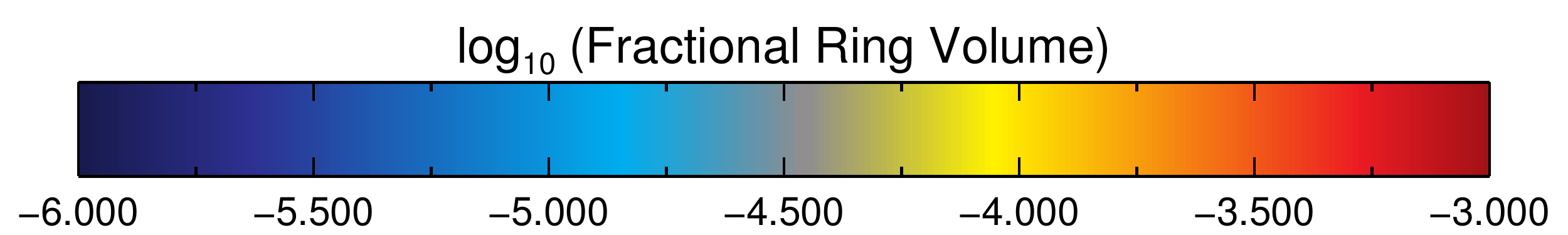}
\end{center}
\caption[]{Time-averaged local correlation between the vertical magnetic flux and the accretion stress within the disk body ($|Z|<H$) for three different radial regions of the disk. The vertical flux is measured in terms of the number of vertical grid cells per fastest unstable mode of the vertical field MRI, $\lambda_{MRI} / \Delta$ and the accretion stress is measured in units of the gas pressure, $W^m_{r\phi} / P_g$. The color contours show the fractional distribution of ring volume over the parameter space, the solid line the average accretion stress at each $\lambda_{MRI} / \Delta$ and the dashed line the cumulative fractional volume of the ring at each $\lambda_{MRI} / \Delta$ (where the numerical value can be read directly from the $y$-axis). From left-to-right, the panels show data for $3\le r/r_S \le 5$, $5 \le r/r_S \le 7$ and $7 \le r/r_S \le 9$. The time-averaging is performed over $\Delta T = 11.5-19P_{orb} (r=15r_S)$ from $3000$ complete three-dimensional data dumps.}
\label{flux_stress} 
\end{figure*}

In local simulations, the total vertical flux threading the domain is set by the initial conditions\footnote{Of course,
in a large shearing box the distribution of flux threading the mid-plane is a dynamically evolved quantity, and
is of physical interest.}. Such
simulations cannot address the question of what would be the self-consistent distribution of magnetic flux
threading different regions of a disk.
In a real disk, the action of the MRI, together with dynamical processes in the disk corona, may combine
to generate local regions of strong flux, even if the disk as a whole has vanishing
net field. Conversely, vertical flux in a disk initially threaded by a net field
way well be transported over large radial distances, or expelled from the disk entirely \citep{Beckwith:2009}.
Global simulations are needed to address these possibilities. Two questions are of particular import. First, what
is the form of (and dispersion in) the flux-stress relation measured globally, and on what spatial scales does
it apply? It is necessary to check whether the relation is causal -- in the expected sense of the flux influencing
the stress rather than the other way around -- since this is not necessarily true globally. Second, is the
distribution of vertical flux threading the disk sufficient as to boost the accretion stress, over and beyond
the expectations of a model in which the disk has zero net flux on all relevant local scales? It is possible to imagine
a situation in which the self-consistent coronal field has a dominant feedback effect on the dynamics
at the disk mid-plane -- increasing the accretion stress from the $\langle \langle \alpha \rangle \rangle \sim 10^{-2}$
values measured in many simulations to values closer to those observationally inferred -- but the existence
of such a regime has yet to be demonstrated.

\cite{Sorathia:2010} made an initial examination of these issues, using data from a series of global disk
simulations that were initialized with small loops of poloidal field. They analyzed the model disk by
considering the flux and stress present in co-moving patches of a single, fixed spatial scale, concluding
both that a causal flux-stress relation existed and that the distribution of vertical flux was strong enough
as to be dynamically interesting. Here, we revisit these questions. We take advantage of the fact
that our simulation is initialized with a net toroidal field, which means that {\em any} flux-stress relationship
within this simulation must arise due to vertical magnetic field generated from turbulence arising from the MRI.
Compared to the simulations analyzed by \cite{Sorathia:2010}, we also attain substantially better resolution
(in the final, saturated state) of the most unstable linear MRI modes. This permits a separation between
numerical effects that are known to occur on the grid scale, and the physical effects at larger scales that
are of primary interest.

To assess whether or not a flux-stress relationship exists in our simulation, we first determine the fractional volume
of the disk that is instantaneously threaded by a given vertical flux and that has a given accretion stress. By working
with this two-dimensional distribution as our fundamental quantity, we avoid the need to make an essentially
arbitrary choice of spatial scale over which to average the flux and the stress. Operationally,
we divide the disk into three rings spanning $5 \le r/r_S \le 7$, $7 \le r/r_S \le 9$ and $9 \le r/r_S \le 11$. Next, we utilize calculate the ratio of the wavelength of the fastest unstable mode of the vertical field MRI, $\lambda^z_{MRI}$ to the physical grid spacing in the vertical direction, $\Delta^z$ for each cell within the disk body ($|Z| < H$), which we denote by $F^z = \lambda^z_{MRI}/\Delta^z$. We utilize this same data to calculate the magnetic (Maxwell) accretion stress in units of the gas pressure, $\alpha^m_g = -W^{m}_{r \phi} / P_{g}$, again for each cell within the disk body. We use a volume weighted binning procedure to create a distribution function for each ring describing the relationship between vertical flux and accretion stress, which we normalize to the total volume of the ring. This procedure yields a volume weighted distribution function describing the relationship between vertical flux and accretion stress, which we denote by $f (F^z,\alpha^{m}_{g})$:
\begin{equation}
f (F^z,\alpha^{m}_{g}) \equiv
\frac{ \int^{\alpha^{m}_{g} + d\alpha^m_g}_{\alpha^{m}_{g} - d\alpha^m_g} {
\int^{F^z + dF^z}_{F^z - dF^z} {
\delta V (F^z, \alpha^m_g) \; d F^z} \; d \alpha^m_g } }
{ \int \int \int dV \; d F^z \; d \alpha^m_g }
\end{equation}
where $\delta V (F^z, \alpha^m_g)$ is the volume element of a cell threaded by vertical flux $F^z$ and an accretion stress $\alpha^g_m$ and $\int \int \int dV dF^z d\alpha^m_g$ is the total ring volume multiplied by the area of the $(F^z,\alpha^m_g)$ plane. We can use this distribution function to find the mean accretion stress, $\left\{ \alpha^m_g \right\} $ associated with a given vertical flux:
\begin{equation}
\left\{ \alpha^m_g (F^z) \right\} \equiv
\frac{ \int \alpha^m_g f (F^z,\alpha^{m}_{g}) \; d \alpha^m_g }
{ \int f (F^z,\alpha^{m}_{g}) \; d \alpha^m_g }
\end{equation}
and also the fraction of the ring volume, $\left\{ \delta V \right\}$ that is threaded by a given vertical flux:
\begin{equation}
\left\{ \delta V (F^z) \right\} \equiv \frac{ \int f (F^z,\alpha^{m}_{g}) \; d \alpha^m_g }
{ \int d \alpha^m_g }
\end{equation}
These data are shown in Figure \ref{flux_stress}, time-averaged over $\Delta T = 11.5-19 P_{orb} (r=15r_S)$ using $20$ dumps per ISCO orbit (approximately $3000$ dumps in all). In these figures, the color contours show the volume-weighted distribution function, $f (F^z,\alpha^{m}_{g})$, the dashed white line the fraction of the ring volume threaded by a given vertical flux, $\left\{ \delta V (F^z) \right\}$  (where $\left\{ \delta V \right\}$ can be read from the $y$-axis scale) and the solid white line the mean accretion accretion stress associated with a given vertical flux, $\left\{ \alpha^m_g (F^z) \right\}$.

\begin{figure}
%\leavevmode
\begin{center}
\includegraphics[width=0.48\columnwidth]
{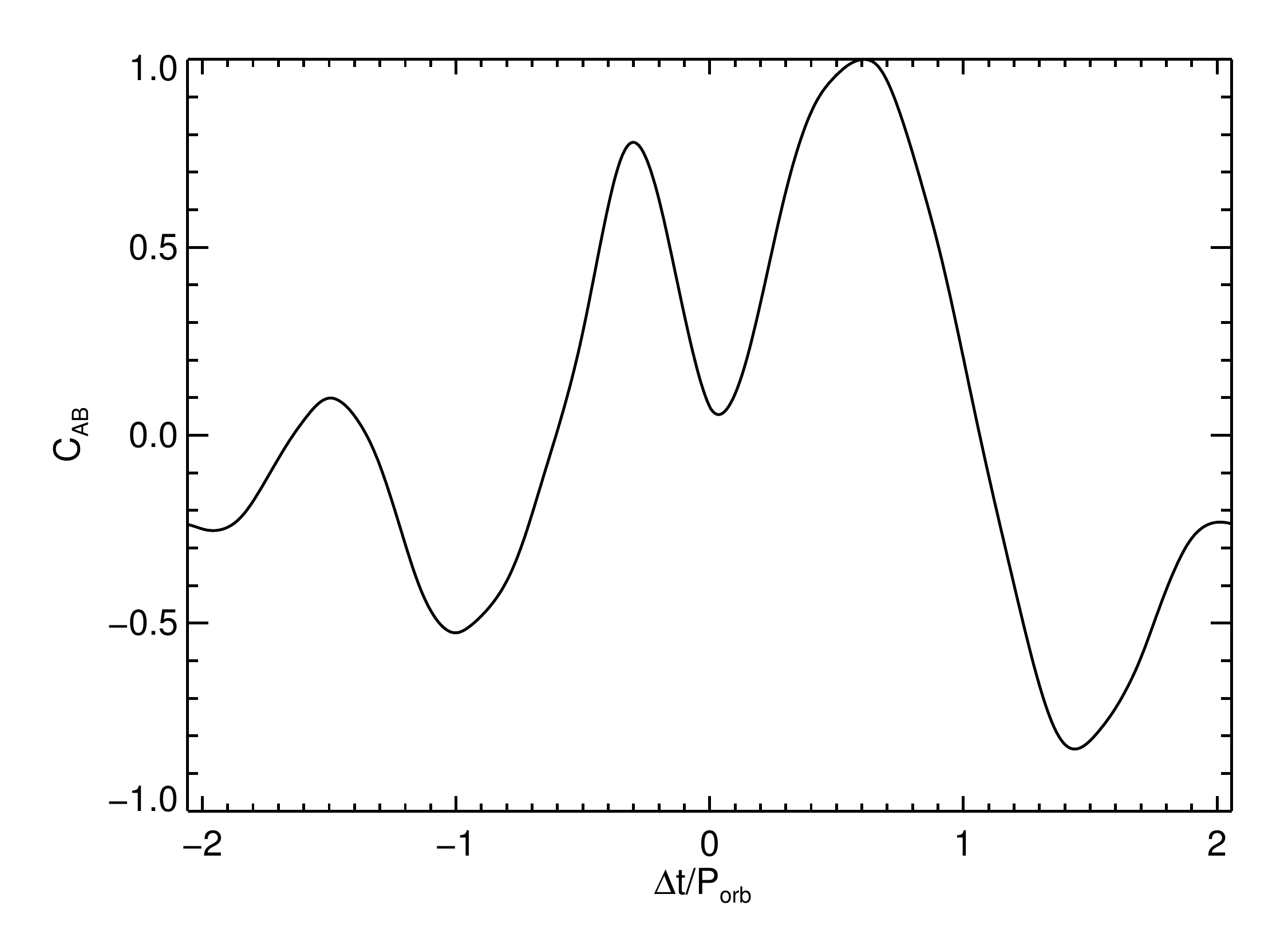}
\includegraphics[width=0.48\columnwidth]
{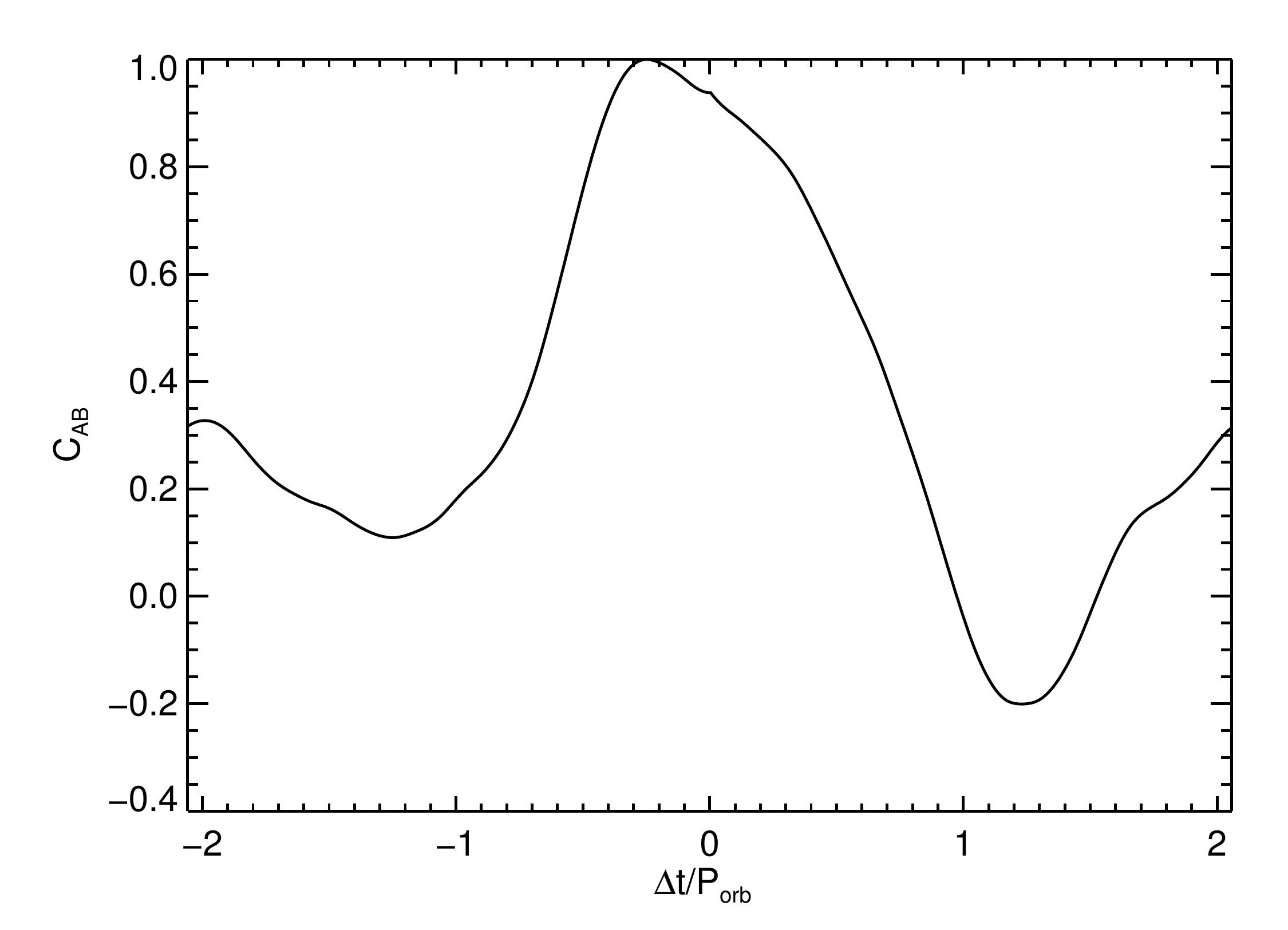}
\includegraphics[width=0.48\columnwidth]
{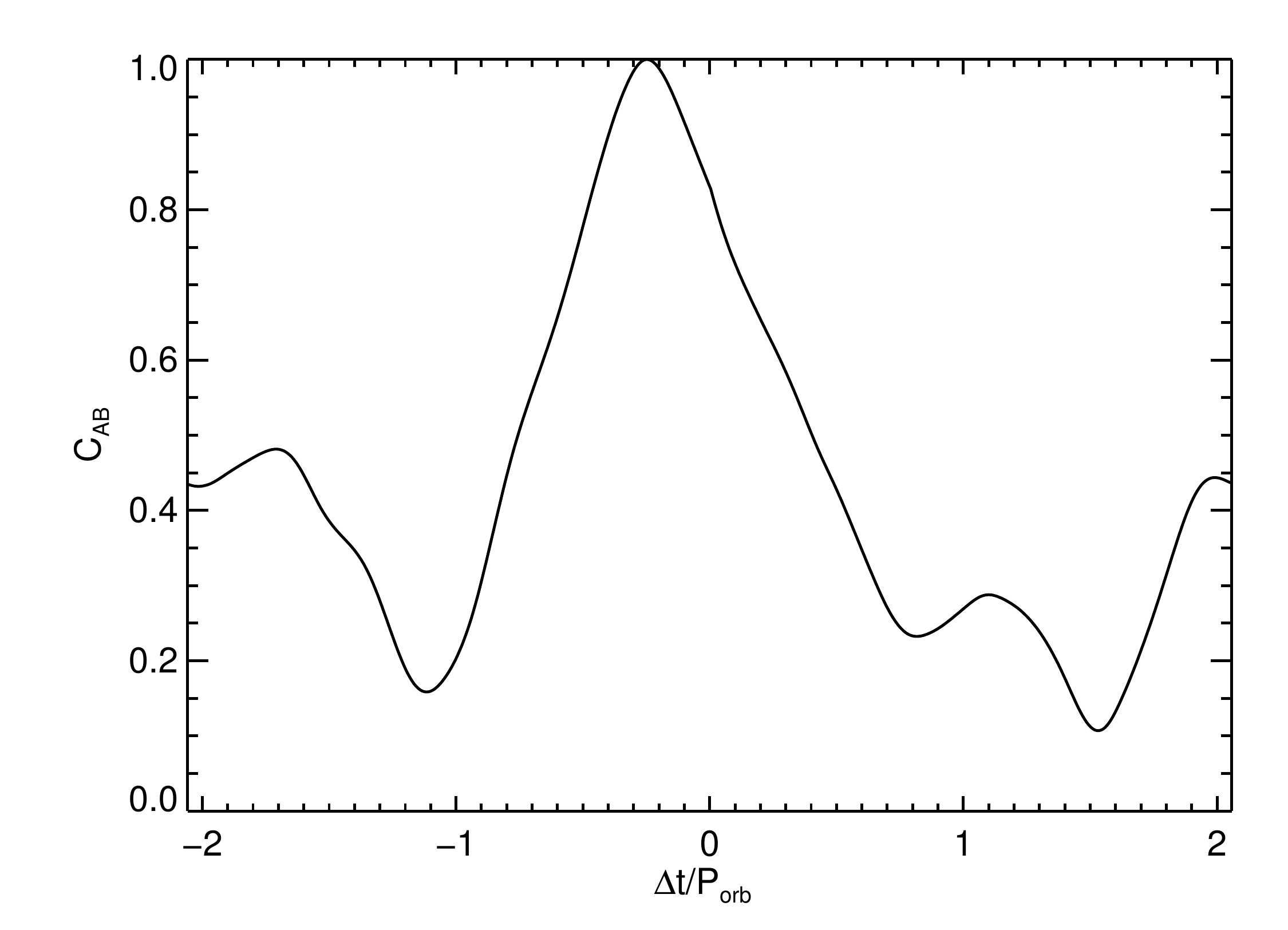}
\includegraphics[width=0.48\columnwidth]
{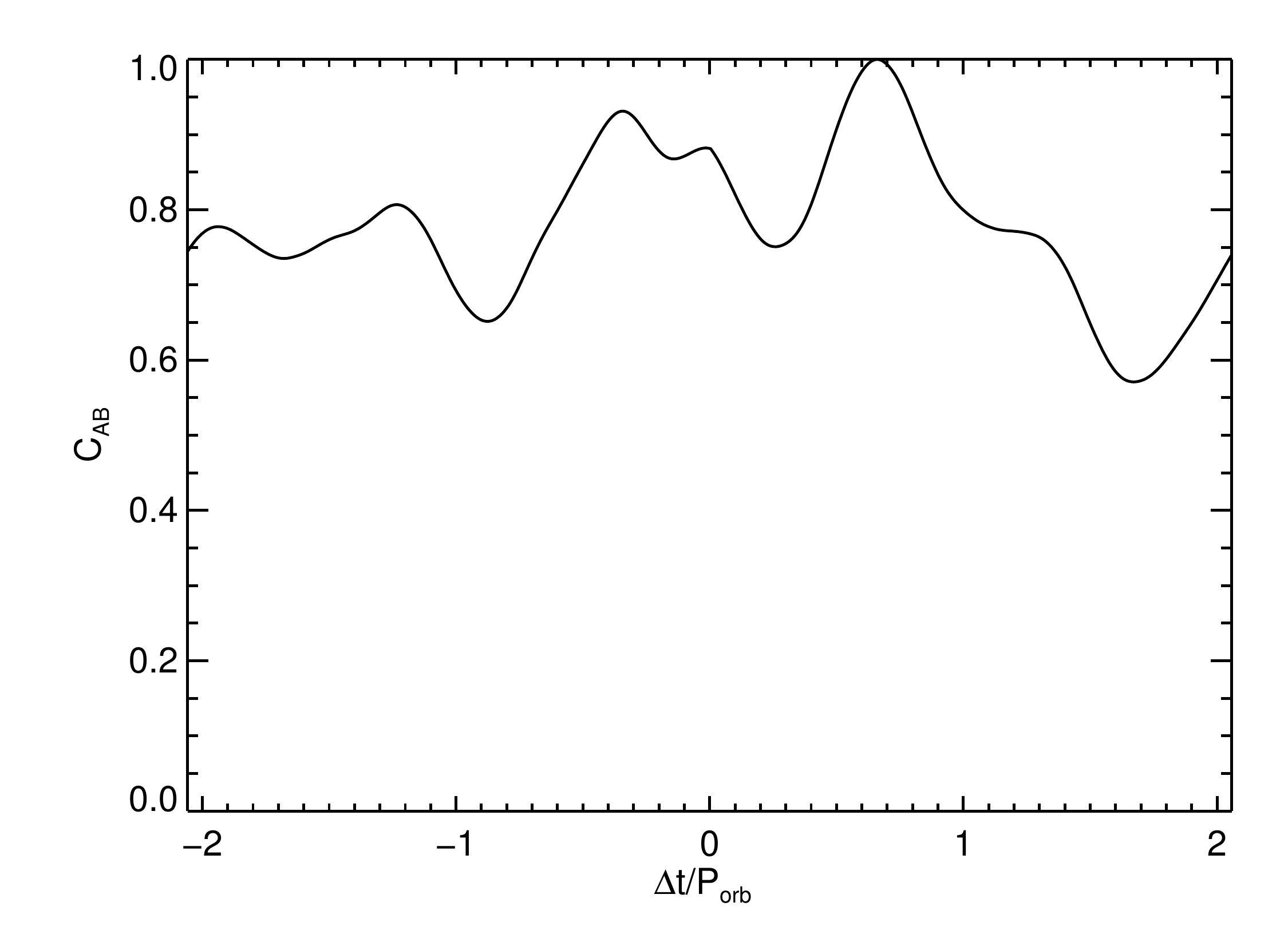}
\end{center}
\caption[]{Integrated cross-correlation, $\left< \left< \left< C_{F^z \; W^m_{r \phi}}(\Delta t) \right> \right> \right>$ calculated as described in \S\ref{power_spectra} time-averaged between orbits $10-12$ (top left), $12-14.5$ (top right), $14.5-16$ (bottom left) and $16-19$ (bottom right) where the orbital period is measured at $r=15r_S$. The cross-correlation function is defined such that if $\left< \left< \left< C_{F^z \; W^m_{r \phi}}(\Delta t) \right> \right> \right>$ is maximized at negative (positive) $\Delta t$, then fluctuations in $F^z(r,\phi,t)$ lead (trail) fluctuations in $W^m_{r \phi} (r,\phi,t)$.}
\label{flux_stress_t_cint} 
\end{figure}

The data of Figure \ref{flux_stress} demonstrates that each of the rings listed above exhibit flux-stress relationships that are broadly similar. The majority of each ring is threaded by significant vertical flux, $F^z > 1$, a result we have confirmed by direct inspection of simulation data. There is a broad range ($3-4$ orders of magnitude) of accretion stresses associated with a given $F^z$. Nevertheless, the mean accretion stress associated with a given vertical flux, $\left\{ \alpha^m_g (F^z) \right\}$ displays a behavior that is broadly consistent with the expectation described above, namely that for weak vertical fields threading a cell, the mean accretion stress is approximately independent of the strength of the vertical field, whilst for strong vertical fields threading a cell, the mean accretion stress is roughly proportional to the strength of the vertical field. Notably, the transition between these two regimes takes place where the wavelength of the vertical field MRI is resolved by approximately $5$ cells, consistent with the results of \cite{Hawley:1995}, where it was found that approximately this number of cells per fastest unstable mode was necessary to reproduce numerically the expectation for the linear growth rate arising from analytic theory. This result in particular gives us confidence that there is a physical relationship between vertical flux and accretion stress, as we do not need to resort to arguments regarding the growth rate of long wavelength modes \citep{Sorathia:2010}.

\begin{figure}
%\leavevmode
\begin{center}
\includegraphics[width=0.48\columnwidth, viewport=0 0 480 445,clip]
{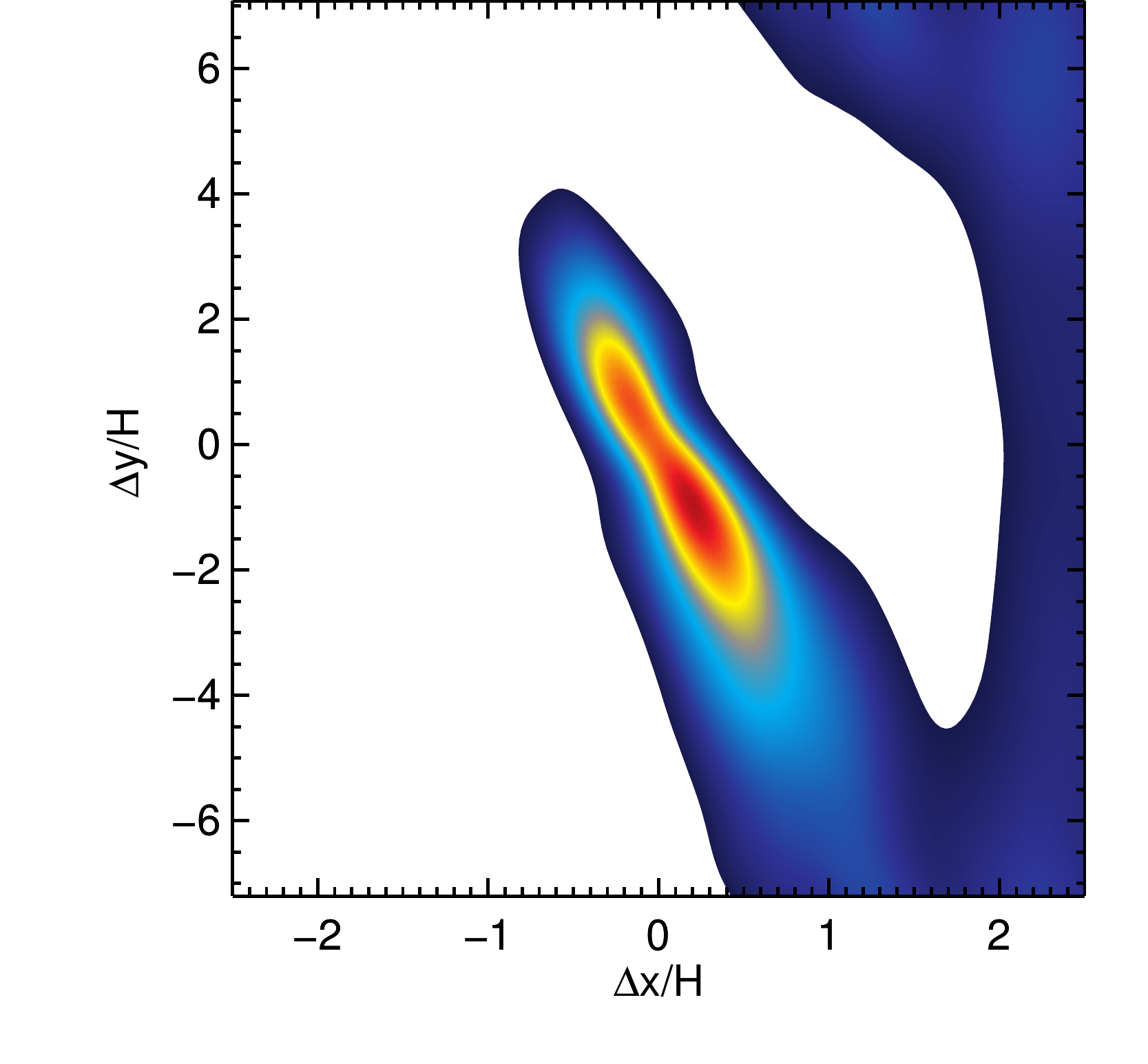}
\includegraphics[width=0.48\columnwidth, viewport=0 0 480 445,clip]
{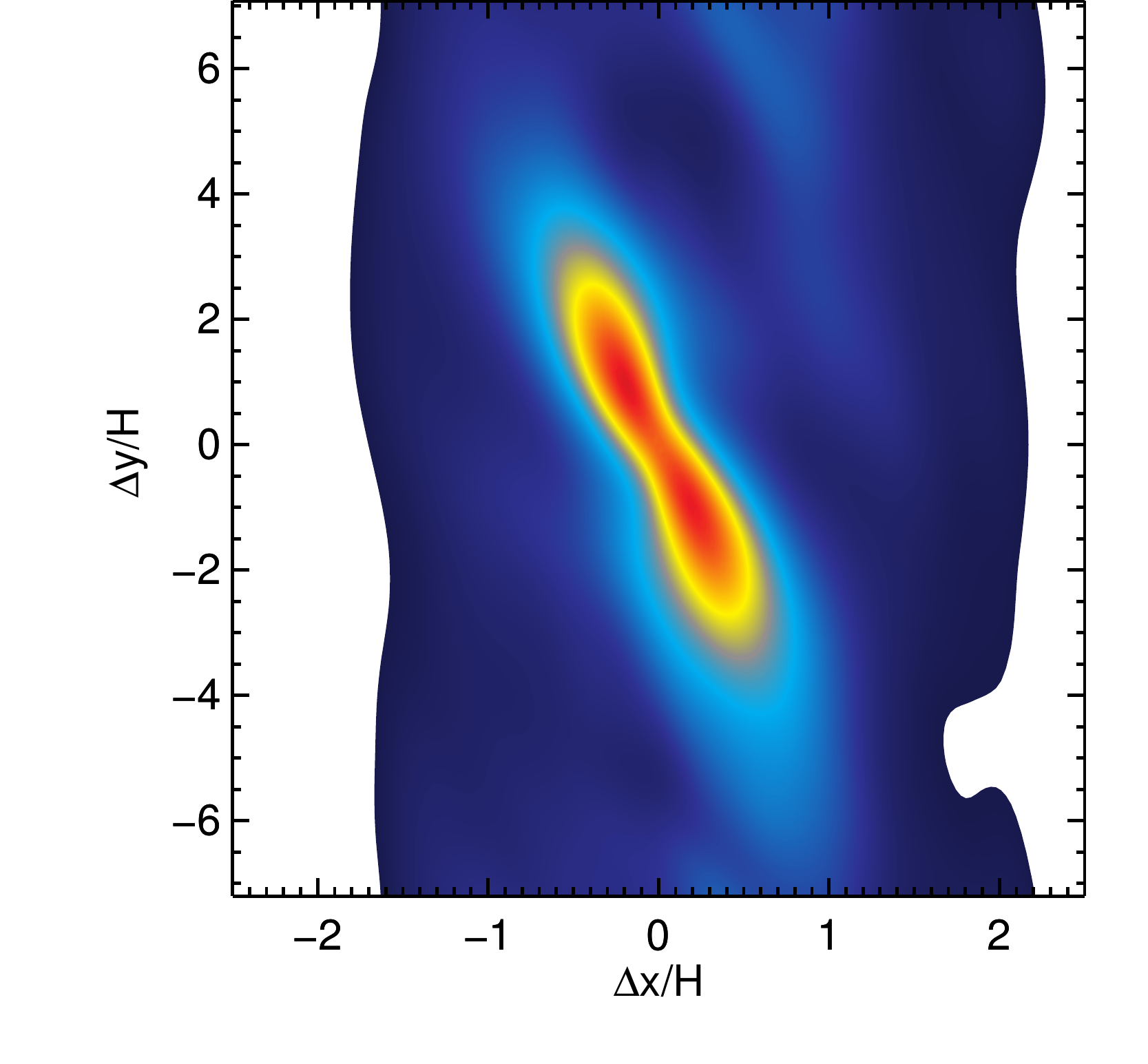}
\includegraphics[width=0.48\columnwidth, viewport=0 0 480 445,clip]
{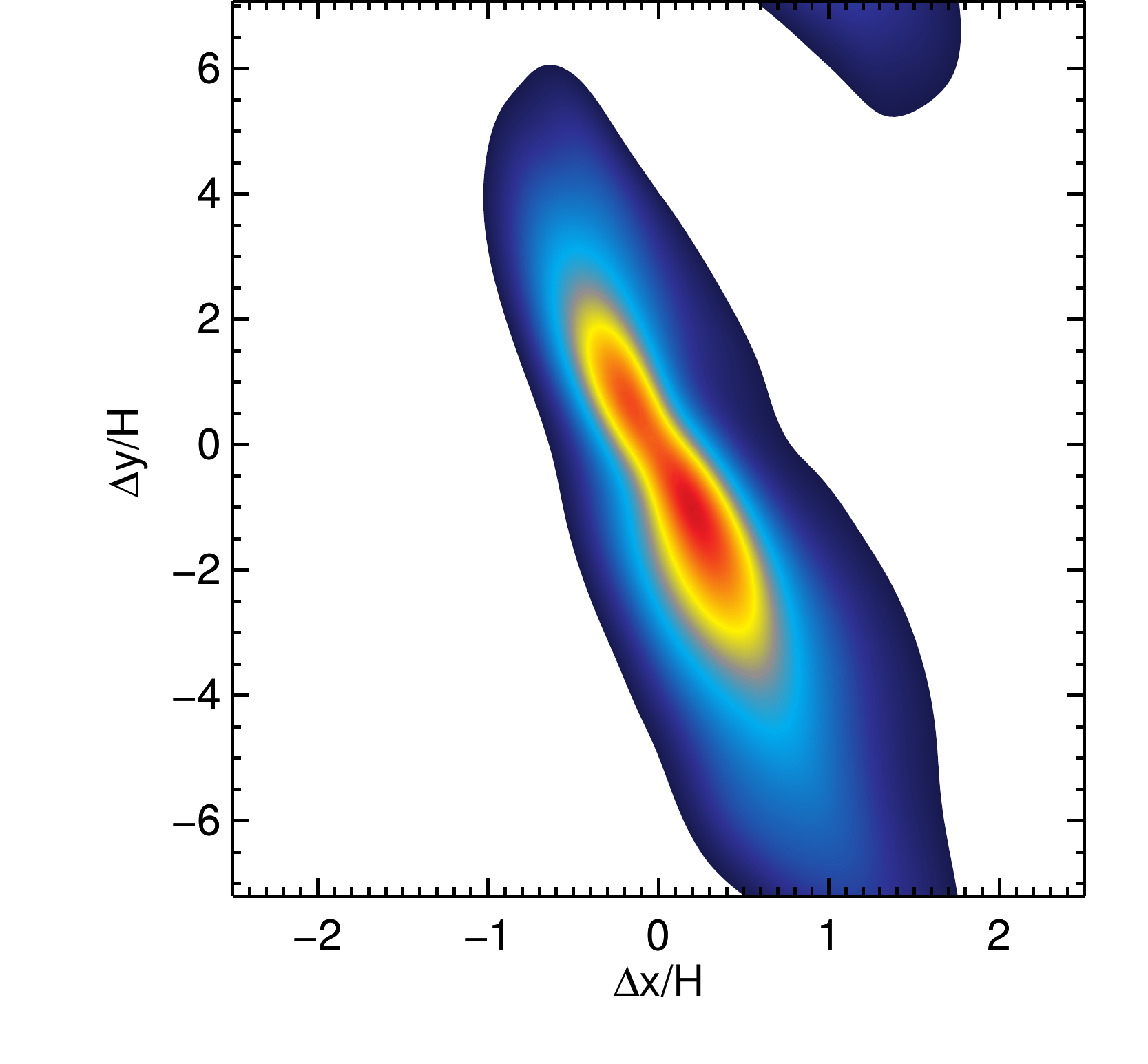}
\includegraphics[width=0.48\columnwidth, viewport=0 0 480 445,clip]
{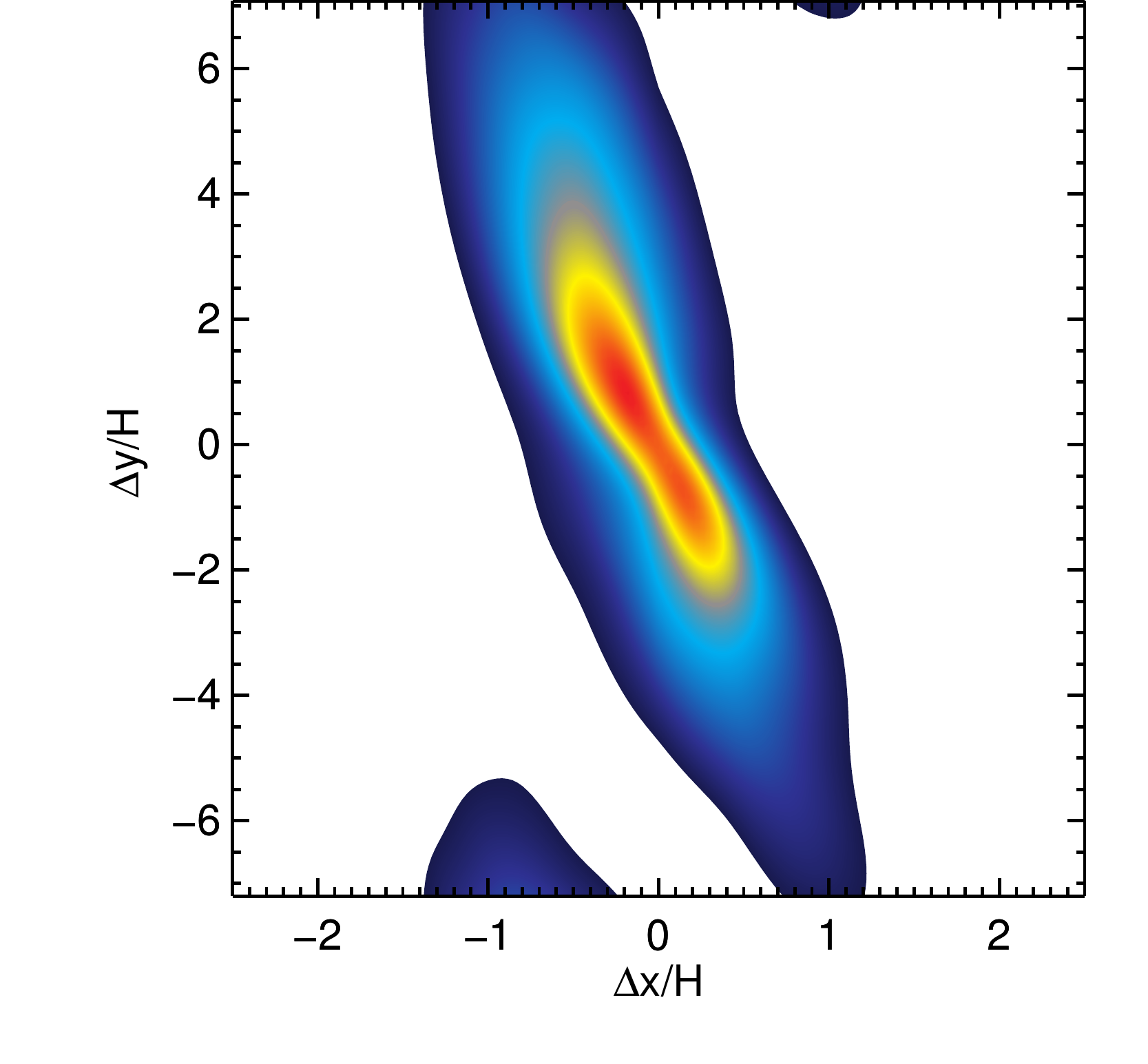}
\includegraphics[width=0.96\columnwidth]{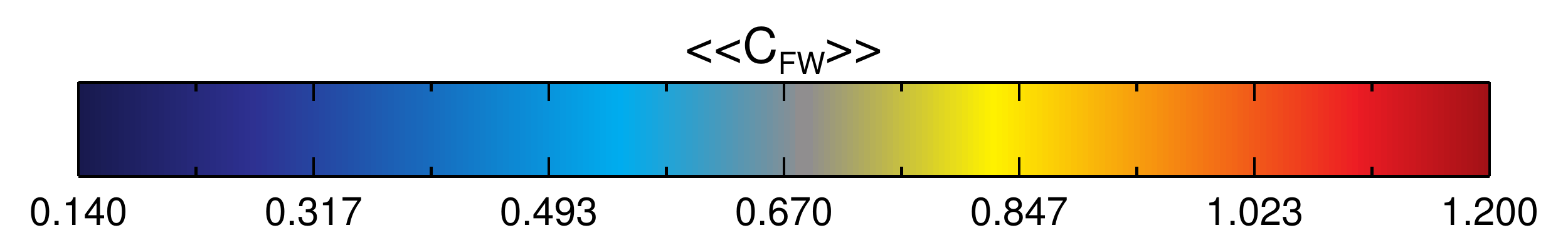}
\end{center}
\caption[]{Structure of $\left< \left< C_{F^z \; W^m_{r \phi}}(\Delta x, \Delta y, \Delta t) \right> \right>$ at $\Delta t = 0$ calculated as described in \S\ref{power_spectra} and time-averaged between orbits $10-12$ (top left), $12-14.5$ (top right), $14.5-16$ (bottom left) and $16-19$ (bottom right) where the orbital period is measured at $r=15r_S$.}
\label{flux_stress_xy_corr} 
\end{figure}

The preceding discussion suggests that there is a local relationship between vertical flux and accretion stress operating within the disk. It does not, however, describe the sense of that relationship, i.e does a fluctuation in the vertical flux lead to a fluctuation in the accretion stress or vice versa? Determining which of these possibilities is the case is clearly crucial in deciding whether or not this flux-stress relationship is physical. \cite{Sorathia:2010} use co-moving wedges to calculate the temporal correlation between vertical flux and accretion stress, where the azimuthal velocity of the fluid is used to track a given co-moving wedge between timesteps. This approach however, assumes that fluctuations in the magnetic field propagate purely in the azimuthal direction with the fluid rotation velocity, which may not be appropriate in a turbulent magnetized disk. We instead address the causal relationship between vertical flux and accretion stress using the three-point cross-correlation function approach outlined in \S\ref{power_spectra}, an approach which bypasses the need to consider a co-moving wedge as it calculates the correlation between all points within the domain of the correlation function simultaneously. The cross-correlation functions are calculated over the radial region $5 \le r/r_S \le 11$ using vertically integrated data within the accretion disk body ($|Z| \le H$) for $2$ orbits at the center of the radial domain, $r=8r_S$ using $250$ frames. We time-average the cross-correlation functions at four different times within the evolution, chosen to coincide with oscillations in the dynamo cycle discussed in \S\ref{dynamo}, namely orbits $10-12$, $12-14.5$ $14.5-16$ and $16-19$ where the orbital period is measured at $r=15r_S$. Each time-average is computed using $70$ dumps per orbit at $r=15r_S$.

%\newpage

Figure \ref{flux_stress_t_cint} shows the total amplitude of the cross-correlation function as a function of the orbital time at $8r_S$ for each of these different periods in the dynamo cycle.  Here, biasing of the correlation function to negative (positive) $\Delta t / P_{orb}$ indicates that fluctuations in the vertical flux (accretion stress) lead fluctuations in accretion stress (vertical flux). Clearly, at orbits $12-14.5$ and $14.5-16$, we see that fluctuations in the vertical flux lead fluctuations in the accretion stress by approximately $0.2 P_{orb} (r=8r_S)$, a period consistent with the results of \cite{Simon:2009} for the typical lifetime of a turbulent fluctuation. The causal nature of the relationship between vertical flux and accretion stress at orbits $10-12$ and $16-19$ is less clear. Here, the correlation functions are double peaked, with the peaks located on either side of $\Delta t = 0$. This result suggest that at these points in the dynamo cycle, fluctuations in the vertical flux lead to fluctuations in the toroidal (or radial) magnetic field, which results in a fluctuation in the accretion stress. The toroidal (or radial) magnetic field fluctuation then leads to a new vertical flux fluctuation some time $\Delta t \sim 0.6 P_{orb} (r=8r_S)$ later, a process reminiscent of the simple model for a magnetic dynamo in an accretion disk described by \cite{Tout:1992}. These results would suggest then that during orbits $10-12$ and $16-19$, we should see rapid rearrangement of the magnetic field as vertical, toroidal (and presumably radial fields) couple together, whilst during orbits $12-14.5$ and $14.5-16$, the structure of the magnetic field should be more stable. Figure \ref{bphi_xz} suggests that this is indeed the case, at orbits $12-14.5$ and $14.5-16$, we see a relatively ordered toroidal magnetic field within the simulation domain, whilst at orbits $10-12$ and $16-19$, the toroidal magnetic field is somewhat less structured.

A final probe of the flux-stress relationship within the disk body is the shape of the cross-correlation function on the $(\Delta x, \Delta y)$-plane at zero temporal offset, $\Delta t = 0$, calculated from the same data as used to create the temporal correlation functions shown in Figure \ref{flux_stress_t_cint}. These data are shown in Figure \ref{flux_stress_xy_corr}, again for $10-12$, $12-14.5$ $14.5-16$ and $16-19$ where the orbital period is measured at $r=15r_S$. Comparing the data of this figure with that of Figure \ref{xy_corr}, we see that the flux-stress cross-correlation function has approximately the same tilt angle with respect to the $\Delta y$ axis as found for $\left< C_{\rho}(\Delta \mathbf{x}) \right>$, etc and that the correlation length along the major axis lies approximately inside that for fluctuations in the density. Whilst the autocorrelation function in a quantity must be symmetric with its maxima located at zero-offset by definition, no such requirement exists for the cross-correlation function between two quantities. We find that the cross-correlation function between vertical flux and accretion stress is double peaked on the $(\Delta x, \Delta y)$-plane, with the peaks lying along the major axis of the correlation function centered on $|\Delta x| \sim 0.25H$. At orbits  $10-12$ and $14.5-16$, the radially outer peak is greatest in amplitude, whilst at $14.5-16$ the peaks are approximately equal in amplitude and orbits $16-19$, the radially inner peak is greatest in amplitude. Overall, this suggests that there are vertical fieldlines penetrating the disk body which link together adjacent radii in a manner reminiscent of that suggested by \cite{Tout:1996}. Combining this result with those discussed in the preceding paragraph suggests that as toroidal field rises out of the disk body (presumably due to magnetic buoyancy), vertical field is created that penetrates the disk midplane. If this vertical field is well resolved in terms of the fastest unstable mode of the vertical field MRI, then we measure an enhanced accretion stress associated with the operation of the vertical field MRI which also creates new toroidal and radial magnetic fields within the disk. These new magnetic fields eventually become buoyantly unstable and the process repeats. This process is strongly reminiscent of the dynamo model described by \cite{Tout:1992}, which one would hope could provide a detailed analytic framework in which to describe the results presented here. We leave such calculations to future work.

\section{Summary, Discussion and Conclusions}\label{conclusion}

Global simulations of magnetized thin accretion disks are required to study the physics of the MRI on large scales (both spatial and temporal), and to make predictions for the structure of turbulent accretion disks that can be tested observationally. To be useful, such calculations must ideally be computed at resolutions which approach that of local simulations, which has only recently become feasible. Here, we have presented an analysis of a simulation of a global, magnetized, thin ($H/R\simeq0.07$) accretion disk designed to investigate the properties of MRI-driven turbulence. The simulation was initialized with a moderately strong net toroidal field 
\citep[similar to, e.g.][]{Hawley:2002,Fromang:2006,Beckwith:2008a}, but it rapidly loses memory of its initial conditions and reverts to a state that is consistent with models initialized with zero net flux. The computation used a second order Godunov scheme with accurate fluxes at a poloidal resolution comparable to moderately well-resolved local simulations \cite[see e.g.][]{Simon:2010}. Our algorithmic choices have been shown to capture the linear growth stage of the MRI accurately \citep{Flock:2009}, and likely yield improved accuracy at fixed spatial resolution over prior simulations. The results allow us to make a quantitative assessment of the structure and locality of the resulting turbulence, and inform a qualitative discussion of the implications for observations and simplified models of disk dynamos.

Our results for the locality of MRI-driven disk turbulence suggest a nuanced picture, in which some aspects of the turbulence are well-described by a local model, whereas others require a global treatment. From an analysis of the spatial two-point correlation functions, we find that accretion disk turbulence, whilst subsonic, contains significant correlations on scales $>H$. The longest correlation lengths exist within the density, for which $\lambda_{maj} = 3H$, followed by the magnetic energy ($\lambda_{maj} = 2H$) and then the kinetic energy ($\lambda_{maj} = 1H$), implying that the largest scales within the turbulence are controlled by the density. Prominent spiral density waves are observed in these simulations and the correlation length along the major axis of the density correlation function suggests that it is these structures that set the size of turbulent fluctuations within the disk \cite[see e.g.][]{Nelson:2010}. If so, then this implies that correctly capturing the formation of these structures is essential to understand the properties of turbulence within the disk. Across a range of spatial scales, the fluctuations in the magnetic and kinetic energies on the poloidal plane are arranged in a fashion consistent with expectations arising from homogeneous isotropic turbulence \cite[i.e. $|k|^{-11/3}$, see e.g.][]{Hawley:1995}.

Although the spatial power spectrum of the turbulent fields is consistent with a simple incompressible turbulence model, the temporal behavior evidences greater complexity. At the most basic level, structures on large scales within the turbulence have lifetimes significantly longer than structures on small scales, as one might expect. However, we also observe that the {\em structure} of the variability varies significantly with spatial scale, breaking the self-similar assumption that underlies simple turbulence models. We find that temporal fluctuations in the magnetic field exhibit different properties at large ($|k|H/2\pi < 1$) versus small ($|k|H/2\pi > 1$) spatial scales. This is in contrast to the kinetic energy, where similar temporal power spectra are found at both spatial scales. We tentatively attribute this behavior to the lack of a clean separation between the energy injection scale and the dissipation scale, which can lead to a non-local (in $k$ space) transfer of energy in MHD turbulence \citep{Lesur:2010}. Unfortunately, determining robustly the range of scales over which the MRI taps the shear energy of the disk is not possible given our resolution, or with any resolution feasibly attainable in a global calculation. Large shearing boxes remain the best numerical setups for studying such questions.

For comparison with real systems, the most basic diagnostic of the properties of accretion disk turbulence is the magnitude of the time-averaged accretion stress, which can be inferred observationally from modeling of thermally unstable disks in dwarf novae and X-ray binaries. For systems where the disk is gas pressure dominated, and hot enough that ideal MHD applies, observational estimates suggest $\langle \langle \alpha \rangle \rangle \approx 0.1$ \citep{Lasota:2001,King:2007}. We measure a value $\langle \langle \alpha \rangle \rangle \approx 2.5 \times 10^{-2}$ from our simulations, that is larger than that derived from most prior calculations without net vertical field, but still formally inconsistent with observations. In principle, this discrepancy could point to a physical effect (disks in binaries could be threaded by, or spontaneously develop, net vertical fields), but it could also be a numerical artifact (the convergence of global simulations has not been demonstrated), or be due to a flawed comparison between simulations and models of outbursting disks computed using classical disk theory. Further work is needed to address each of these possibilities. The vertical distribution of the accretion stress is approximately constant within $\sim 2H$ of the disk midplane. This is consistent with the results of well-resolved vertically stratified shearing box models \citep{Simon:2010}, but is in contrast to previous global simulations \cite[see e.g.][]{Fromang:2006,Sorathia:2010} where a double peaked stress profile was measured. We have found that approximately $80\%$ of the total accretion stress is located at toroidal angular scales $>40^\circ$. We attribute the contrasts in the total accretion stress and vertical stress distribution between this work and \cite{Fromang:2006,Sorathia:2010} as being due the use of toroidal angular domains of $\pi/4$ and $\pi/3$ by these authors respectively.
This leads us to regard the use of toroidal domains of extent $\pi/2$ \emph{essential} in order to correctly capture the physics of angular momentum transport driven by the MRI.

Measurements of the turbulent velocity field in disks \citep{Horne:1994,Carr:2004} are potentially more powerful probes of disk physics than single-point comparisons of measured and simulated accretion stress, provided that a separation of turbulent motion from other non-Keplerian flow is possible. Protoplanetary disks currently represent the most promising observational targets \citep{Hughes:2011}.  In the ideal MHD limit, we find that the vertical distribution of velocity fluctuations steepens from $\sim0.1c_s$ in the disk midplane to $\sim0.2c_s$ in the corona. This implies that magnetized turbulence within disks is characterized by turbulent line widths that are between $0.1$ and $0.2$ of the local sound speed. The fluctuation amplitude that we observe is consistent with previous calculations of MHD turbulence in protoplanetary disks \cite{Fromang:2006}. It also matches recent observations of such systems \cite{Hughes:2011}, although the importance of non-ideal MHD effects for protoplanetary disks  means that a quantitative comparison requires more realistic simulation work than that presented here.

Largely for reasons of numerical convenience, our simulation utilized a pseudo-Newtonian potential that results in an innermost stable circular orbit near the inner boundary of the disk. We studied the structure of the disk near and within the ISCO, whose detailed properties are important for observational attempts to measure the spin of accreting black holes \citep{Zhang:1997,Brenneman:2006,Done:2007,Steiner:2010}. We found consistency between the measured radial turbulent accretion stress distribution within the disk, and the expectations of models that assume a stress free condition at the ISCO. At and inside the ISCO, accretion stresses are due to large scale correlations in radial and toroidal magnetic fields and the flow dynamics is controlled by `flux-freezing' rather than turbulence. Our results support a picture in which, for thin disks, it is the level of net vertical flux at the ISCO that is crucial in determining the stress levels there. If the net vertical flux is small or zero, then stresses are negligible, whereas significant non-zero vertical flux is associated with stresses that could be of observational importance \citep{Agol:2000}. A number of simulations -- both fully relativistic and pseudo-Newtonian -- lend credence to this scenario \citep{Reynolds:2001,Beckwith:2008b,Penna:2010,Noble:2010}.

Numerical simulations -- whether they be local or global -- cannot follow disks over the very long timescales that are characteristic of many interesting observational phenomena. It is therefore important to understand whether there are features of simulated disks that can be abstracted for use in simpler models of disk evolution or disk dynamos. One interesting question is whether or not a local connection between vertical flux and accretion stress \citep{Hawley:1995,Pessah:2007} persists in global simulations \citep{Sorathia:2010}.  We find that such a relationship \emph{does} exist within the turbulence and that provided the vertical flux threading a given cell is sufficiently strong (here determined by the criterion that the wavelength of the fastest unstable mode of the vertical field MRI associated with the flux threading a given cell is resolved), the vertical flux acts as an accurate predictor of the accretion stress. We further find, that in a time-average sense, the majority of the disk body is threaded by vertical fluxes that are well-resolved by this criterion. By use of two-point space-time cross-correlation functions between the vertical flux and the accretion stress, we find that causal sense of the flux-stress relationship depends on which point in the dynamo cycle the correlation function is calculated. When the toroidal field is well-ordered on the poloidal plane, there is a causal connection between vertical flux and accretion stress. Less ordered toroidal field configurations are associated with causal vertical flux-accretion stress that are less well-defined. Speculatively, this behavior suggests that as toroidal field emerges from the disk body during the dynamo cycle, vertical fields thread the disk body for sufficient lengths of time  that they become unstable to the vertical field MRI and thereby determine the local accretion stress. This process is reminiscent of the dynamo model described by \cite{Tout:1992}.

\newpage

\section*{Acknowledgements}
We thank Mitch Begelman, Chris Reynolds, Kareem Sorathia, Jim Stone and John Hawley for useful discussions and advice. We also thank an anonymous referee for useful comments on an earlier draft of this paper. This work was supported by the NSF under grant numbers AST-0807471 and AST-0907872, and by NASA under grant numbers NNX09AB90G and NNX11AE12G. This research was supported in part by the NSF through TeraGrid resources provided by Texas Advanced Computing Center under grant number TG-AST090106. The authors acknowledge the Texas Advanced Computing Center at The University of Texas at Austin for providing HPC and visualization resources that have contributed to the research results reported within this paper.

\label{lastpage}

\end{document}